\documentclass[prd,preprintnumbers,showkeys,nofootinbib]{article}
\usepackage{bm, graphicx,amsmath,amssymb,epsf}

\newcommand{\tr}{{\rm tr}} 

\begin{document}

\title{\Large \bf  The triple Pomeron vertex in large-$\bm N_c$ QCD\\ and the pair-of-pants  topology }
\author{\large J.~Bartels and M.~Hentschinski 
\bigskip \\
{\it  II. Institute for Theoretical Physics, Hamburg University, Germany} \\
 }
\maketitle

\begin{abstract}
  We investigate the high energy behavior of QCD for different surface
  topologies of color graphs.  After a brief review of the planar
  limit (bootstrap and gluon reggeization) and of the cylinder
  topology (BFKL) we investigate the $3 \to 3$ scattering in the
  triple Regge limit which belongs to the pair-of-pants topology.  We
  re-derive the triple Pomeron vertex function and show that it
  belongs to a specific set of graphs in color space which we identify
  as the analogue of the Mandelstam diagram.
\end{abstract}

\section{Introduction}
\label{sec:intro}

Starting from the classical paper by 't Hooft \cite{'tHooft:1973jz}
the large $N_c$ limit of gauge theories has remained in the center of
attention for more than 25 years. In high energy QCD-phenomenology for
instance, the large-$N_c$ expansion proves its usefulness as a
simplifying tool, if one attempts to resum higher order contributions,
enhanced by large logarithms. With the general structure of the color
factors being often too complex, the large $N_c$ limit allows for
resummation with an acceptable reduction of accuracy.

A new attraction of the large $N_c$ expansion results nowadays from
the fact that it organizes the color structure of Feynman diagrams in terms
of topologies of two-dimensional surfaces which resemble the loop
expansion of a closed string theory. With the advance of the AdS/CFT
correspondence \cite{adscft} which, in the limit of large $N_c$,
connects $N=4$ Super-Yang-Mills theory with a closed string theory in
Anti-de-Sitter space and with string coupling proportional to
$N_c^{-1}$ this idea is more prevailing than ever.

The leading term of the large $N_c$ expansion is given by color
factors that have the topology of the sphere or equivalently the
plane, and one usually refers to these leading contributions as
'planar' diagrams.  Planar diagrams contribute, for example, to
gluon-gluon scattering amplitudes or to multi-gluon production
amplitudes.  On the other hand, there exist also processes where the
$N_c$ leading color factor has not the topology of the plane.  This is, for
instance, the case for the scattering of two electromagnetic currents
or virtual photons at high energies where the center of mass energy
squared $s$ is far bigger than the momentum transfer squared and the
virtualities of the photons.  In this case, the interaction between
the scattering currents is mediated by the exchange of gluons between
two quark loops, and,   with quarks in QCD in the fundamental representation of the gauge group $SU(N_c)$,
 the leading color factor has the topology of a
sphere with two boundaries, a cylinder.  Within perturbative QCD such
processes, at leading \cite{bfkl} and next-to-leading order accuracy
\cite{nlabfkl}, are described by the BFKL-Pomeron, which is a bound
state of two reggeized gluons. Higher order unitarity corrections are
expected to involve bound states of more than 2 reggeized gluons, so
called BKP-states \cite{bkp}, which for the cylinder topology have
been found to be integrable \cite{integrable}.
 
In the present study we go one step further and consider the class of
process whose leading color factor has the topology of a sphere with
three boundaries.
\begin{figure}[htbp]
  \centering
  \begin{minipage}[h]{5cm}
    \includegraphics[width=4.5cm]{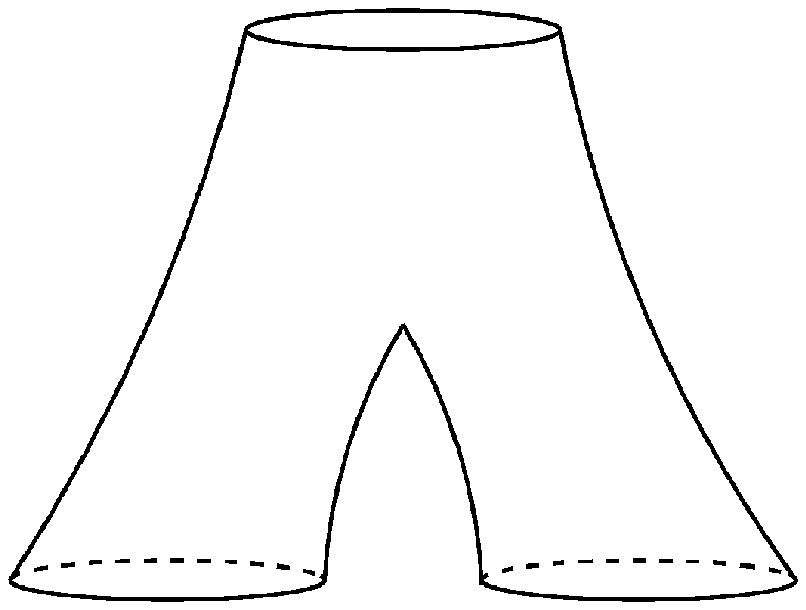}
  \end{minipage}
  \begin{minipage}[h]{.5cm}
$\,$
  \end{minipage}
\begin{minipage}[h]{8.5cm}
\parbox{4cm}{\includegraphics[height = 4.5cm]{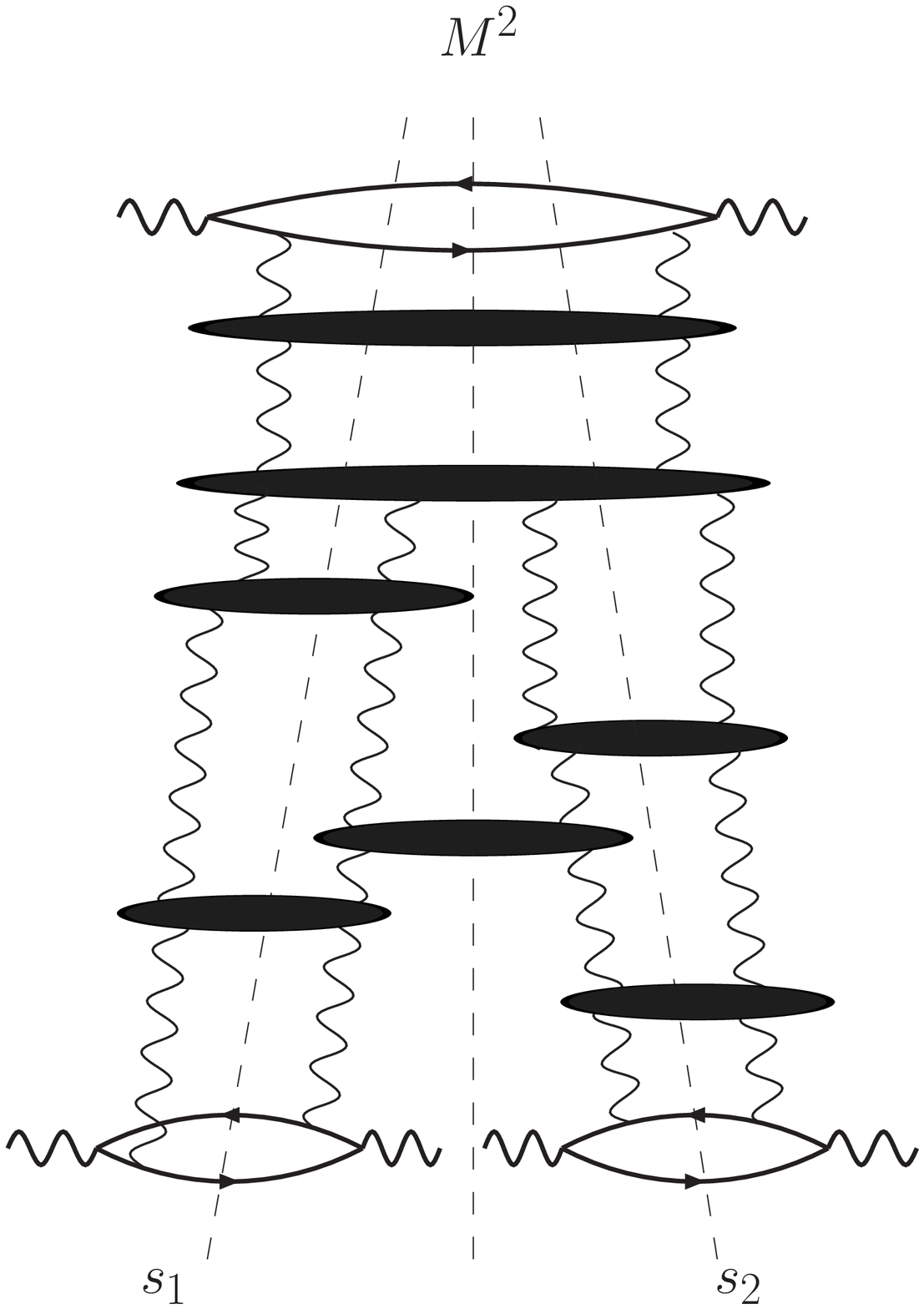}}
\parbox{4cm}{\includegraphics[height = 4.5cm]{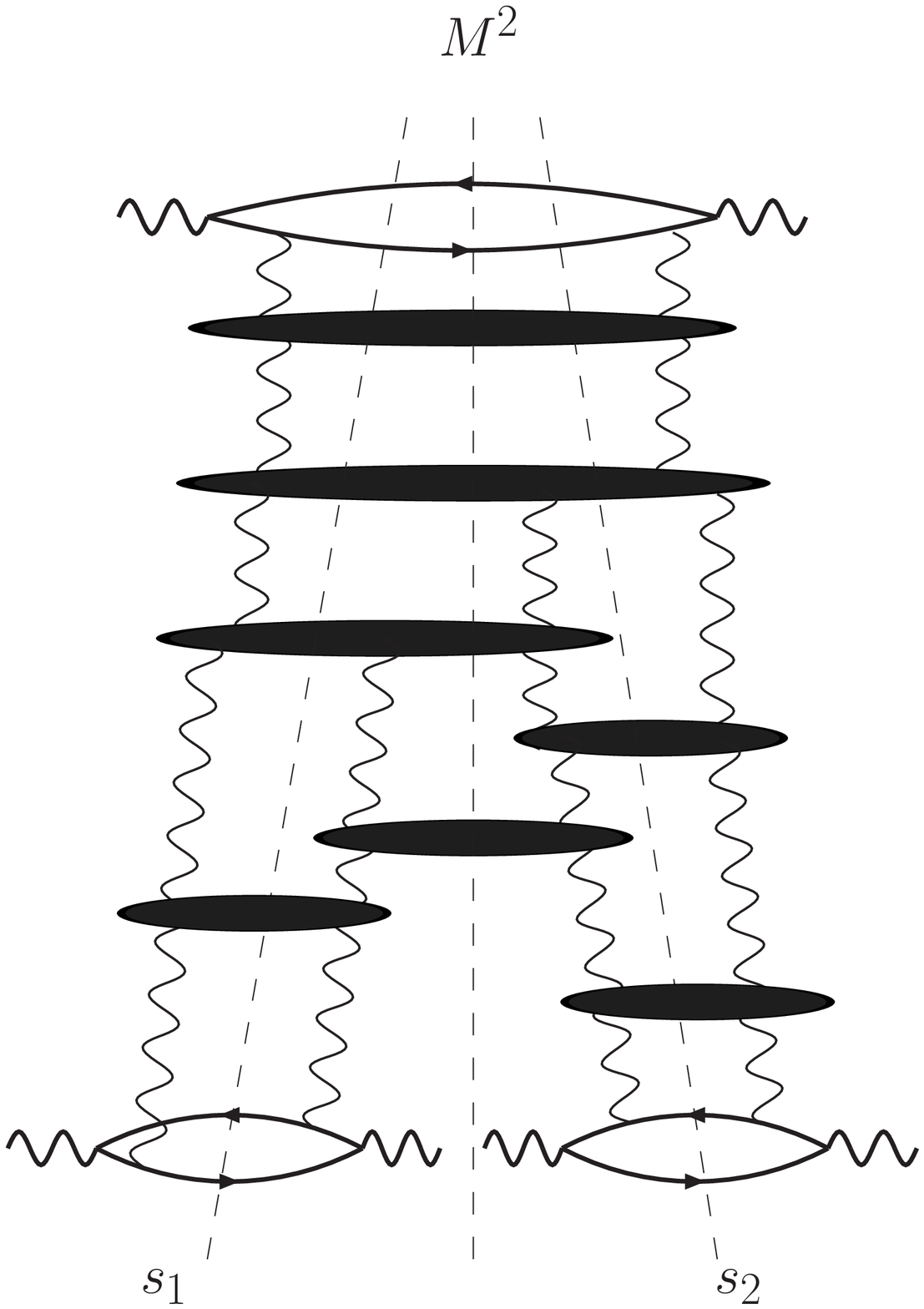}}
  \end{minipage}
\\
\begin{minipage}[h]{5cm}
\caption{\small The 'pair of pants' topology}
  \label{fig:trouserr}
\end{minipage}
 \begin{minipage}[h]{.5cm}
$\,$
  \end{minipage}
\begin{minipage}[h]{8.5cm}
\caption{\small Typical contributions to the triple energy discontinuity for scattering of three virtual photons }
\label{fig:triple_discon}
\end{minipage}
\end{figure}
Such a surface is depicted in Fig.\ref{fig:trouserr} and it is usually
referred to as the topology of a 'pair of pants'. As a suitable
example we consider the three-to-three process which describes the
scattering of a highly virtual photon on two virtual photons in the
triple Regge limit. Similar to the BFKL-Pomeron on the cylinder, we
resum all contributions which are maximally enhanced by logarithms of
energies. The class of diagrams selected in this way is illustrated in
Fig.\ref{fig:triple_discon}: the three photon impact factors introduce
three boundaries, and thus belong to the topology depicted in
Fig.\ref{fig:trouserr}.  Compared to the simple BFKL cylinder, the new
feature is the splitting of one cylinder into two cylinders which is
related to the 'triple Pomeron vertex' \cite{Bartels:1994jj}.  Within
the AdS/CFT correspondence, the electromagnetic current corresponds to
the R-current, and the high energy behavior of the 6-point amplitude,
on the string side, is expected to exhibit the triple graviton vertex.

In QCD this $3 \to 3$ process has been investigated before for $N_c=3$
in \cite{Bartels:1994jj}, the triple Pomeron vertex has been derived,
and it has been shown to be invariant under two-dimensional M\"obius
transformations \cite{Bartels:1995kf}.  It is straightforward to
repeat the analysis for arbitrary $N_c$ and to take the large-$N_c$
limit of these calculations.  The system of four reggeized gluons and
the triple Pomeron vertex have been further investigated particularly
for the limit $N_c \to \infty$ in \cite{Braun:1995hh,Braun:1997nu}.
However the connection of these results with the expansion in terms of
topologies of two-dimensional surfaces is not apparent.  Instead of
the pair-of-pants, the color factor rather seems to correspond to
three disconnected cylinders.  In the present paper, we therefore
demonstrate that by summing only diagrams with the topology of the
pair-of-pants, Fig.\ref{fig:trouserr}, one obtains the result of the
large-$N_c$ limit of \cite{Bartels:1994jj}.  In particular we find
that the 'reggeizing' and 'irreducible' terms of \cite{Bartels:1994jj}
can be attributed to distinct classes of diagrams on the surface of
the pair-of-pants.

Our paper is organized as follows. In section \ref{sec:elastic} we
briefly review the planar and cylinder topologies. In the high energy
limit, the planar diagrams satisfy the bootstrap condition of the
reggeizing gluon, whereas the cylinder diagrams lead to the famous
BFKL amplitude.  In section 3 we turn to the analytic form of the $3
\to 3$ scattering amplitude and give a general description of the
diagrams which need to be summed. In Section 4 we study the color
lines on the surface of the pants, arriving at the definition of two
distinct classes of diagrams in color space (named 'planar' and
'non-planar'). We also review the different momentum space kernels
which describe the interactions of reggeized gluons. In section 5 we
sum the 'planar' diagrams and arrive at the 'reggeized' amplitudes
introduced in \cite{Bartels:1994jj}, whereas in the section 6 we
investigate the 'non-planar' diagrams and re-derive the triple Pomeron
vertex of \cite{Bartels:1994jj}. In section 7 we briefly summarize our
results, and the final section 8 contains a few conclusions.

\section{Elastic amplitudes}
\label{sec:elastic}
In this section we begin by recalling the definition of the large
$N_c$ expansion, following the classical paper by 't Hooft
\cite{'tHooft:1973jz}. As a preparation for the study of the
pair-of-pants topology, we then reconsider the simplest examples of
diagrams, whose color factor have the topology of the plane and of the
cylinder. In the Regge-limit they yield the reggeized gluon and the
BFKL-Pomeron, respectively.

\subsection{The large $N_c$ expansion}
\label{sec:NCexp}

To study the large $N_c$ limit as  an expansion of
topologies of the color factor, one is asked to translate the
color factors of a scattering amplitude into the so-called double line notation. To
this end we use the $SU(N_c)$ generators in the fundamental representation, 
$(g^a)^i_j$, in the normalization\footnote{
  Note that this deviates by a factor $2$ from  
  the standard normalization $\tr(t^a t^b) = \delta^{ab}/2$.} 
$\tr(g^ag^b) = \delta^{ab}$, and we make use of the identity
\begin{align}
  \label{eq:structure_const}
f^{abc}  
&=
 \frac{1}{i\sqrt{2}}\big[\tr(g^ag^bg^c) - \tr(g^cg^bg^a) \big]  .
\end{align}
For all inner gluon lines in the adjoint representation, the indices
can then be expressed in terms of (anti-)fundamental indices by means
of the Fierz identity
\begin{align}
  \label{eq:fierz}
    (g^a)^i_j(g^a)^k_l  =  \delta^i_l \delta^k_j - \frac{1}{N_c} \delta^i_j\delta^k_l.
\end{align}
Making use of a diagrammatic notation, where a Kronecker-delta is
represented by a single line with an arrow, indicating the flow from
the upper to the lower index,
\begin{align}
  \label{eq:kronecker_delta}
\delta^i_j = \parbox{2cm}{\includegraphics[width=2cm]{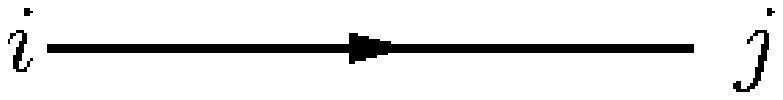}},
\end{align}
the structure constants $f^{abc}$ are  expressed as
\begin{align}
  \label{eq:structure_const1}
    f^{a_1a_2a_3} (g^{a_1})^{i_1}_{j^1} (g^{a_2})_{j_2}^{i^2} (g^{a_3})_{j_3}^{i^3}
 =
\frac{1}{i\sqrt{2}}   \left(\,\, \parbox{1cm}{\includegraphics[height=2cm]{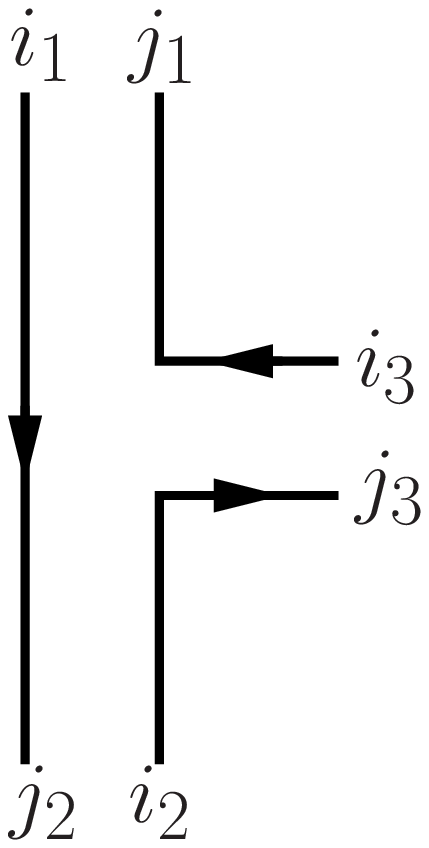}}
    -
    \parbox{1.3cm}{\includegraphics[height=2cm]{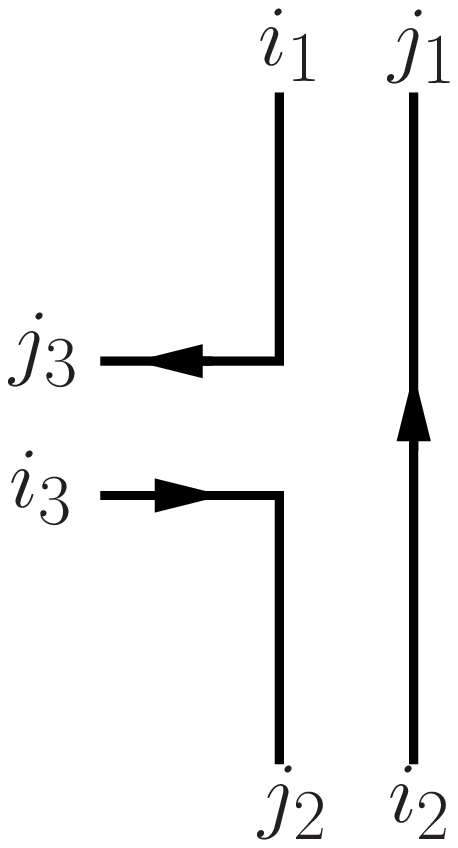}}
\right).
\end{align}
In order to arrive at the large $N_c$ expansion, one drops, in the
Fierz-identity Eq.(\ref{eq:fierz}), the second term. As a consequence,
the gluon is represented by a double line
\begin{align}
  \label{eq:double_line}
\delta^i_l \delta^k_j= \parbox{2cm}{\includegraphics[width=2cm]{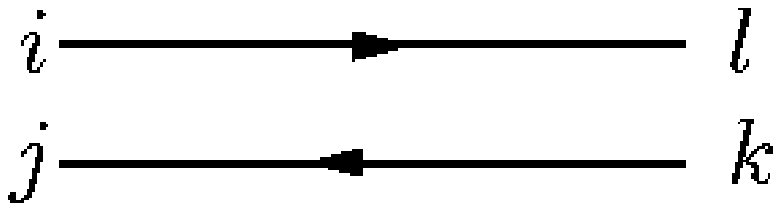}}.
\end{align}
The neglect of second term of the Fierz-identity Eq.(\ref{eq:fierz})
which serves to subtract the trace of the $SU(N_c)$ gluon implies that
we consider an $U(N_c)$ gluon rather than an $SU(N_c)$ gluon. While
this term is suppressed for large $N_c$, dropping this term is only
correct as long as we stay within the leading term of the expansion.
Tracelessness of the $SU(N_c)$ gluon can be taken into account by
introducing an additional $U(1)$ ghost field
\cite{Manohar:1998xv,Matevosyan:2008zz}, which subtracts the trace of
the $U(N_c)$-gluon. Since the $U(1)$ ghost-field commutes with the
$U(N_c)$ gluon-field, it couples only to quarks, but not to gluons.
Within the Leading Logarithmic Approximation, which we will use in the
following, the interaction between scattering objects is mediated only
by gluons, while quarks occur in the coupling of the gluons to the
scattering objects.  Corrections due to the $U(1)$ ghost in this
approximation therefore appear only in low orders of the strong
coupling and are taken into account easily.

Using the double line representation for gluons,
Eq.(\ref{eq:double_line}), and representing quarks in the fundamental
representation by single lines, the color factor of a vacuum Feynman
diagram turns into a network of double and single lines, which can be
drawn on a two-dimensional surface with Euler number $\chi = 2 -2h -b
$, where $h$ is the number of handles of the surface, and $b$ the
number of boundaries or holes.  A closed color-loop always delivers a
factor $N_c$. With quarks being represented by single lines, a closed
quark-loop, compared to a corresponding gluon-loop, is $1/N_c$
suppressed and leads always to a boundary.  For an arbitrary vacuum
graph $T$ one arrives at the following expansion in $N_c$
\begin{align}
  \label{eq:vacuumgraph}
T = \sum_{h,b}^\infty N_c^{2 - 2h -b} T_{h,b}(\lambda).
\end{align}
where
\begin{align}
  \label{eq:thooft}
\lambda &= g^2N_c 
\end{align}
is the 't Hooft-coupling which is held fixed, while $N_c$ is taken to
infinity.  The expansion Eq.(\ref{eq:vacuumgraph}) matches the loop
expansion of a closed string theory with the string coupling $1/N_c$.
For $N_c \to \infty$, the leading diagrams are those that have the
topology of a sphere: zero handles and zero boundaries, $h=b=0$. If
quarks are included, the leading diagrams have the topology of a disk,
i.e. the surface with zero handle and one boundary, $h=0, b=1$. The
disk fits on the plane, with the boundary as the outermost edge.
Diagrams with two boundaries and zero handles can be drawn on the
surface of a cylinder, those with three boundaries on the surface of a
pair-of-pants. Boundaries are also be obtained by removing, from the
sphere, one or more points: removing one point, one obtains the disk,
which can be drawn on the plane, and by identifying the removed point
with infinity, the graphs can be drawn on the (infinite) plane.
Removing two points we obtain the cylinder and so on. By definition,
the expansion Eq.(\ref{eq:vacuumgraph}) is defined for vacuum graphs.
However, from the earliest days on \cite{'tHooft:1974hx}, the large
$N_c$-limit has been also applied to the scattering of colored
objects.  In order to consider the topological expansion of an
amplitude with colored external legs, one needs to embed it into a
vacuum graph which then defines the topological expansion of an
amplitude with colored external legs.

In the following it will be convenient to define modified couplings 
\begin{align}
  \label{eq:modified_thooft}
\bar{g} = \frac{g}{\sqrt{2}},\hspace{3cm}  \bar{ \lambda} = 
\bar{g}^2N_c = \frac{g^2N_c}{2},
\end{align}
which absorb the factor $1/\sqrt{2}$. Due to our normalization of $SU(N_c)$
generators which differs from the more standard one, $\tr(t^at^b) =
\delta^{ab}/2$,
such a factor $1/\sqrt{2}$ arises for each quark-gluon coupling
and, because of Eq.(\ref{eq:structure_const1}), also for each   
gluon-gluon coupling.

\subsection{Planar amplitudes: Reggeization of the gluon}
\label{sec:reggeization}

In the present paragraph, we shall discuss the Regge-limit of planar
amplitudes (first addressed in \cite{Braun:1997ax}) which satisfy
the bootstrap condition of the reggeized gluon, the basic ingredient
for the further studies.  In order to study reggeization of the gluon,
one considers the scattering of colored objects.  To be definite, we
consider the scattering of a quark on an anti-quark. The disk-topology
of the color factors becomes apparent, if we connect color lines of
the incoming quark and antiquark with each other and of the outgoing
quark and antiquark (note that, if instead we would connect the color
lines of the ingoing and outgoing quark with each other and of the
incoming and outgoing antiquark, we would discover the cylinder
topology).  As far as the momentum part of the diagrams is concerned,
our method is the following: We consider the Regge limit $s \gg -t$ of
the elastic amplitude and make use of the Leading Logarithmic
Approximation (LLA), where we select all diagrams that are maximally
enhanced by a logarithm in $s$, i.e. that are proportional to
$\frac{\lambda}{N_c}(\lambda \ln s)^n$, and sum them up to all orders
in $\lambda$. Furthermore, in this approximation, all $t$-channel
particles are gluons (quark exchanges are power suppressed).

We start by considering tree- and one-loop-level diagrams within the
LLA, and convince ourselves that these leading order result, for
planar color graphs, satisfy the bootstrap condition and are thus in 
agreement with the reggeization of the gluon.  At
tree level, the (anti-)quarks interact by exchange of a single gluon.
In the standard adjoint notation, the color factor of the diagram is given
by $(g^{a})^i_j (g^{a})^l_k$, which in the double-line notation turns into
$\delta^i_k\delta^l_j$, while the $U(1)$-ghost can be disregarded for
the plane. We therefore have
\begin{align}
  \label{eq:tree_quarks}
 \parbox{1.5cm}{\includegraphics[width=1.5cm]{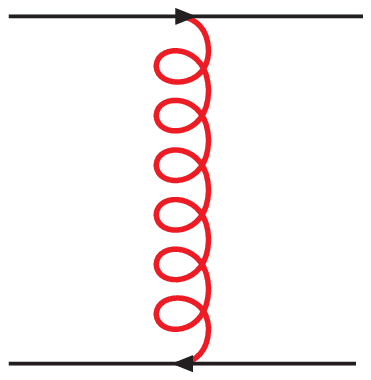}} & = \parbox{1.5cm}{\includegraphics[width=1.5cm]{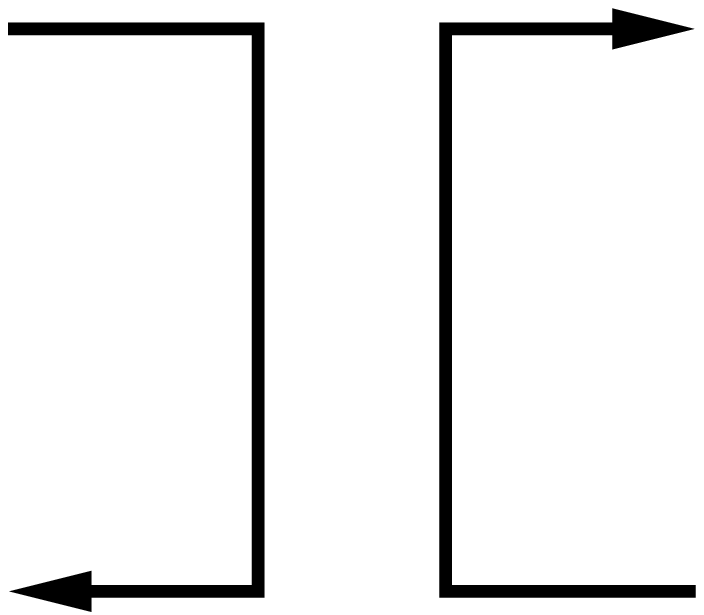}} \times T^{\text{tree}}_{q\bar{q}}(s,t), &
 T^{\text{tree}}_{q\bar{q}}(s,t) &= 2 \bar{g} \frac{-s}{t} \bar{g}.
\end{align}
From now on we adopt the following notation: in a Feynman 
diagram gluons are usually denoted by curly lines, and in this notation the diagram 
is understood to represent both the color factors and the momentum part.
Alternatively, when drawing a diagram with the double line notation for 
each gluon, this diagram represents only the color part, and the momentum 
part has to be written separately. 

\begin{figure}[htbp]
  \centering
  \begin{minipage}{.5\textwidth}
\parbox{1.7cm}{\includegraphics[width=1.7cm]{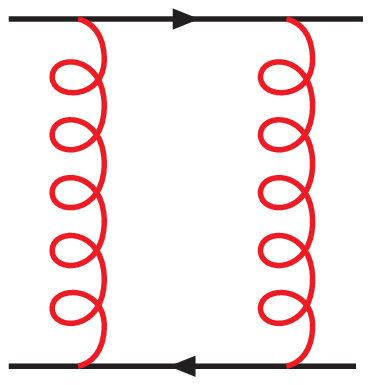}}  : $\quad$
     \parbox{2cm}{\includegraphics[height=1.2cm]{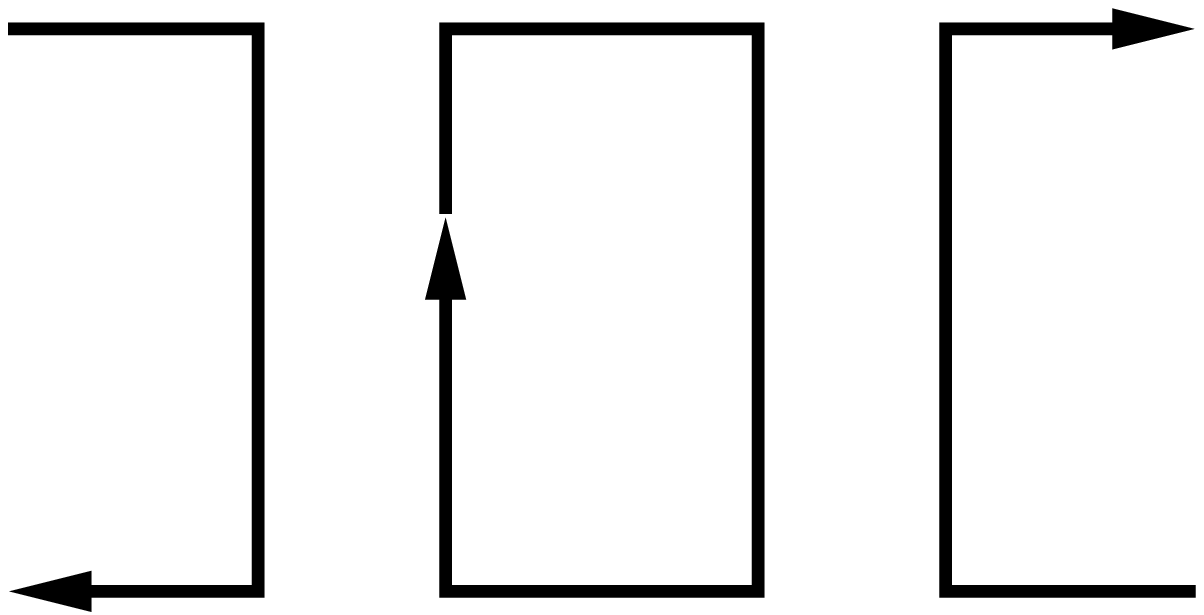}}
= {\large $N_c$}  \parbox{1cm}{\includegraphics[height=1.2cm]{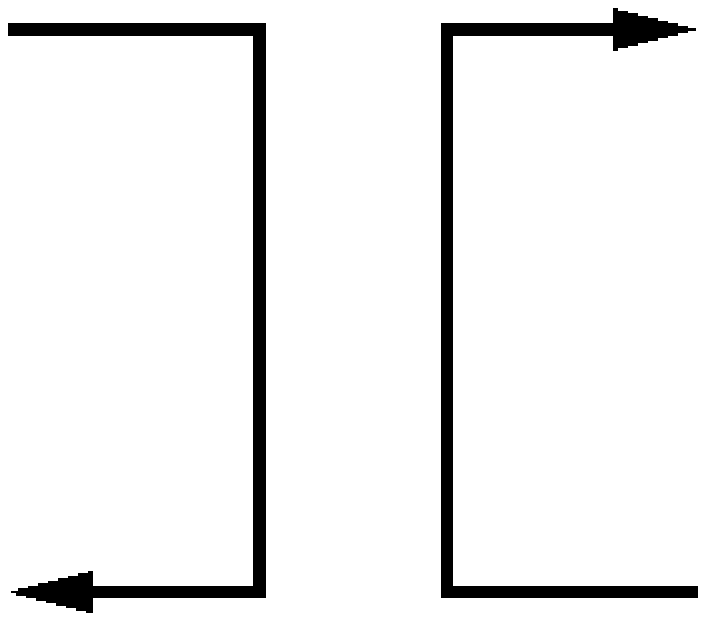}}
  \end{minipage}
  \begin{minipage}{.1\textwidth}
$\,$
  \end{minipage}
  \begin{minipage}{.3\textwidth}
\parbox{1.7cm}{\includegraphics[width=1.7cm]{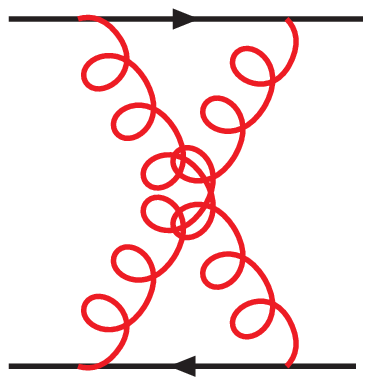}} : $\quad$
     \parbox{1.8cm}{\includegraphics[height=1.2cm]{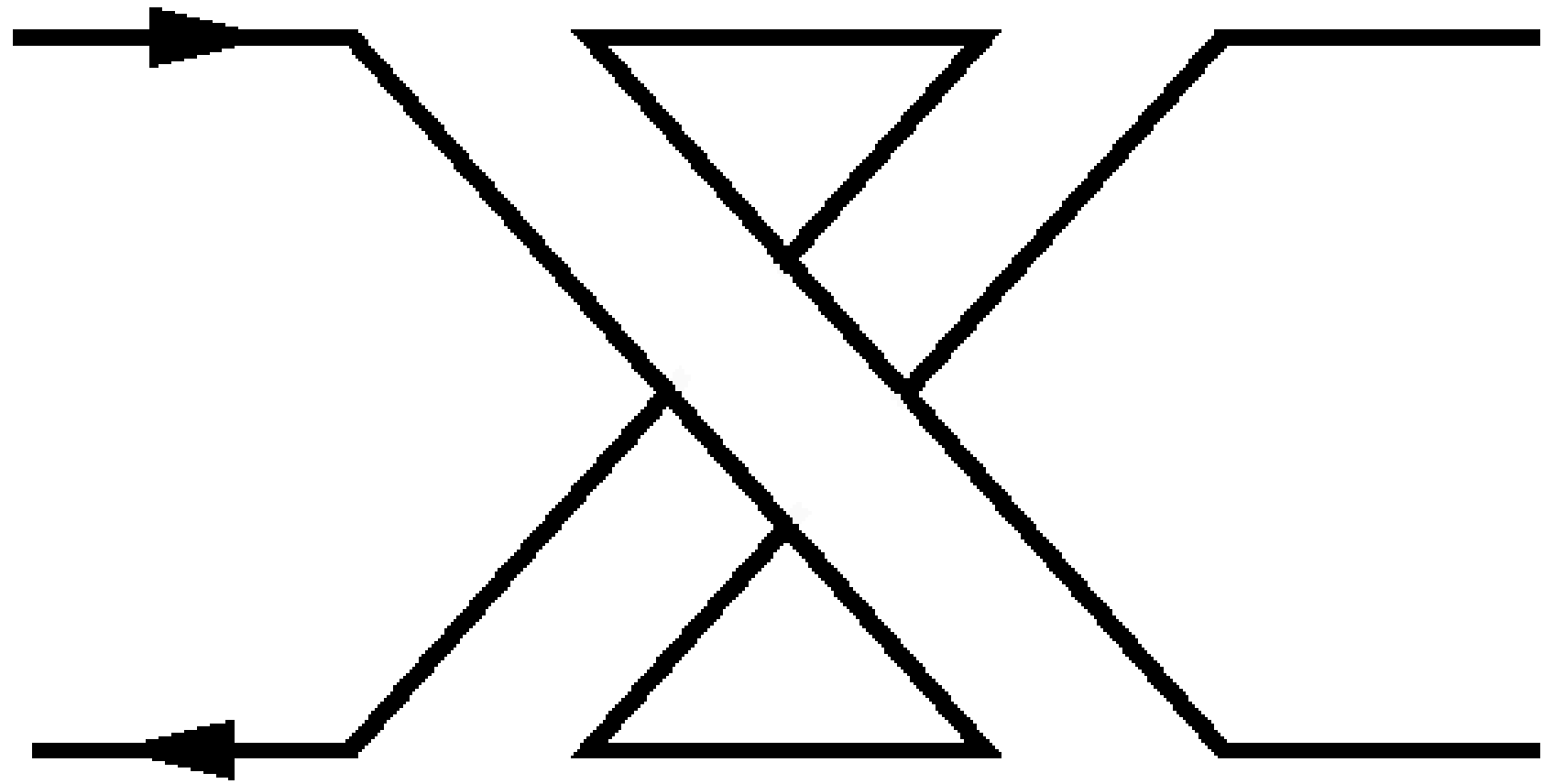}}
  \end{minipage}
  \caption{\small One-loop corrections to  quark-antiquark scattering. 
On the left hand side, 
the color factor of the planar Feynman diagram is $N_c$ times the Born color 
factor. On the right hand side,
the non-planar diagram does not fit onto the plane and is $N_c$ suppressed.}
  \label{fig:col_corr_1gl}
\end{figure}

Turning to higher order graphs, we have the
restriction that, with every insertion of an additional internal loop 
coming with a factor $g^2$, we have to 
produce a closed color loop, yielding a factor $N_c$ which then
combines to the 't Hooft coupling $\lambda = g^2 N_c$. Only such 
higher order term will be included into the planar approximation discussed in this 
section. 
This restriction holds also for other topologies in the expansion
Eq.(\ref{eq:vacuumgraph}): every insertion of an additional gluon
into the Born diagram provides automatically a factor $g^2$ and must
be compensated by a closed color loop in order to stay within the
considered coefficient $T_{h,b}$ of the expansion.  
For our planar loop, for each higher order graph the tensorial structure of the
color factor will be proportional to the one of the Born term.   
As far as the momentum part is concerned, the  1-loop diagrams for
quark-antiquark scattering that contribute within the LLA are shown  in Fig.\ref{fig:col_corr_1gl}.
For both diagrams, the momentum part is of the same order of magnitude,
i.e. it is proportional to $g^2 s \ln(-s)$ and $g^2 u \ln (-u)$,
respectively, with $u \simeq -s$ in the Regge limit. However, 
when counting closed color loops, one sees that only the
first diagram is leading in $N_c$. Also, when closing the color lines of the 
incoming quark and antiquark and of the outgoing particles, 
we observe that for the first diagram the 'closed' color factor, indeed, 
fits on the
disk with $h=0, b=1$, while the second 'closed' color factor has an
additional handle $h=1, b=1$.  We therefore find for the quark-antiquark
scattering amplitude at 1-loop:
\begin{align}
  \label{eq:beta_func}
T_{q\bar{q}}^{\text{1-loop}} (s, t) &= \ln (-s)\beta(t)T_{q\bar{q}}^
{\text{tree}} (s, t)
&& \mbox{with}&
\beta(- {\bf q}^2) &= \bar{\lambda} \int \frac{d^2 {\bf k}}{(2\pi)^3}\frac{- {\bf q}^2}{{\bf k}^2({\bf q} - {\bf k})^2},
\end{align}
where the color factor $N_c$ has already been included into 
the gluon trajectory function, and bold letters denote Euclidean momenta 
perpendicular  to the light-like momenta $p_1$ and $p_2$ of the scattering 
quark and antiquark; in particular  $t = -{\bf q}^2$. Eq.(\ref{eq:beta_func}) can be
understood as the order $\mathcal{O}(g^4)$ term of the expansion of the exchange of
a planar reggeized gluon:
\begin{align}
  \label{eq:qqbar_reggeized}
T_{q\bar{q}} (s, t) &= 2 \bar{g} (-s)^{1 +\beta(t)} \frac{1}{t} \bar{g}. 
\end{align}
For our further analysis we use the following analytic representation of the planar elastic amplitude in the
Regge limit:
\begin{align}
  \label{eq:elastic_rep}
 T_{2\to2}(s,t) =  s\int_{\sigma -i\infty}^{\sigma + i\infty} \frac{d \omega}{2\pi i} s^\omega \xi(\omega) \phi(\omega,t).
\end{align}
The partial wave amplitude $\phi(\omega, t)$ is a real-valued function, 
and phases are contained in the signature factor $\xi(\omega)$: 
\begin{align}
  \label{eq:sigfac_reggeon}
\xi(\omega) = -\pi\frac{e^{-i\pi\omega}}{\sin(\pi\omega)}.
\end{align}
Note the difference from the 'usual' signature factor $\xi \sim {e^{-i\pi\omega}} \pm 1$ 
which contains both right and left hand energy cuts:  
at one loop we have shown that only the Feynman diagram with the $s$-channel 
discontinuity belongs to the planar approximation whereas the $u$ discontinuity is absent. 
For quark-antiquark-scattering, this also holds for higher order terms.
We then take  the discontinuity in $s$ 
\begin{align}
  \label{eq:disc_elastic}
 \Im\text{m}_s  T_{2\to2}(s,t) = \mathrm{disc}_s  T_{2\to2}(s,t) =  \pi s\int \frac{d \omega}{2\pi i} s^\omega  \phi(\omega,t),
\end{align}
which, by unitarity 
\begin{align}
2 \Im\text{m}_s  T_{2\to2}(s,t) = \sum \int T_{2\to n}  T_{2\to n}^*
\label{unitarity}
\end{align}
relates the Mellin transform of the partial wave amplitude
$\phi(\omega, t)$  to the sum of production processes. To
leading order in $g$, the partial wave has the form
\begin{align}
  \label{eq:reggeon_PWlo}
\phi^{(0)}(\omega, t) = \frac{2}{\omega} \phi_{(2;0)} \otimes \phi_{(2;0)}
= \frac{2\bar{g}^2}{\omega t} \beta(t)
\end{align}
where $\phi_{(2;0)} = \bar{g}^2 N_c^{1/2}$ denotes the non-singlet two gluon impact 
factor for quark and antiquark, and the convolution symbol is defined as 
\begin{align}
\otimes = \int\frac{d^2 {\bf k}}{(2\pi)^3 {\bf
    k}_1^2 {\bf k}_2^2}
\end{align}
with ${\bf q} = {\bf k}_1+{\bf k}_2$ and $t= - {\bf q}^2$.
Higher order corrections involve diagrams where 
additionally real gluons are produced. To leading logarithmic
accuracy, real particle production takes place within the
Multi-Regge-Kinematics (MRK), where the produced particles are widely
separated in rapidity.  The leading order diagram, with one additional
$s$-channel gluon, is illustrated in Fig.\ref{fig:sdisc_real3}: 
\begin{figure}[htbp]
   \centering 
   \parbox{4cm}{\includegraphics[height=2cm]{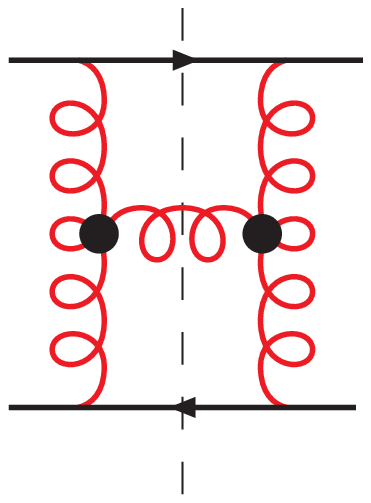}}
   \parbox{2.5cm} { \includegraphics[width=2.5cm]{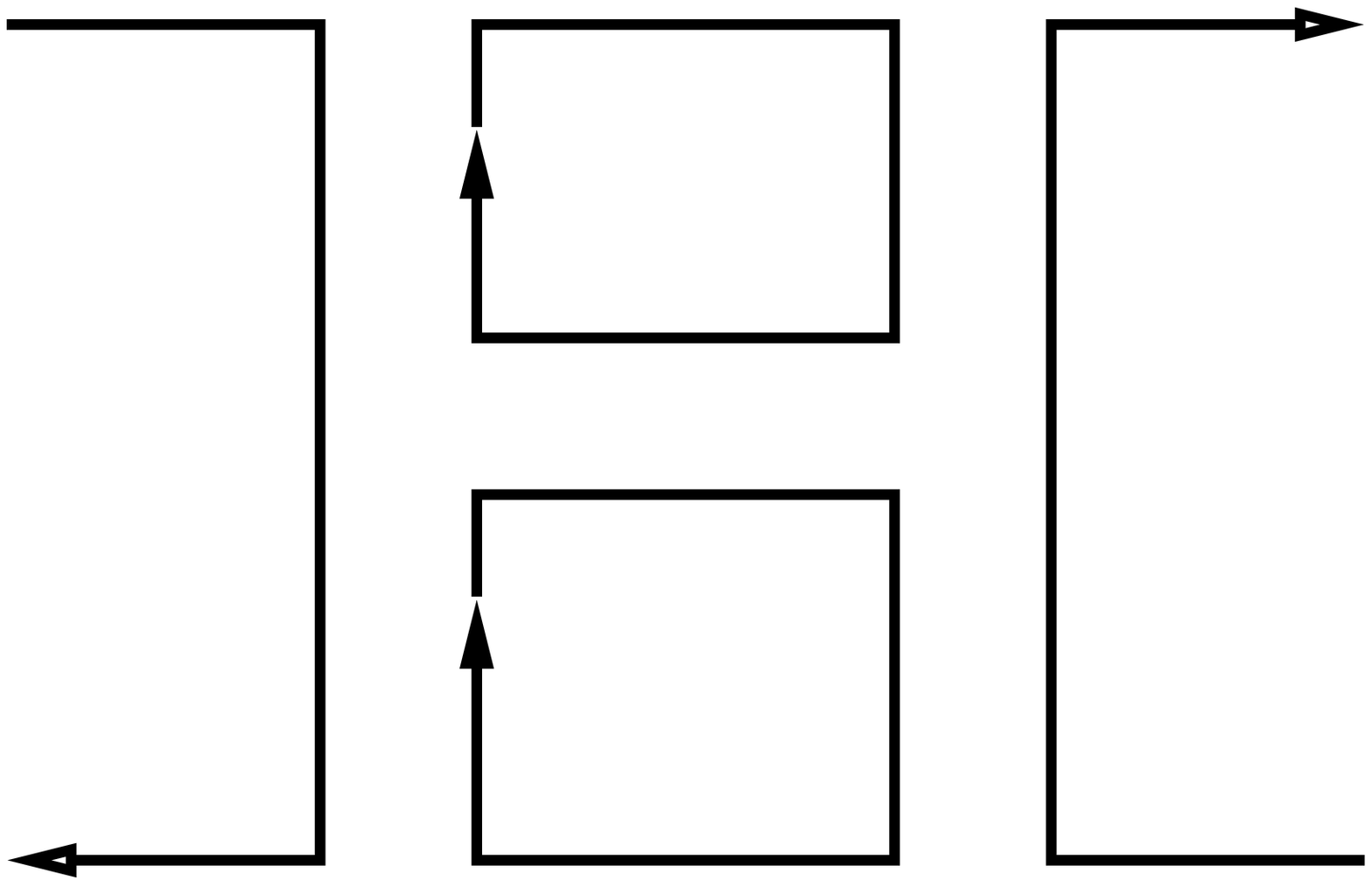}}
   \caption{\small $s$-channel discontinuity with three real particles. To the right  the corresponding planar color factor.}
   \label{fig:sdisc_real3}
\end{figure}
Here the particle production vertex (depicted by a dot) is an effective vertex for the production 
of one real gluon.
This production vertex is build by the following Feynman diagrams:
\begin{align}
  \label{eq:prod_vertex}
\parbox{1.5cm}{\includegraphics[width=1.5cm]{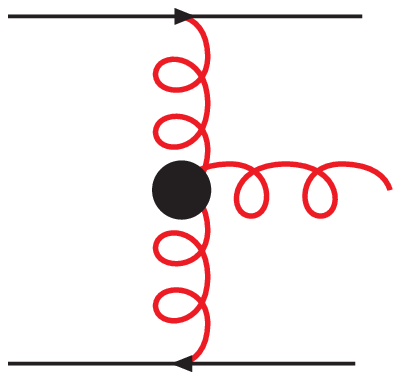}}
        & =         \parbox{1.5cm}{\includegraphics[width=1.5cm]{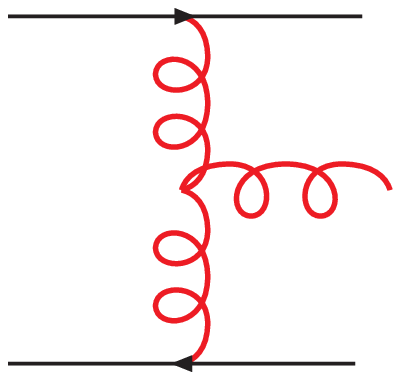}} 
         &+&
         \parbox{1.5cm}{\includegraphics[width=1.5cm]{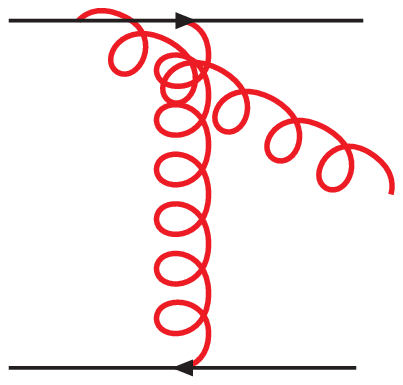}}
         &+ &
         \parbox{1.5cm}{\includegraphics[width=1.5cm]{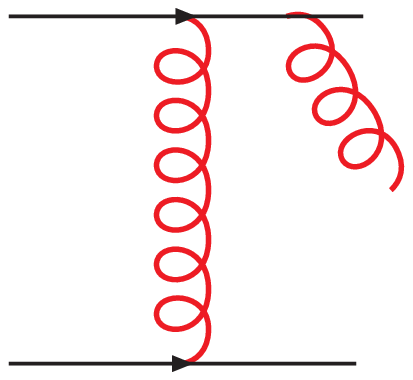}} 
         &+&
         \parbox{1.5cm}{\includegraphics[width=1.5cm]{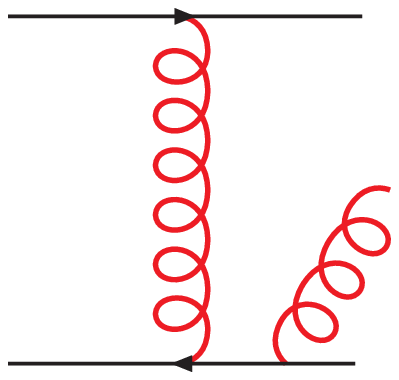}} 
         &+&
         \parbox{1.5cm}{\includegraphics[width=1.5cm]{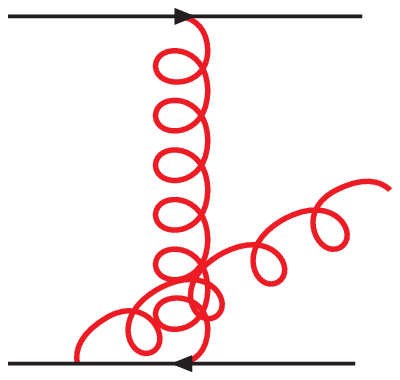}} 
\notag \\
&\frac{1}{i}\bigg(
         \parbox{1cm}{ \includegraphics[width=1cm]{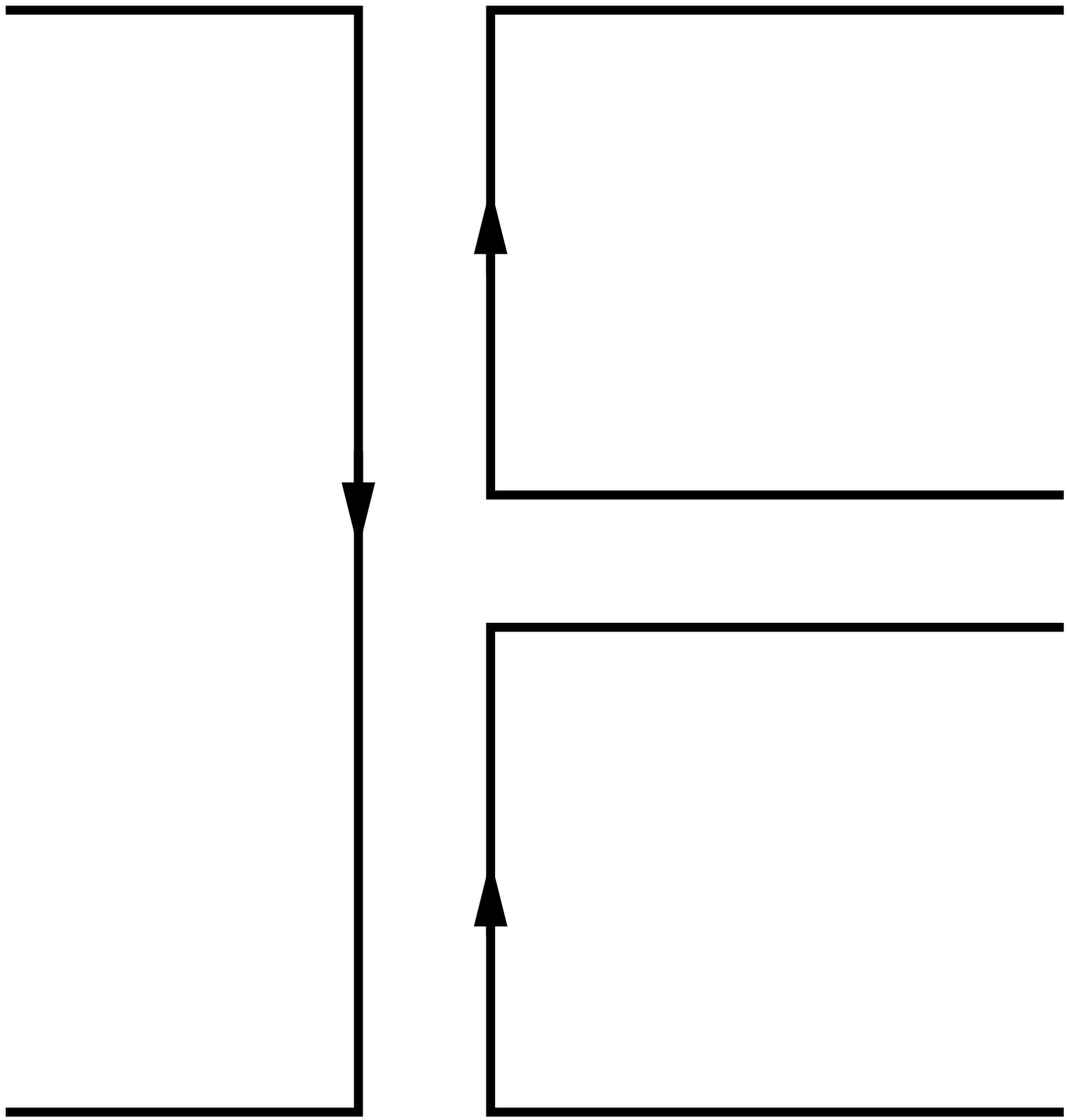}}
         -
         \parbox{1cm}{ \includegraphics[width=1.3cm]{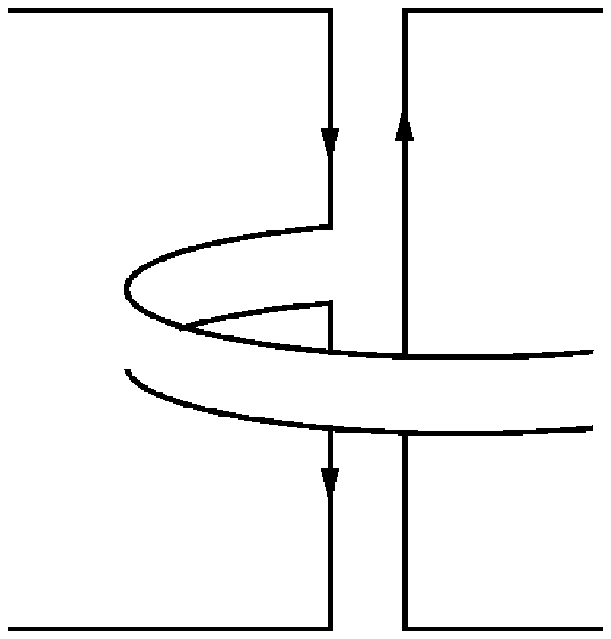}} 
         \bigg) 
&&
        \frac{-1}{i}   \parbox{1.2cm}{  \includegraphics[width=1cm]{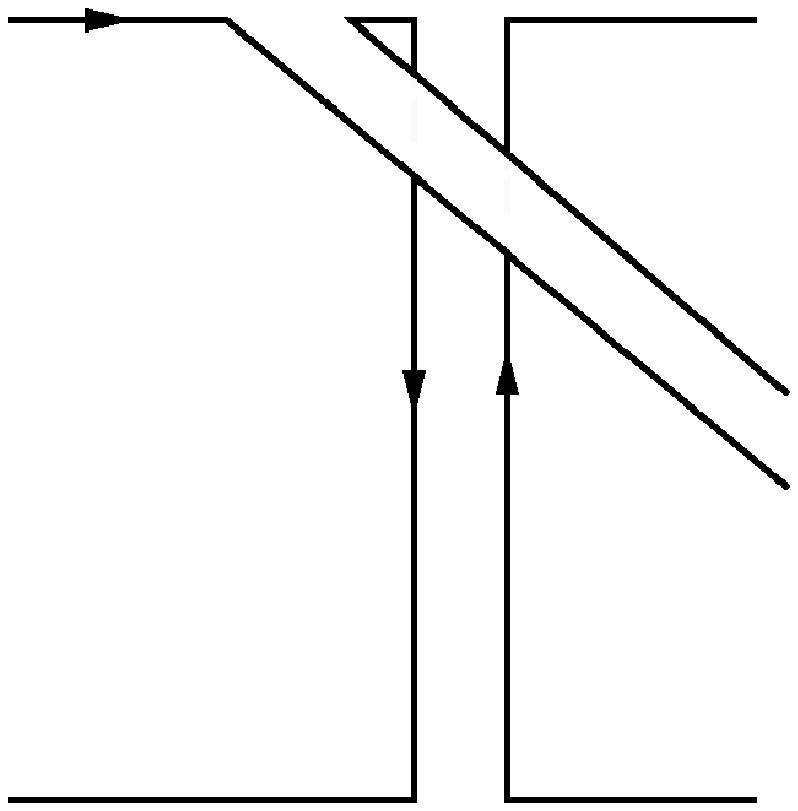}} 
        && 
\frac{1}{i}
        \parbox{1cm}{\includegraphics[width=1cm]{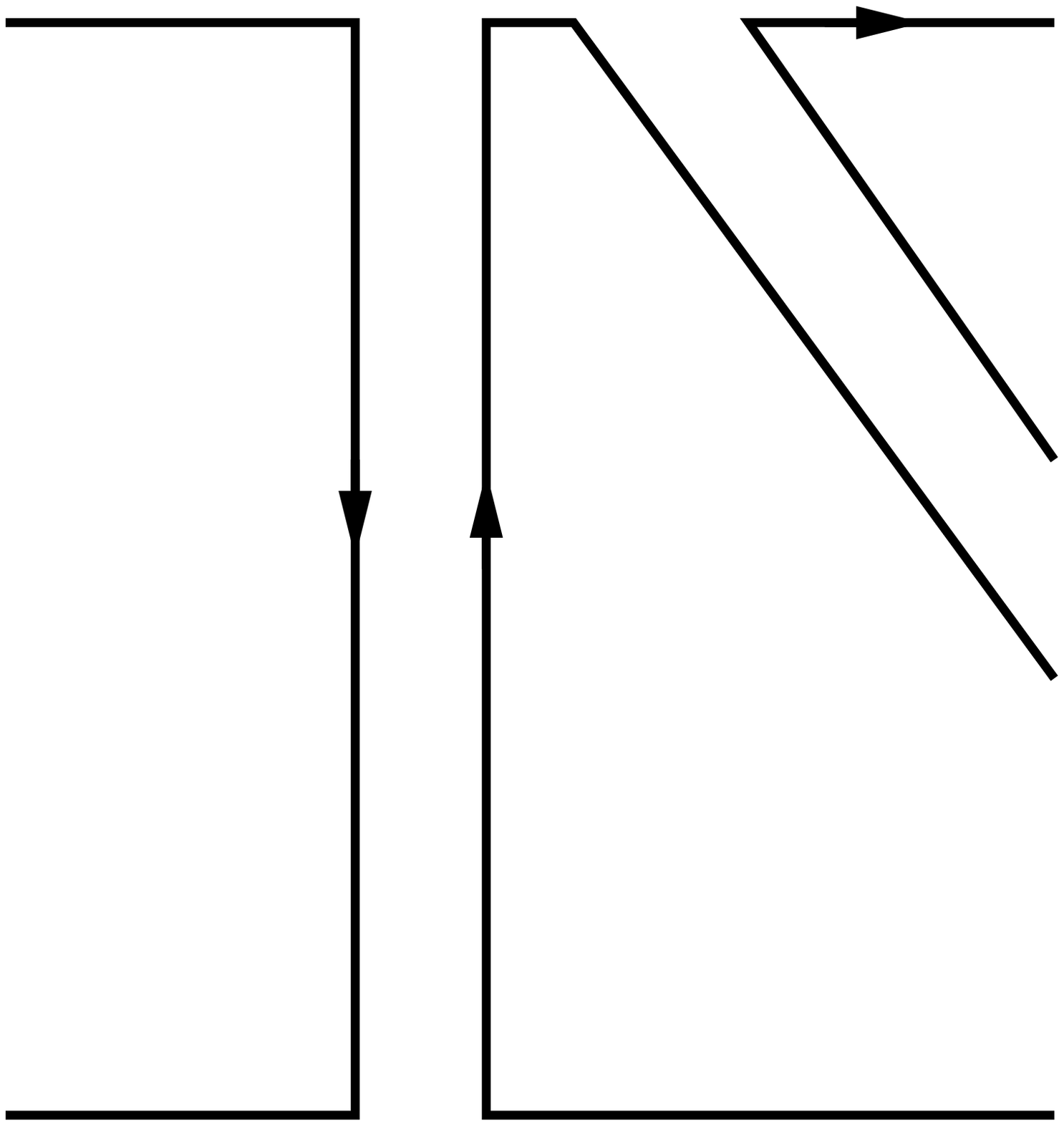}} 
        &&
 \frac{1}{i}
        \parbox{1cm}{\includegraphics[width=1cm]{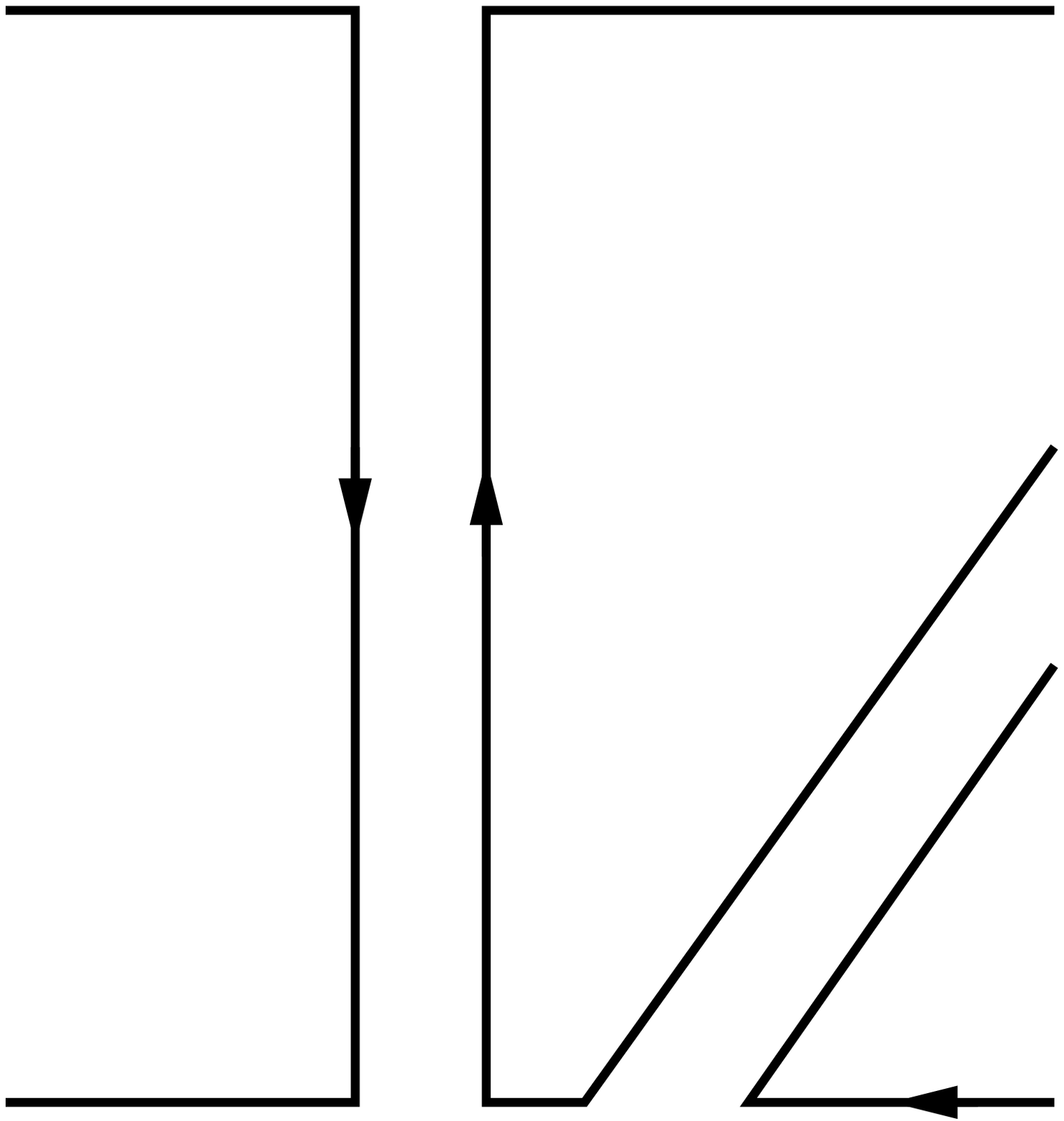}}
&&  \frac{-1}{i}
        \parbox{1cm}{\includegraphics[width=1cm]{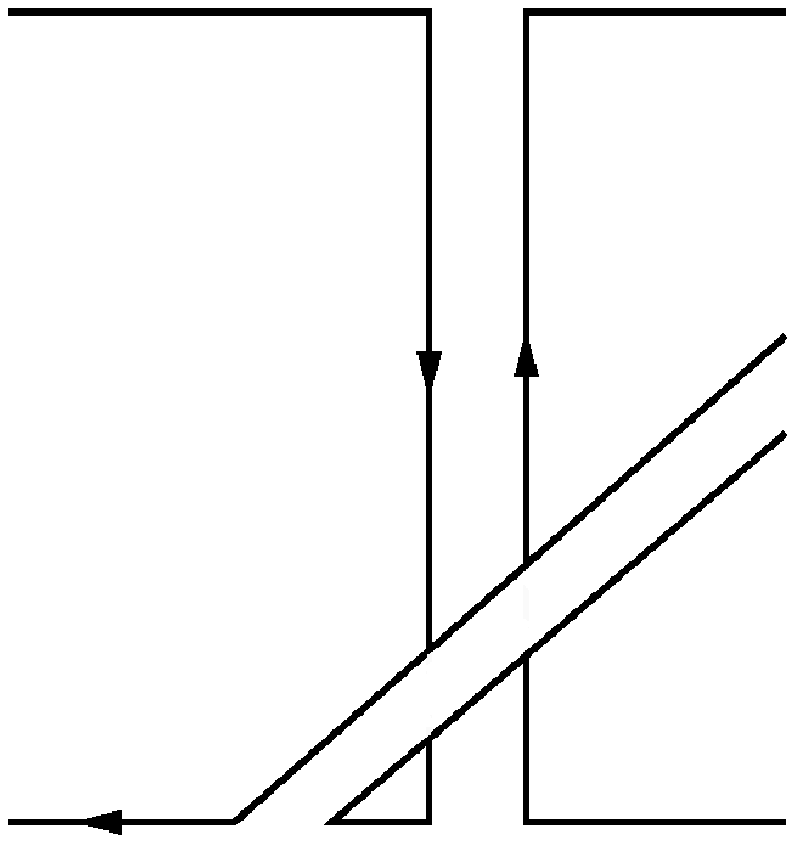}.}
\end{align}
In the second line we show, for each of the five different
Feynman-diagrams, the corresponding color factor in the double-line
notation. For the last four diagrams on the right hand side, where the
real gluon is emitted from the quark and the antiquark respectively,
we have shifted factors $i$ and $(-i)$ from the momentum part to the
color factor. With such a shift, the (remaining) momentum parts of the
second and the third diagram and the momentum parts of the fourth and
the fifth diagram coincide in the considered kinematical regime with
each other. Color factors and momentum parts of the production vertex
can therefore be written in a factorized form, and we obtain
\begin{align}
  \label{eq:lipatov_factor}
\parbox{1.5cm}{\includegraphics[width=1.2cm]{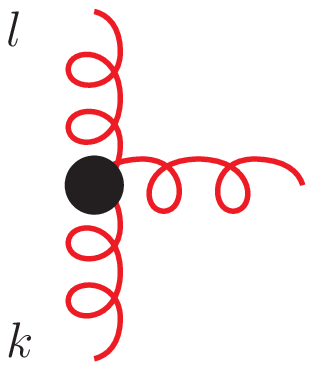}}
        & =
 \frac{{\bar g}}{i} 
     \left(\quad
      \parbox{1cm}{\includegraphics[height=1.2cm]{prod_gross1.eps}} \quad
      - \quad
      \parbox{1cm}{ \includegraphics[height=1.2cm]{prod_gross1b_bow1.eps}} 
     \quad  \right) \quad \times C(l, k)\cdot \epsilon_{(\lambda)},
\end{align}
where $\lambda$ denotes the helicity of the produced gluon. 
With the light-like momenta of the quark and the antiquark given by $p_1$ and $p_2$ respectively we have with $s \simeq 2p_1\cdot p_2$,
\begin{align}
  \label{eq:production_vertex}
        C^\mu(l, k) =    \left(\alpha_l  + 2\frac{{\bf   l}^2}{ \beta_k s}      \right)p_1^\mu  + \left(\beta_k  + 2 \frac{{\bf  k}^2}{ \alpha_l s}      \right) p_2^\mu  - (l_\perp^\mu + k_\perp^\mu), 
\end{align}
where  $ \alpha_l s = 2 p_2\cdot l $
and $\beta_k s = 2 p_1\cdot k $.  When squaring the production
amplitude, we find 
for the color part that only one of the four
possible combinations fits on the plane, namely the one shown on the
right hand side of Fig.\ref{fig:sdisc_real3}. 
From the point of view of Feynman-diagrams, the planar approximation 
includes only a sub-set of the diagrams in Eq.(\ref{eq:prod_vertex}), namely 
those with the color structure
\begin{align}
  \label{eq:prod_vertex_NC}
  \parbox{2cm}{\center \includegraphics[width=1.2cm]{prod_gross1.eps}} 
         +
         \parbox{2cm}{\center\includegraphics[width=1.2cm]{prod_gross2.eps}}
         + 
         \parbox{2cm}{\center \includegraphics[width=1.2cm]{prod_gross3.eps} }
\end{align}
However due to the factorization of momentum and color parts in the
MRK, Eq.(\ref{eq:lipatov_factor}), we still recover the complete
production vertex $C_\mu$ and therefore the BFKL kernel.  Making use
of our re-defined couplings $\bar{g}$ and $\bar{\lambda}$, and
including the factors $1/i$ of Eq.(\ref{eq:lipatov_factor}), we find,
for the plane, the $2\to2$ Reggeon transition kernel $K_{2 \to2}$ as:
\begin{align}
  \label{eq:22kernel_momis}
\parbox{2cm}{\includegraphics[width=1.8cm]{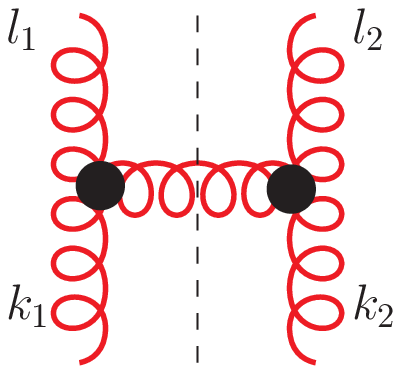}}  = \bar{\lambda} K_{2 \to 2}({\bf l}_1, {\bf l }_2;{\bf k}_1, {\bf k}_2) 
= - \bar{\lambda} \big[
 ({\bf k}_1 + {\bf k}_2)^2 - \frac{{\bf l}_1^2 {\bf k}_2^2}{({\bf k}_1 - {\bf l}_1)^2}  - 
\frac{{\bf l}_2^2 {\bf k}_1^2}{({\bf k}_1 - {\bf l}_1)^2} \big]
\end{align}
with the constraint ${\bf l}_1 + {\bf l}_2 ={\bf k}_1 + {\bf k}_2 ={\bf q}$.

\begin{figure}[htbp]
  \centering
  \parbox{4cm}{\center \includegraphics[height = 3.5cm]{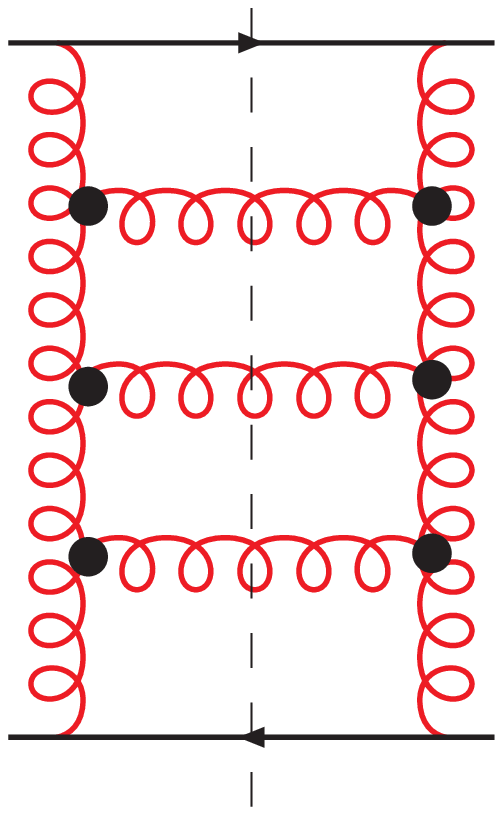}}
  \parbox{1cm}{$\,$}
  \parbox{4cm} { \center \includegraphics[height = 3cm]{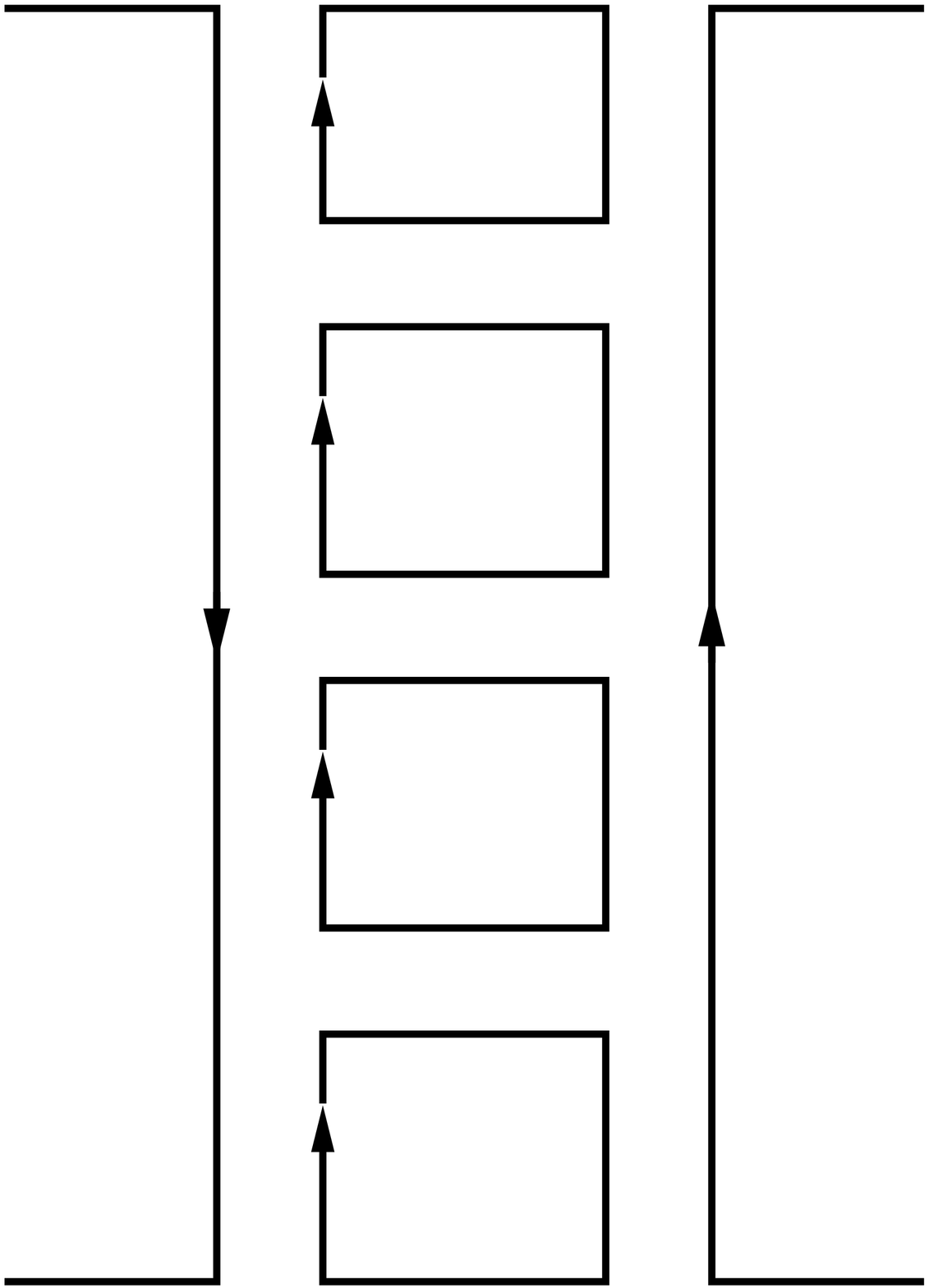}} 
  \caption{\small The s-channel discontinuity of the elastic quark-antiquark scattering amplitude for three gluon production.
On the right, the color structure of the planar diagrams.}
  \label{fig:elastic_disc}
\end{figure}

Higher order terms with the production of $n$-gluons within the MRK are taken into
account in a similar way. Making use of Regge-factorization of the production amplitudes, 
it can be shown that each emission of an additional real gluon is
taken into account by insertion of the effective production vertex 
Eq.(\ref{eq:lipatov_factor}), as illustrated in
Fig.\ref{fig:elastic_disc} for the squared production amplitude.  Similar 
to the production of one-gluon, for each of the inserted production vertices
only one of the four possible combinations of color factors fits onto the plane, 
and the resulting color factor can be found on the right-hand-side of Fig.\ref{fig:elastic_disc}.

In order to obtain the complete expression of order $\lambda^n$ of the partial wave $\phi$,  
we need to include the reggeization of the gluon. This is done in the same way as 
in the finite-$N_c$ case, except that at each step we have to restrict ourselves to the 
planar approximation. In lowest order we return, on the rhs of the unitarity equation (\ref{unitarity}), to the 
elastic intermediate state and include, in one of the two factors $T_{2\to2}$, the one loop result 
(\ref{eq:beta_func}). Together with the real one-gluon production discussed above, this provides, in 
(\ref{eq:qqbar_reggeized}), the complete imaginary part to order $g^6$ (or $\lambda^3$). Compared to the 
tree diagram expression (\ref{eq:tree_quarks}), (\ref{eq:qqbar_reggeized}) replaces the exchange of 
an elementary $t$-channel gluon by a reggeized gluon. In higher order $g$, the analogous replacement has to be done, 
on the rhs of (\ref{unitarity}), for all $t$-channel exchanges in the production amplitudes $T_{2 \to n}$.        

In order to prove the self-consistency of this procedure ,
one has to show that the BFKL bootstrap condition is satisfied in the planar approximation. 
For the partial wave in Mellin-space, reggeization of the $t$-channel gluons within the LLA requires to
introduce, for every $t$-channel state of two gluons, a Reggeon-propagator
$1/(\omega - \beta({\bf k}) - \beta({\bf q} - {\bf k}))$. In order to perform the resummation of all 
production processes in Fig.\ref{fig:elastic_disc}, it is most convenient to formulate the BFKL integral-equation in the planar approximation.  
To this end we first factorize off, at the lower end of the ladder, the antiquark-impact factor $\phi_{(2;0)}$.
We then define, for the scattering of a quark on a reggeized gluon,
the amplitude $ \phi_2(\omega|{\bf k}_1, {\bf k}_2)$, which contains the Reggeon propagator
$1/(\omega - \beta({\bf{k}}_1) - \beta({\bf{k}}_2))$ of the two
reggeized gluons at the lower end. The BFKL-equation for this amplitude reads:
\begin{align}
  \label{eq:bfkl-eq}
\big(\omega - \sum_i^2\beta({\bf{k}}_i)\big) \phi_2(\omega | {\bf k}_1, {\bf k}_2) &= 
 \phi_{(2;0)}( {\bf k}_1 + {\bf k}_2)
+
 \bar{\lambda}  K_{2 \to 2}
\otimes \phi_2(\omega| {\bf k}_1, {\bf k}_2),  
\end{align}
where the kernel coincides with the finite $N_c$ case.
With our impact factor $\phi_{(2;0) }$ which depends only on the sum of
the transverse momenta of the $t$-channel gluons we find for Eq. (\ref{eq:bfkl-eq}) 
the familiar pole solution
\begin{align}
  \label{eq:pole_sol}
\phi_2(\omega|{\bf k}_1 +{ \bf k}_2) =& \frac{\phi_{(2;0)} }{\omega - 
\beta( -({\bf k}_1 +{ \bf k}_2)^2 )}.
\end{align}
The partial wave for quark-antiquark scattering becomes 
\begin{align}
  \label{eq:pole_phi}
\phi (\omega, t) = 2 \int \frac{d^2{\bf l}}{(2 \pi)^3} \frac{1}{{\bf l}_1^2{\bf l}_2^2} \frac{\phi_{(2;0)} \phi_{(2;0)}  }{\omega - \beta(t )} = \frac{1}{t} \frac{2 \bar{g}^2  \beta(t)}{\omega - \beta(t ) }.
\end{align}
Inserting this into Eq.(\ref{eq:elastic_rep}) and using that in LLA $\sin
\pi\omega \simeq \pi\omega$, we find the reggeized
gluon, as proposed in Eq.(\ref{eq:qqbar_reggeized}).
This proves that, in the planar approximation, the bootstrap property is satisfied.

\subsection{Cylinder topology: The BFKL-Pomeron }
\label{sec:cylinder}
Next we turn to the cylinder topology which,
in the Regge-limit, leads to the BFKL-Pomeron.  In the expansion
Eq.(\ref{eq:vacuumgraph}), this corresponds to the term with two
boundaries and zero handles, $b=2, h= 0$.  Even though it would be
possible to carry out this analysis for quark-antiquark scattering, it
is more natural to do so for scattering of two highly virtual photons,
which provides an IR-finite and gauge-invariant amplitude.
Reggeization of the gluon will be an important ingredient in this
analysis, as we will see in short.
\begin{figure}[htbp]
  \centering
  \parbox{4.5cm}{\center \includegraphics[width=4cm]{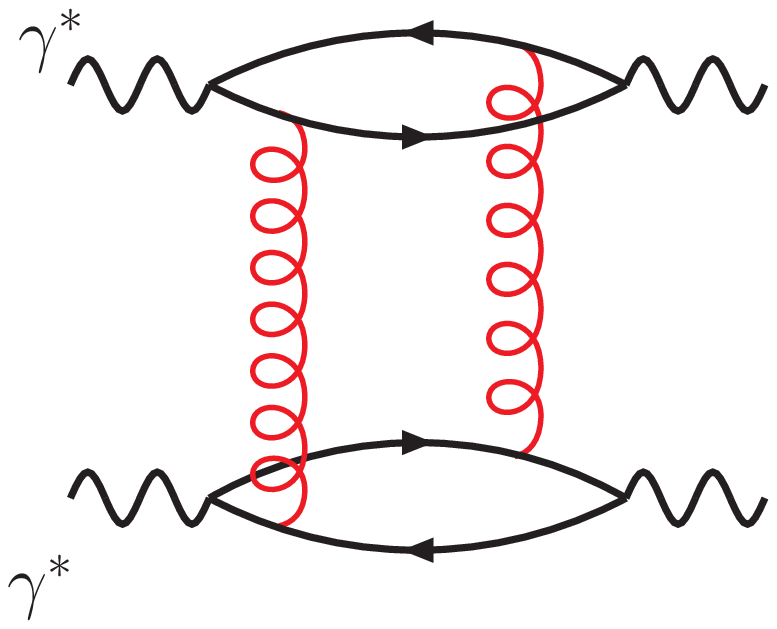}} 
  \parbox{4cm}{\center\includegraphics[width=3.5cm]{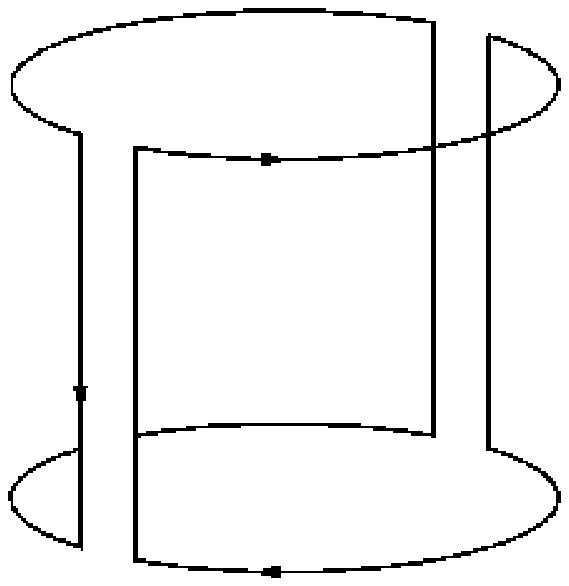}}
  \parbox{3.5cm}{\center \includegraphics[width=2.5cm]{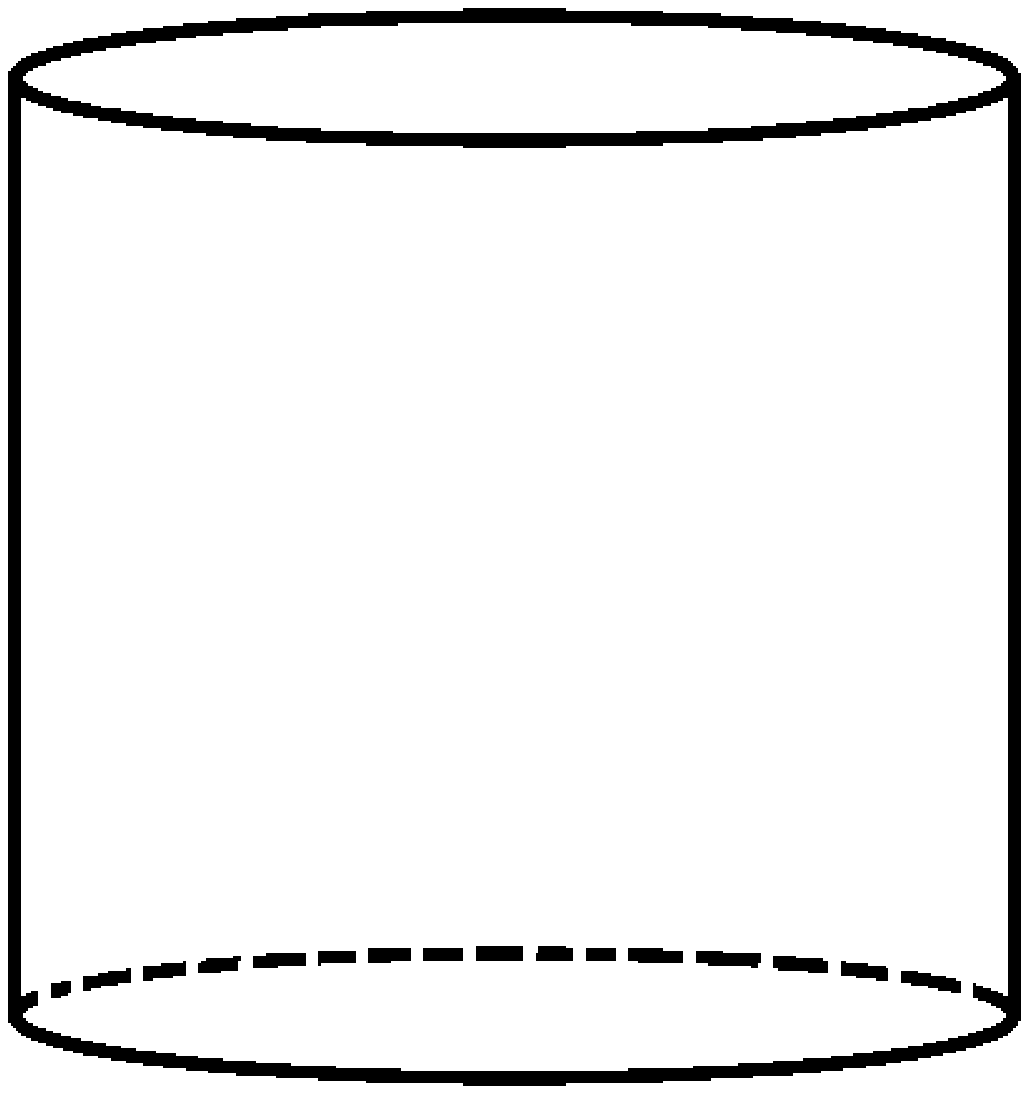}  } 
  \caption{\small To the left, a typical leading order diagram with its double-line color line factor in the middle.  The color factor fits on the cylinder surface, depicted to the right.}
  \label{fig:cyl}
\end{figure}
As a physical process with such a topology we consider the scattering
of two highly virtual photons in the Regge limit $s \gg |t|$, and
again we use the LLA to sum to all orders those diagrams which have
one power of $\ln s$ for each loop. In particular, the interaction
between the two photons is mediated by gluons in the $t$-channel, and
to lowest order in the electromagnetic coupling, each photon couples
to a quark-loop.  The quark-loops provide the two boundaries of the
cylinder. To leading order, the two photons interact by the exchange
of two $t$-channel gluons. A typical leading order diagram is shown in
Fig.\ref{fig:cyl}, together with its color factor. Unlike the planar
case, color factors of diagrams with both discontinuities in the $s$-
and in the $u$-channel fit on the cylinder. The
$t$-channel-interaction has therefore positive signature, similar to
the $N_c$ finite case,

To resum higher order terms we use again the analytic representation Eq.(\ref{eq:elastic_rep})
of the elastic amplitude in the Regge limit. Now the signature factor includes both right and left hand cuts: 
\begin{align}
  \label{eq:sigfac_possig}
\xi(\omega) = -\pi\frac{e^{-i\pi\omega} - 1}{\sin (\pi\omega)}
\end{align}
and belongs to positive signature in the $t$-channel. As in
the previous paragraph, we take the discontinuity in $s$ which allows
to determine the partial wave amplitude by considering real particle
production processes. Coupling of the two gluon state to the virtual
photon is described by the two gluon impact factor of the virtual
photon which is given by the sum of the following four Feynman diagrams:
\begin{align}
  \label{eq:gamma_impact}
{D}_{(2;0)} ({\bf k}_1, {\bf k}_2) = 
        \parbox{2.5cm}{\includegraphics[width=2.5cm]{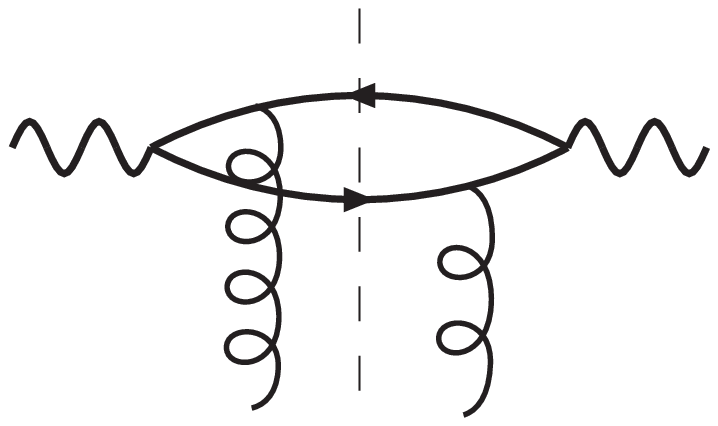}} 
        +
        \parbox{2.5cm}{\includegraphics[width=2.5cm]{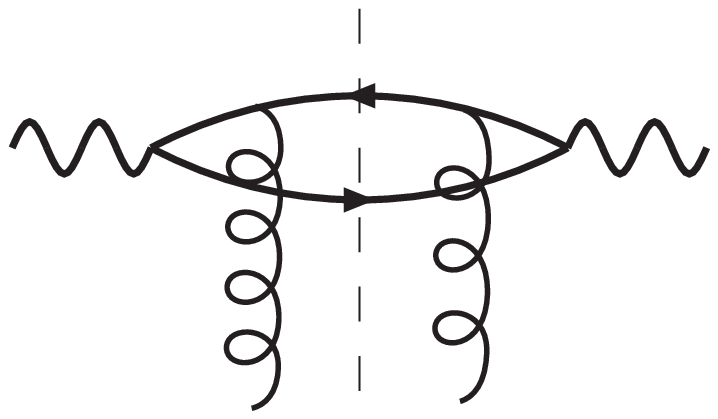}} 
        +
        \parbox{2.5cm}{\includegraphics[width=2.5cm]{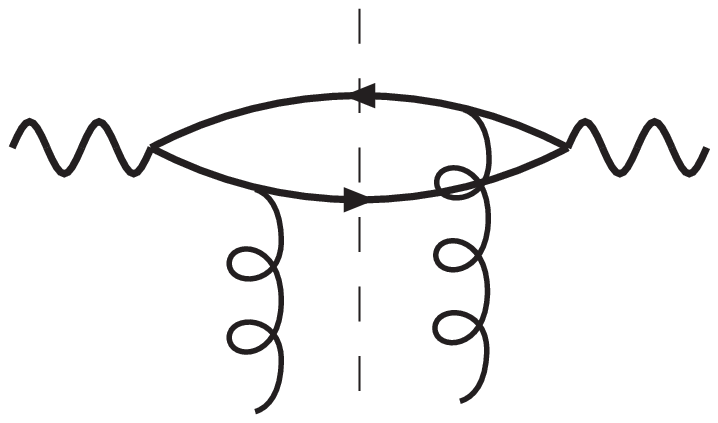}} 
        +
        \parbox{2.5cm}{\includegraphics[width=2.5cm]{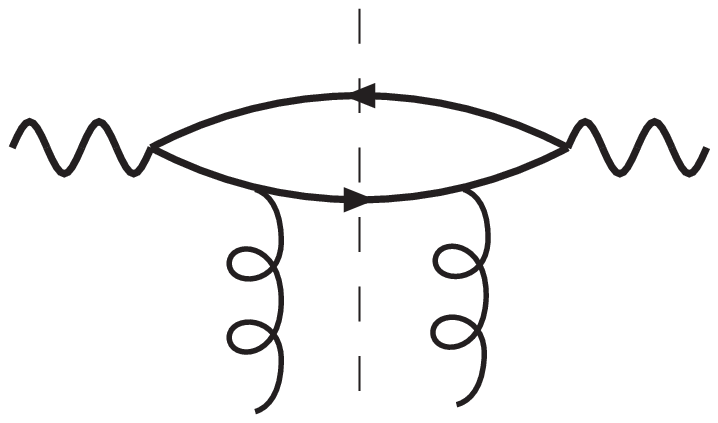}}
\end{align}
with the constraint ${\bf k}_1 + {\bf k}_2 = {\bf q}$.  It has  the important
properties to be symmetric under exchange of its transverse momenta
arguments, and to vanish whenever one of the transverse momenta
goes to zero\footnote{An analytic form of $D_{(2;0)}$ for the forward case can be found in \cite{Bartels:1994jj}. In particular, the  above $D_{(2;0)}$ is  $1/\sqrt{8}$ times  Eq.(2.4) of \cite{Bartels:1994jj}, where the factor $\sqrt{8}$ originates from $\sqrt{N_c^2 -1}$ with $N_c=3$.  }.  For the large $N_c$ treatment it is convenient to absorb a
factor $N_c$ into the impact factor by defining $\mathcal{D}_{(2;0)}
({\bf k}_1, {\bf k}_2) :=N_c{D}_{(2;0)} ({\bf k}_1, {\bf k}_2)$ which
is proportional to the 't Hooft coupling $\lambda$.  To leading
order in $\lambda$, the partial wave is  given by
\begin{align}
  \label{eq:cylind-topol-bfkl}
  \phi^{(0)} (\omega, t) = \frac{2}{\omega} \mathcal{D}_{(2;0)} \otimes \mathcal{D}_{(2;0)}.
\end{align}
Higher order corrections to the $s$-discontinuity are taken into account
by considering real gluon production in the Multi-Regge-kinematics.
Production of real gluons is described, in the same way as in the previous section, by
the particle production vertex in Eq.(\ref{eq:lipatov_factor}).  To
study the color factor on the cylinder, we start with
the Born term, Fig.\ref{fig:cyl} and insert one additional
$s$-channel gluon
\begin{figure}[htbp]
  \centering
  \parbox{4cm}{\includegraphics[height=3cm]{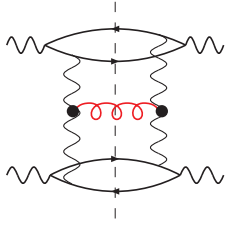}}
  \caption{\small $s$-discontinuity for the scattering of two virtual photons with one additional real    $s$-channel gluon.  Wavy lines for $t$-channel gluons denote reggeized gluons.}
  \label{fig:discgamma2}
\end{figure}
which leads us to the diagram Fig.\ref{fig:discgamma2}. For the color factors on the cylinder we find, 
compared to the plane an important difference: 
in addition to the combinations that fits  on the plane,
\begin{align}
  \label{eq:bfkl_sing1}
   \parbox{2cm}{\includegraphics[height=2cm]{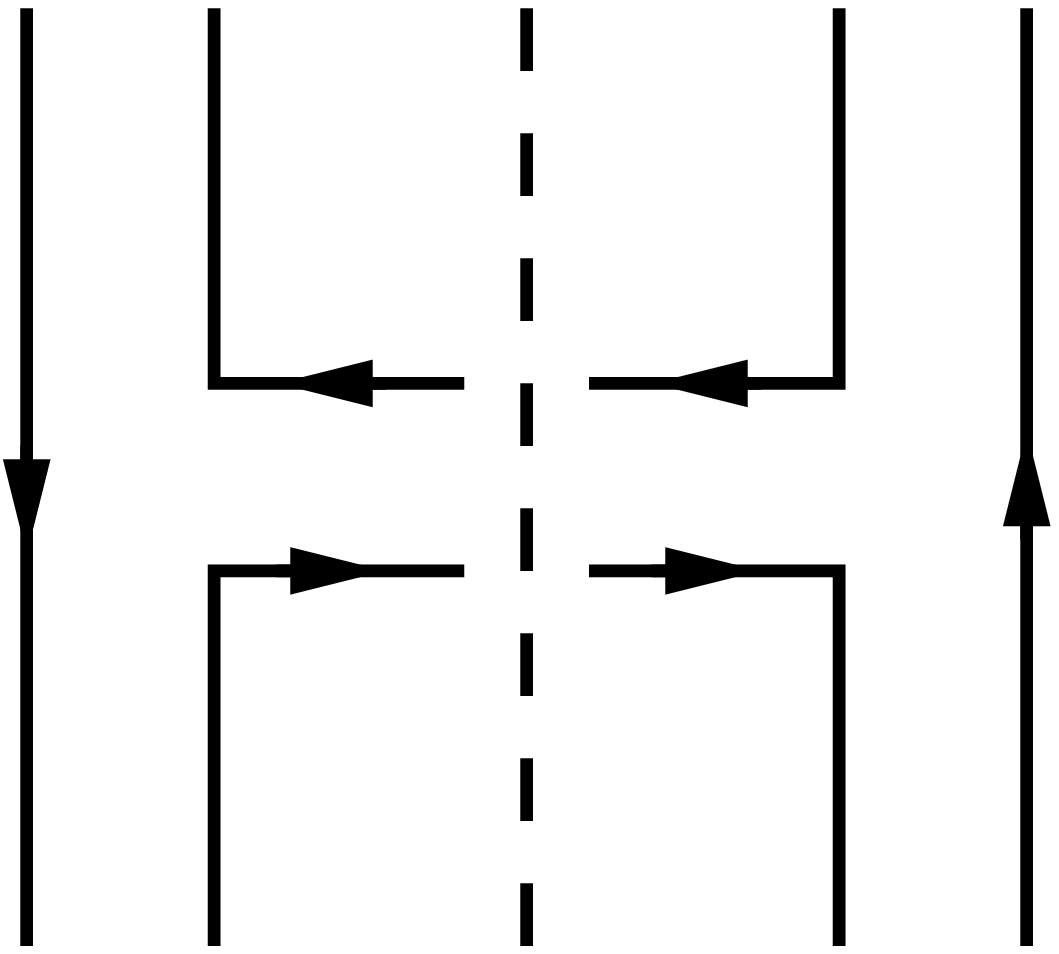}} \qquad \to \qquad  \parbox{3cm}{ \includegraphics[width=2.5cm]{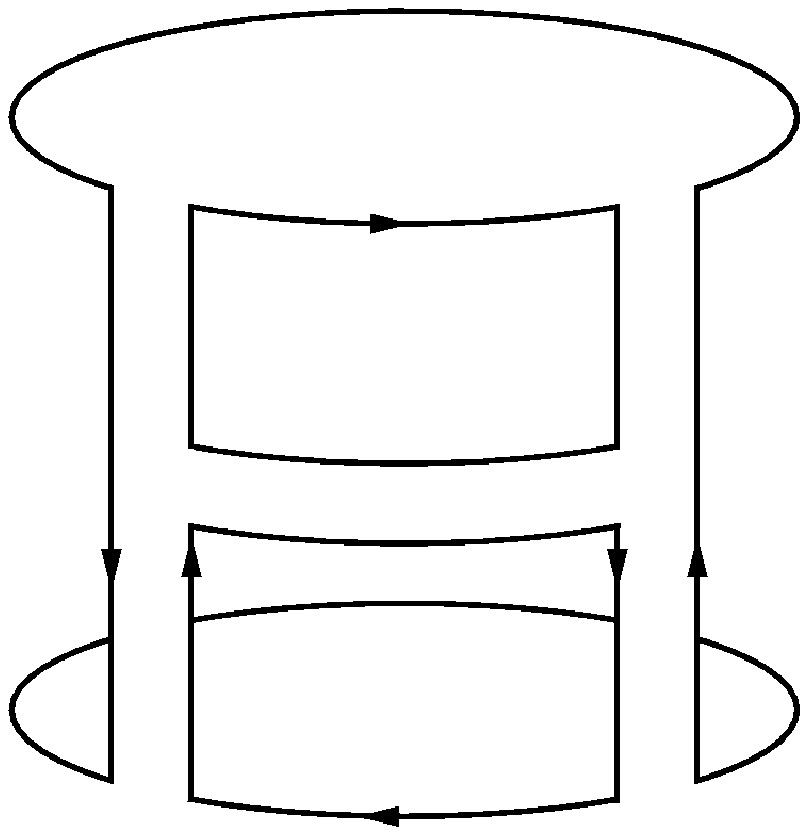}\quad ,}
\end{align}
on the cylinder also the following combination contributes:
\begin{align}
  \label{eq:bfkl_sing2}
 \parbox{3cm}{\includegraphics[height=2cm]{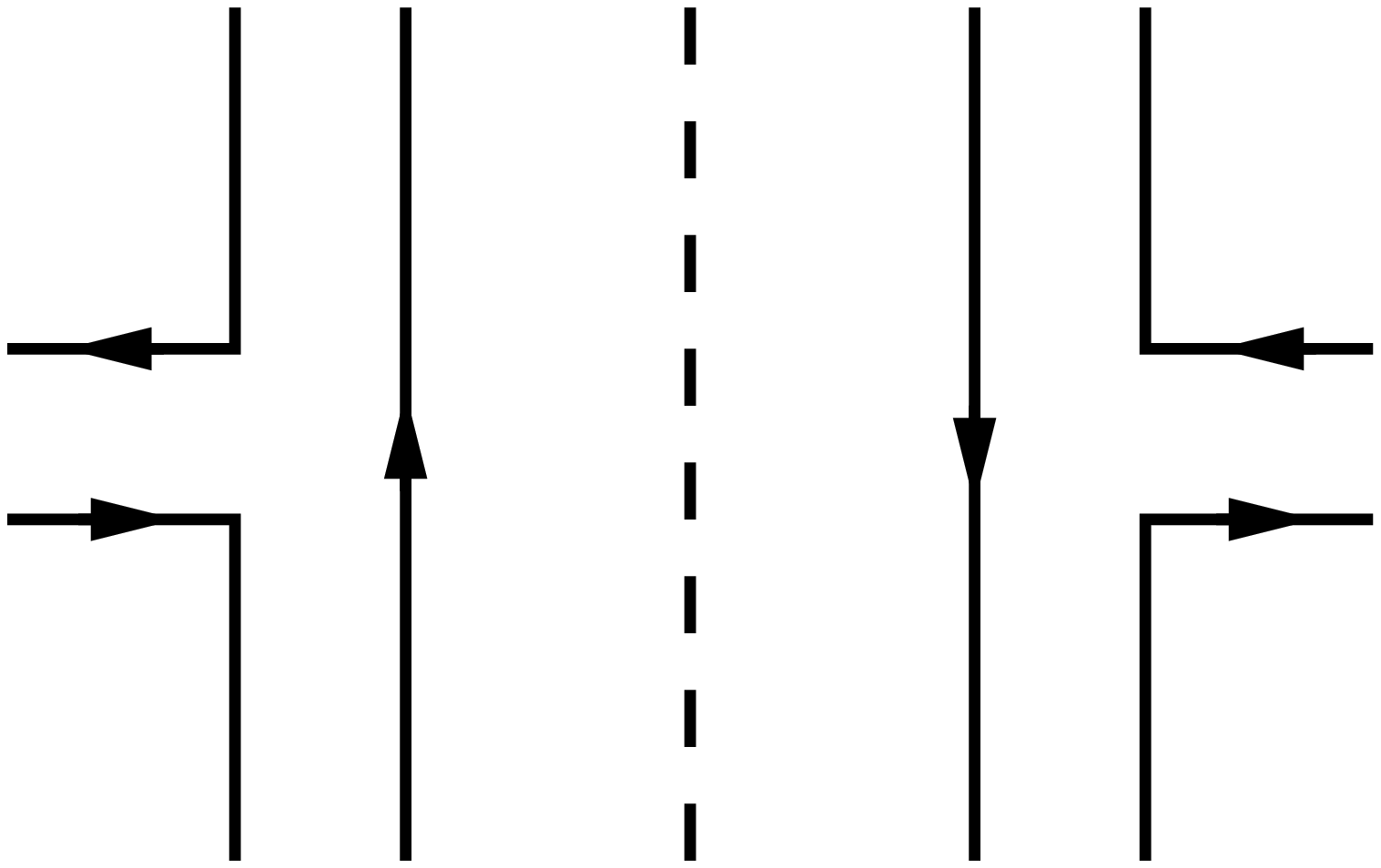}}\qquad \to\qquad \parbox{3.5cm}{ \includegraphics[width=2.5cm]{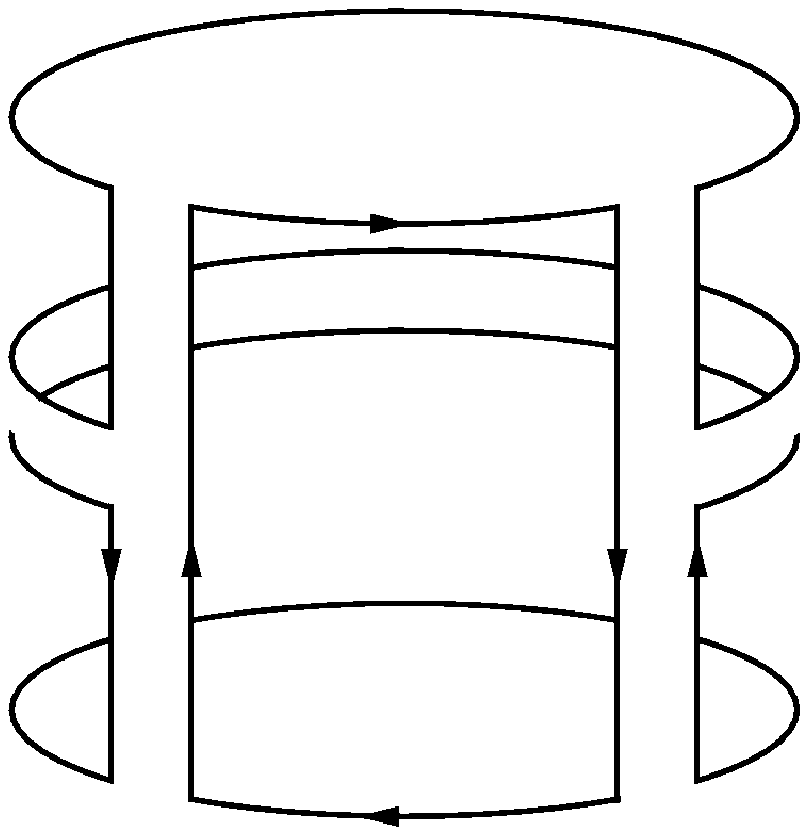}\quad.}
\end{align}
For the two-to-two Reggeon transition on the cylinder we therefore
obtain a factor two compared to the plane.  This counting generalizes
to diagrams with more than one real gluon: For each produced real
gluon the two combinations of color factors Eq.(\ref{eq:bfkl_sing1})
and Eq.(\ref{eq:bfkl_sing2}) need to be added.  An example with three
produced gluons is shown in Fig.\ref{fig:fore_back}:
\begin{figure}[htbp]
  \centering 
  \parbox{5cm}{\includegraphics[width=4cm]{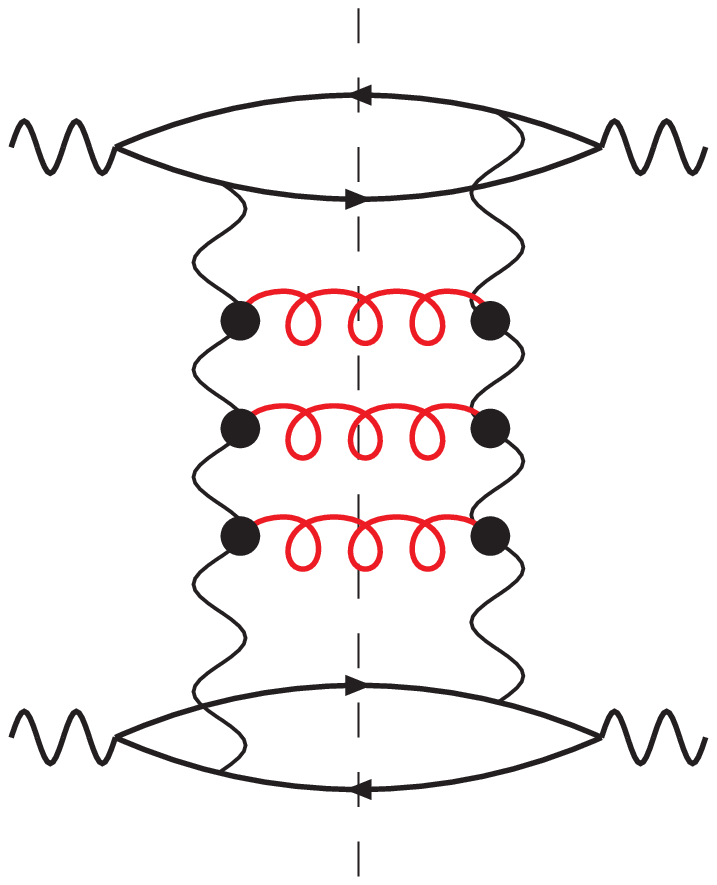}}
  \parbox{5cm}{\center \includegraphics[width=2.5cm]{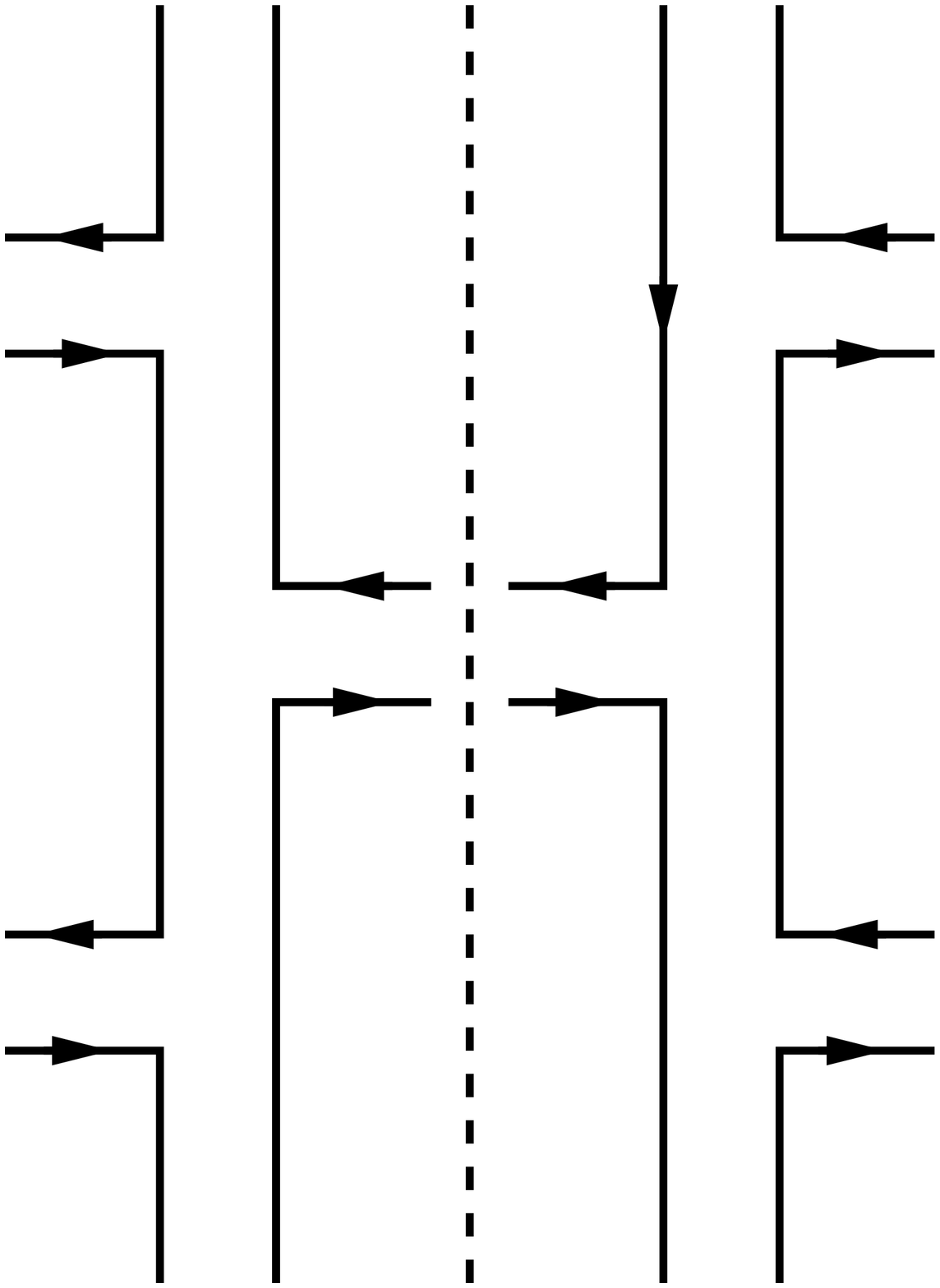}}
  \parbox{5cm}{\center \includegraphics[width=2.5cm]{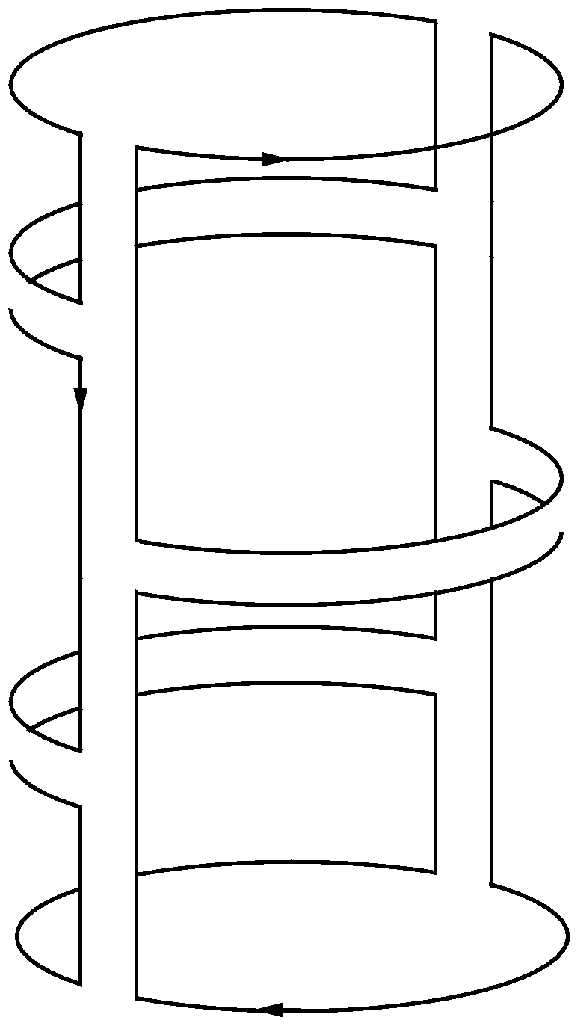}} 
\\
   \centering 
   \parbox{5cm}{\center (a)} 
   \parbox{5cm}{\center  (b)} 
   \parbox{5cm}{\center  (c)}
\caption{ \small Multi-gluon emission within the MRK. (a) The cut Feynman diagram (b) Combination of relevant color diagrams of the production vertex.   (c) The combination of (b) on the cylinder.  }
  \label{fig:fore_back}
\end{figure}
For each rung, we have two possibilities of connecting the $t$-channel
lines of the ladder, either on the forefront and or on the backside of
the cylinder \footnote{ The fact that, on the cylinder, each emission
  has two possibilities, one on the forefront and another one on the
  backside of the cylinder, has been realized also in
  \cite{DelDuca:1993pp}.  However, the approach pursued in this paper
  is quite different form ours: it starts from Park-Taylor
  amplitudes.}  Similar to the planar case discussed before, the
summation over all production processes is done by formulating the
BFKL equation on the cylinder.  This integral equation coincides with
Eq.(\ref{eq:bfkl-eq}), except for the factor $2$ in front of
$K_{2\to2}$, which results from the cylinder topology. Technically, we
again split off the impact factor of the lower virtual photon and
consider the partial wave amplitude $ \mathcal{D}_{2} (\omega|{\bf
  k}_1, {\bf k}_2) $ for the scattering of a virtual photon on two
reggeized gluons, which as before we define to contain the Reggeon
propagator of the external reggeized gluons. We find
\begin{align}
  \label{eq:bfkl-singlet}
(\omega - \sum_i^2\beta({\bf{k}}_i)) \mathcal{D}_{2} (\omega|{\bf k}_1, {\bf k}_2) &= 
  \mathcal{D}_{2,0} ({\bf k}_1, {\bf k}_2)
+2\bar{\lambda}
  K_{2 \to 2}\otimes \mathcal{D}_{2} (\omega|{\bf k}_1, {\bf k}_2) .
\end{align}
The kernel coincides with the BFKL kernel. Unlike the planar BFKL-equation Eq.(\ref{eq:bfkl-eq}),
Eq.(\ref{eq:bfkl-singlet}) has no pole solution, but the solution is
known to have a cut in the complex $\omega$-plane. The solution is known
both for the forward $(t=0)$ \cite{bfkl} and the non-forward $(t \neq
0)$ case \cite{Lipatov:1985uk}. Furthermore, the BFKL-Green's
function, which is obtained from $\mathcal{D}_2(\omega)$ by splitting up
the impact factor of the above virtual photon, is invariant under
M\"obius transformations.

\section{Six-point amplitudes and the pair-of-pants -topology}
\label{sec:sixpoint}
\begin{figure}[htbp]
  \centering \parbox{4cm}{\center
    \includegraphics[height=3cm]{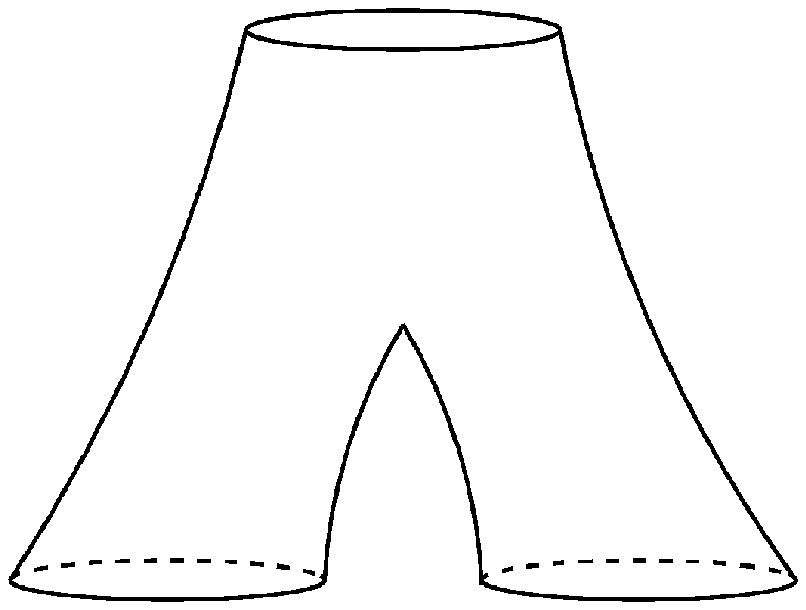}}
  \caption{ \small The sphere with three boundaries   $b=3$  and no handles $h = 0$ which yields the pair-of-pants.}
  \label{fig:pop}
\end{figure}

Returning to the expansion Eq.(\ref{eq:vacuumgraph}), we finally address
the term with three boundaries and zero handles $b=3 , h=0$.  It is
proportional to $1/N_c$ and leads to color factors that fit on the
pair-of-pants, as illustrated in Fig.\ref{fig:pop}.  As we did for
scattering amplitudes on the plane and on the cylinder, also for the
pair-of-pants we identify a QCD-amplitude which, using the LLA, will
be determined in a certain kinematical high energy limit.
As it was the case for the elastic amplitude, within the LLA, quark
loops occur only inside the coupling to external states. In order to
arrive at a surface with three boundaries, we are thus lead to a
six-point amplitude, which we will study in the triple-Regge limit, to
be specified in the following.

\begin{figure}
\centering
 \parbox{8.5cm}{ \includegraphics[height = 4cm]{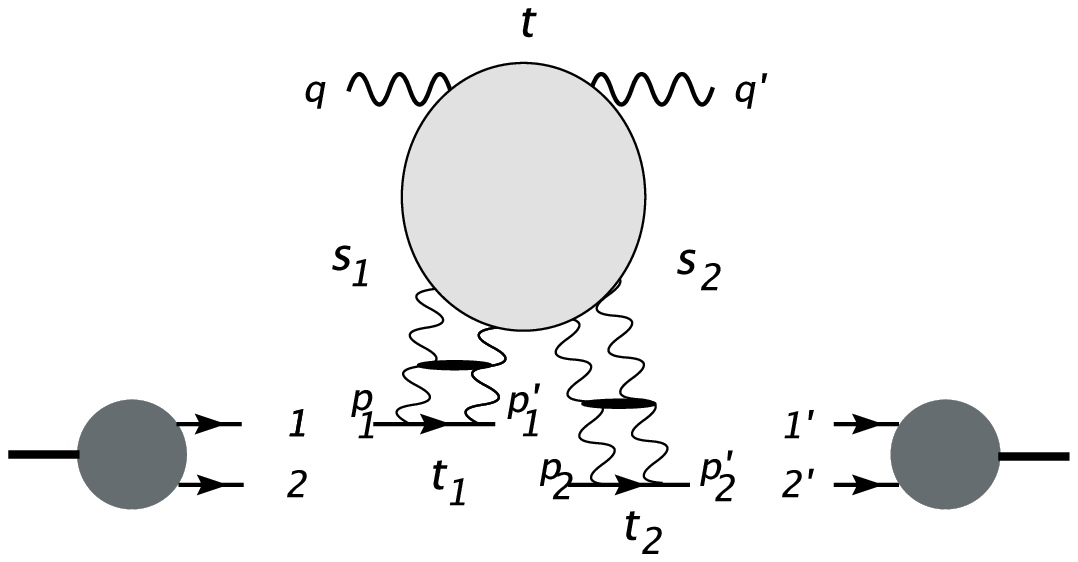}}
 \parbox{6.5cm}{ \includegraphics[height = 4cm]{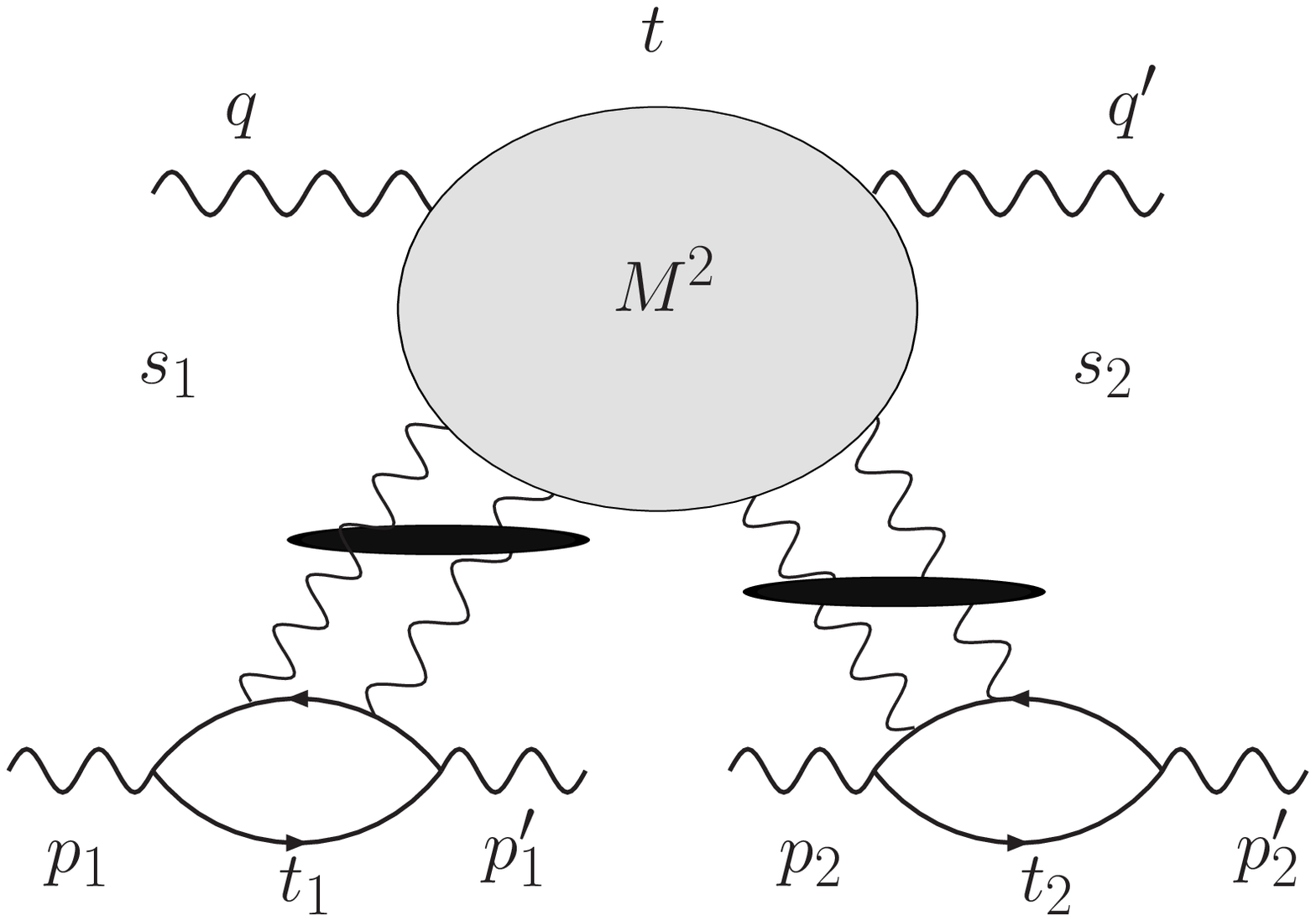}}
\caption{\small Scattering of a virtual photon on a weakly bound nucleus and the corresponding process where the lower nuclei are replaced by two highly virtual photons to the right.  }
\label{fig:nucl_phot}
\end{figure}

Within QCD, scattering amplitudes with more than $4$ external
particles arise naturally in the context of deep inelastic scattering
on a weakly bound nucleus, see  Fig.\ref{fig:nucl_phot}a.
To be definite, let us consider deep inelastic electron scattering on
deuterium. The total cross section of this scattering process is
obtained from the elastic scattering amplitude , $T_{\gamma^*(pn) \to
  \gamma ^* (pn)}$, which describes a $3 \to 3$ process,
\begin{equation}
\label{eq:sigmatot}
\sigma^{tot}_{\gamma^*(pn) \to \gamma ^* (pn)} = \frac{1}{s} \Im{\text m}\, 
T_{\gamma^*(pn) \to \gamma ^* (pn)},
\end{equation} 
where $s$ denotes the square of the total center of mass energy of the
scattering process.  To study the pair-of-pants topology, we replace
the two nucleons by virtual photons which furthermore provides us with
a clean perturbative environment for the study of the six-point
amplitude.  The kinematics is illustrated in Fig.\ref{fig:nucl_phot}.
Large energy variables are $s_1=(q+p_1)^2$, $s_2=(q'+p'_2)^2$, and $M^2
= (q+p_1 - p'_1)^2$ which gives the squared mass of the diffractive
system in which the upper virtual photon dissociates. The total energy
square is given by $s=(q+p_1+p_2)^2$.  Furthermore, we have the
momentum transfer variables $t=(q-q')^2$, $t_1=(p_1-p'_1)^2$, and
$t_2=(p_2-p'_2)^2$.  The triple Regge-limit is given by $ s_1 = s_2
\gg M^2 \gg t, t_1, t_2$.  The investigation of such a process, for
finite $N_c$, has been started in \cite{Bartels:1994jj}, in the
context of diffractive dissociation in deep inelastic scattering, and
in the following we will stay close to the methods used there.

Let us begin with finite $N_c$.  The $3 \to3 $ amplitude in the triple
Regge limit has the following analytic representation
\begin{align}
T_{3 \to 3}(s_1, s_2, M^2| t_1,t_2,t)=
\frac{s_1s_2}{M^2} \int  \frac{d\omega_1 d\omega_2 d\omega}{(2\pi i)^3}
&
 s_1^{\omega_1}{s_2}^{\omega_2} (M^2)^{\omega-\omega_1-\omega_2}
 \xi({\omega_1})  \xi({\omega_2})  \xi({\omega,\omega_1,\omega_2}) \notag \\
&\cdot
 F(\omega, \omega_1, \omega_2| t, t_1,t_2) .   
\label{tripleregge}
\end{align}
All three $t$-channels carry positive signature, and the signature factors
are given by
\begin{align}
  \label{eq:sig_facs}
            \xi(\omega) &= -\pi\frac{e^{-i\pi\omega}-1}{\sin(\pi\omega)}  
&\mbox{and}& &
 \xi({\omega,\omega_1,\omega_2}) &= -\pi\frac{e^{-i\pi(\omega - \omega_1 -\omega_2)} - 1}{\sin\pi(\omega - \omega_1-\omega_2)}.
\end{align}
The partial wave $F(\omega_1,\omega_2,\omega| t_1,t_2,t)$ has no phases and is
real valued. In analogy to the treatment of $T_{2 \to 2}$, we take
the triple energy discontinuity in $s_1$, $M^2$, and $s_2$,
\begin{equation}
\mathrm{disc}_{s_1} \mathrm{disc}_{s_2} \mathrm{disc}_{M^2} T_{3 \to 3} =
 \pi^3\frac{s_1s_2}{M^2} \int  \frac{d\omega_1 d\omega_2 d\omega}{(2\pi i)^3}
 s_1^{\omega_1}  {s_2}^{\omega_2}  (M^2)^{\omega-\omega_1-\omega_2}
\cdot
 F(\omega_1, \omega_2, \omega|t_1,t_2,t),
\label{tripledisc}
\end{equation}
which via a triple Mellin transform relates the partial wave
$F(\omega_1,\omega_2,\omega| t_1,t_2,t)$ to the real valued triple
energy discontinuity.

\begin{figure}[htbp]
  \centering
  \parbox{6cm}{\includegraphics[height = 3cm]{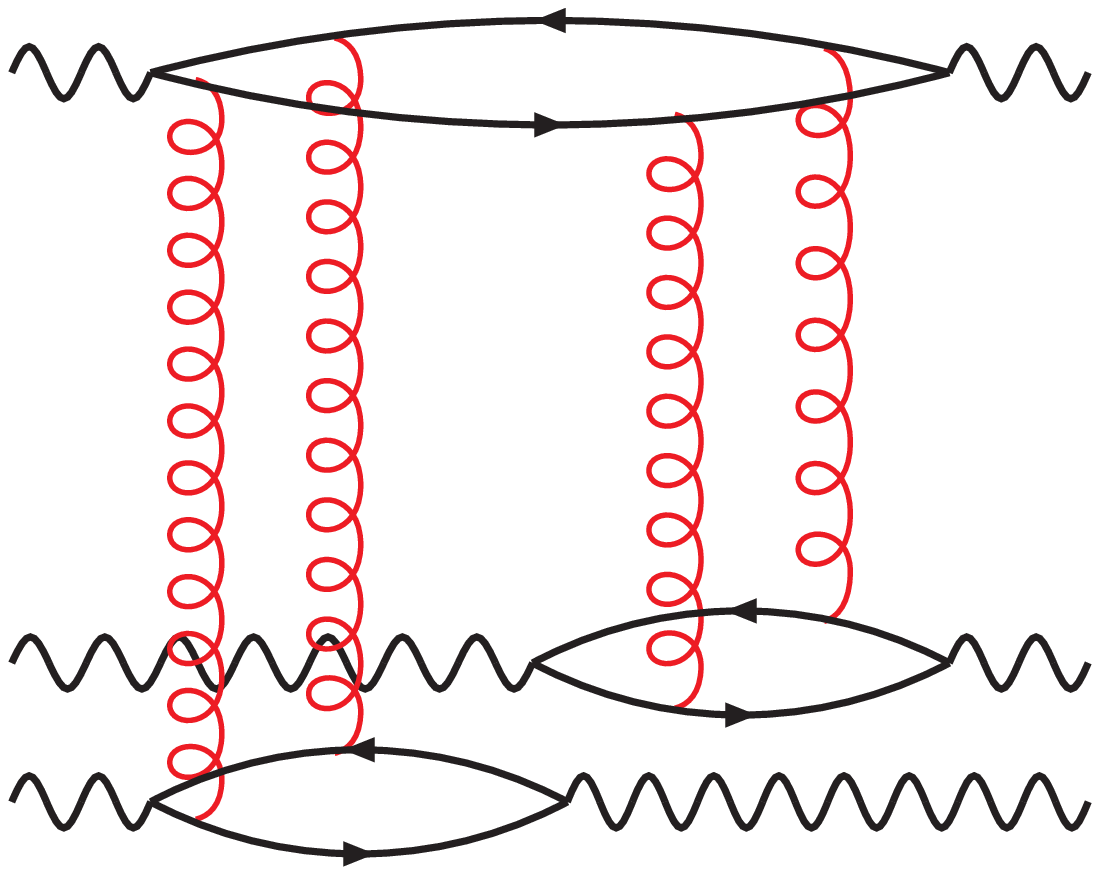}}
   \parbox{4cm}{\includegraphics[height = 3cm]{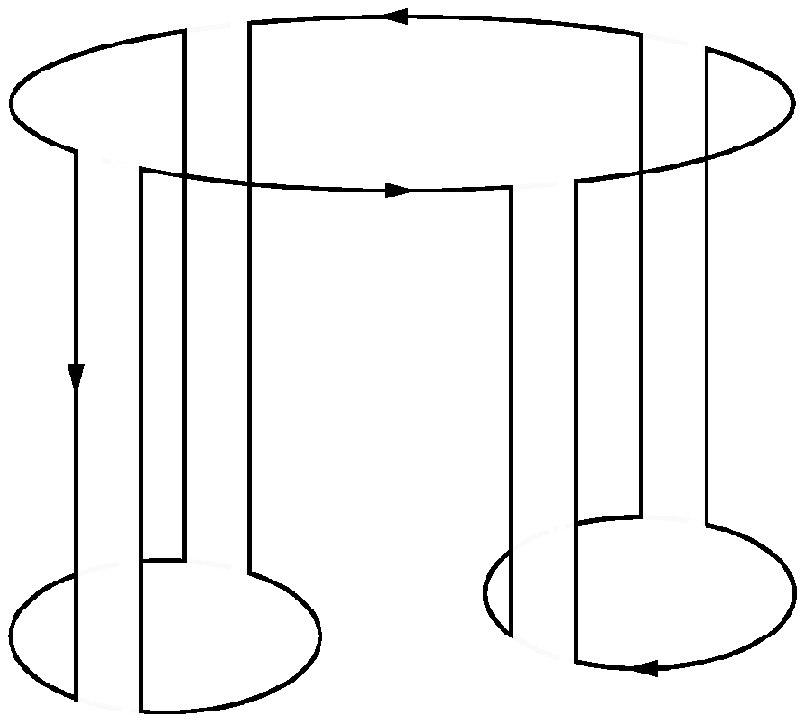}}
  \caption{ \small An example of a leading order diagram for the six-point amplitude with  its color factor depicted to the right. The diagram is proportional to $g^8N_c^3 = \lambda^4N_c^{-1}$ as required for the pair-of-pants.}
  \label{fig:trouser}
\end{figure}

\begin{figure}[ht]
\centering
\parbox{6cm}{\includegraphics[height = 5cm]{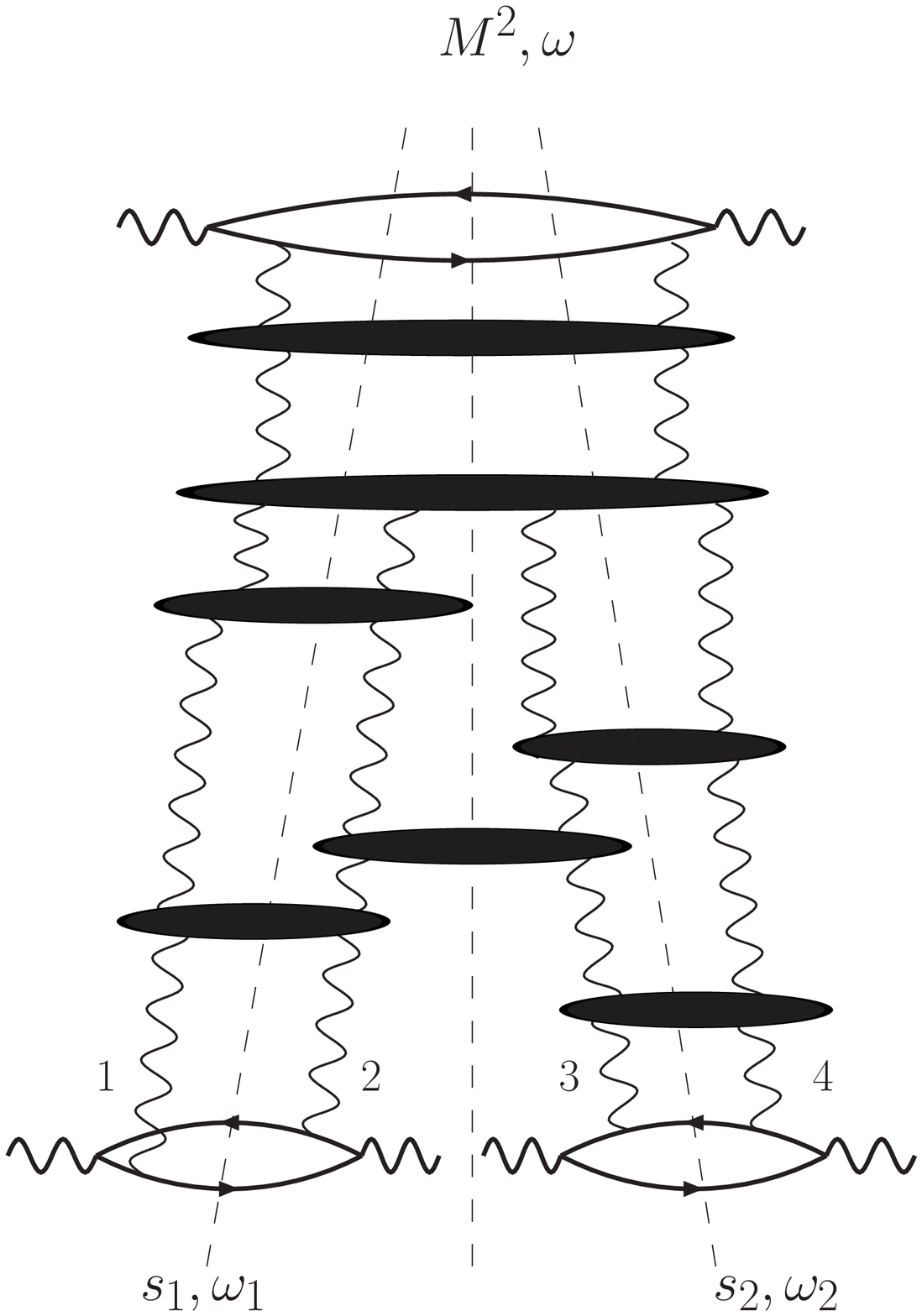}}
\parbox{6cm}{\includegraphics[height = 5cm]{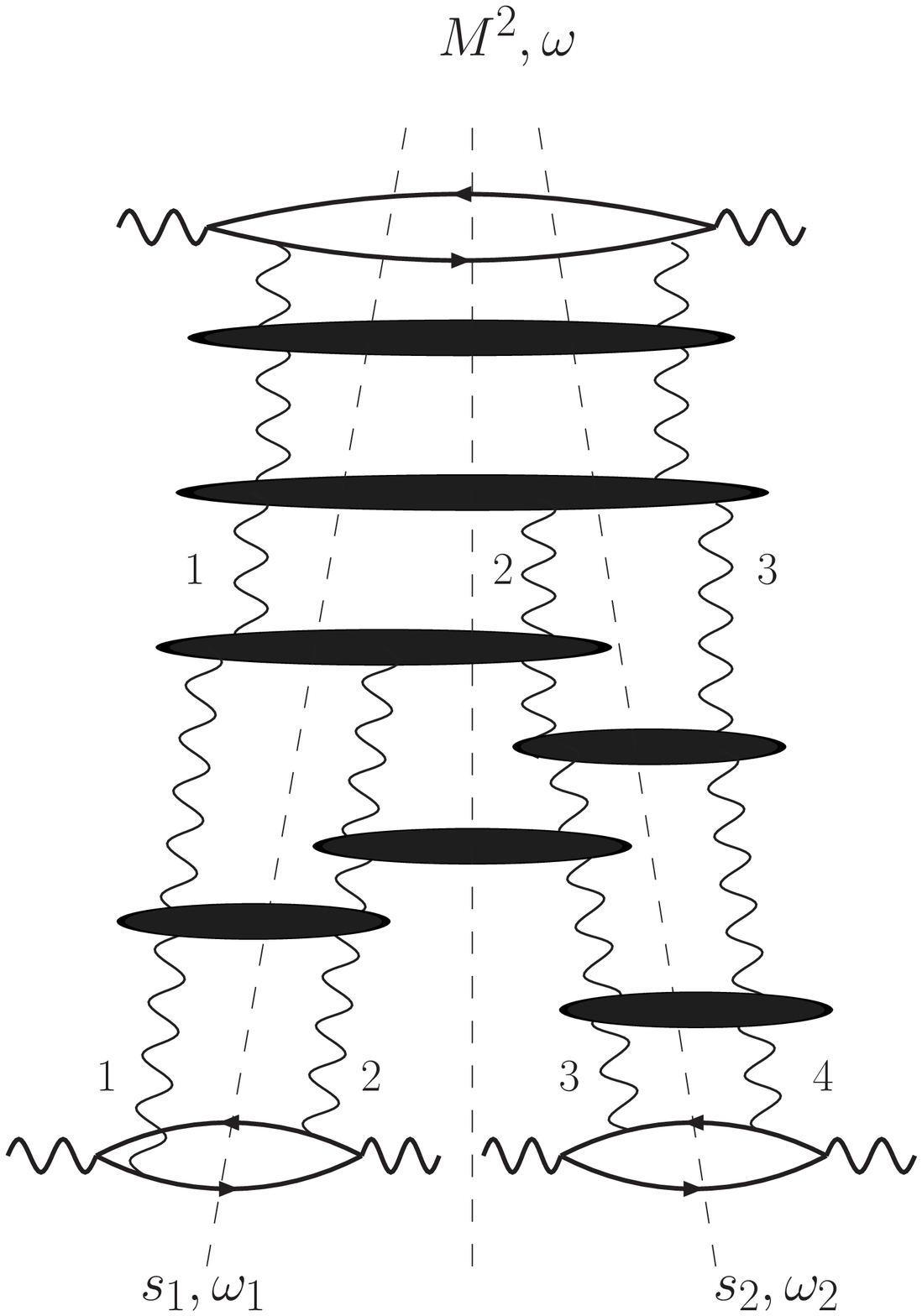}}
\caption{\small A few contributions to the triple energy discontinuity of Eq.(\ref{tripleregge}) }
\label{fig:triple_disc}
\end{figure}

Within the LLA, each of the three virtual photons couples to the
$t$-channel gluons by a quark-loop (which in the topological expansion
provides the three boundaries of the pair-of-pants). To leading order
in $g^2$, four $t$ channel gluons couple to the upper quark-loop, and
two gluons to each of the two lower quark-loops, which yields diagrams
like the one shown in Fig.\ref{fig:trouser}.
When taking the triple energy discontinuity, all intermediate states
between the $t$ channel gluons are on mass-shell.  For higher order
corrections to these diagrams, we make extensive use of unitarity and
compute sums over products of production processes.  Some examples are
shown in Fig.\ref{fig:triple_disc}.
Due to the triple discontinuity, $t$-channel gluons do not intersect,
and it is therefore useful to enumerate them from the left to right,
with the most left gluon carrying the index $1$. In order to form,
within the LLA, color singlet states which at the top couple to the
virtual photon and at the bottom to two virtual photons, we are lead
to QCD diagrams with four $t$-channel gluons at the lower and two,
three or four $t$-channel gluons at the upper end.  In contrast to our
previous discussion of scattering processes on the plane and on the
cylinder, the number of $t$-channel gluons can change, i.e. a $t$
channel gluon can emerge also from a $s$ channel produced gluon.
However, in LLA there is the restriction that, when moving downwards,
the number of reggeized $t$ channel gluons never decreases.

A special role in our analysis is taken by the diffractive mass $M^2$
as it defines the size of the upper 'cylinder' of the pair-of-pants.
In particular, the lowest intermediate state inside the
$M^2$-discontinuity (descending from the top to the bottom of the
diagram) defines the last interaction between the two 'legs' of the
pants, i.e. the branching point of the upper into the two lower
cylinders.  We use this branching point to factorize the partial wave
$F(\omega,\omega_1,\omega_2)$ into a convolution of three different
amplitudes. With $s_1, s_2 \gg M^2 \gg t, t_1, t_2$, the diagrams
below the branching point, within the LLA, do not depend on the
details of the dissociation of the upper virtual photon, and are
therefore described by two independent amplitudes
$\mathcal{D}_2(\omega_1)$ and $\mathcal{D}_2(\omega_2)$ which describe
scattering of two $t$-channel gluons and their coupling to the lower
photons.  The functions $\mathcal{D}_2(\omega_1)$ and
$\mathcal{D}_2(\omega_2)$ are given by BFKL-Pomeron Green's functions,
convoluted with the impact factors of the corresponding lower virtual
photon. In the following we will confirm that, in our topological
approach, on the pair-of-pants the required color factors work out
correctly.  The part above the the branching point is resummed by the
amplitude $\mathcal{D}_4(\omega)$, which describes the scattering of the upper
virtual photon on the four $t$-channel gluons.  In the following, our
interest will mainly concentrate on this part of our scattering
amplitude.

\section{Color factors with pair-of-pants topology}
\label{sec:prodtrous}

After our general outline of which QCD diagrams contribute to the
leading log approximation of the $3 \to 3$ process we have to select
now the contributions which belong to the large-$N_c$ limit on the
pair-of-pants. This will be done by attributing to each QCD diagram a
'color diagram' on the pair-of pants surface, where each gluon is
drawn by a pair of color lines with opposite directions, as done for
the plane and the cylinder in Sec.\ref{sec:elastic}.

\subsection{Color factors at Born-level}
\label{sec:pop_tree}
\begin{figure}[htbp]
  \centering
    \begin{minipage}{.9\textwidth}
    \parbox{3cm}{\center   \includegraphics[width=2cm]{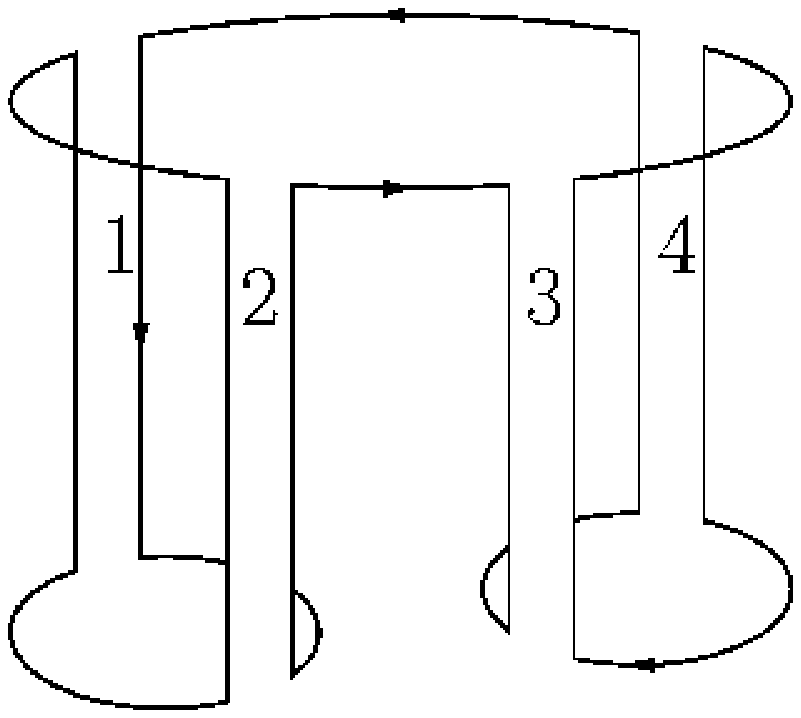}} 
    \parbox{3cm}{\center   \includegraphics[width=2cm]{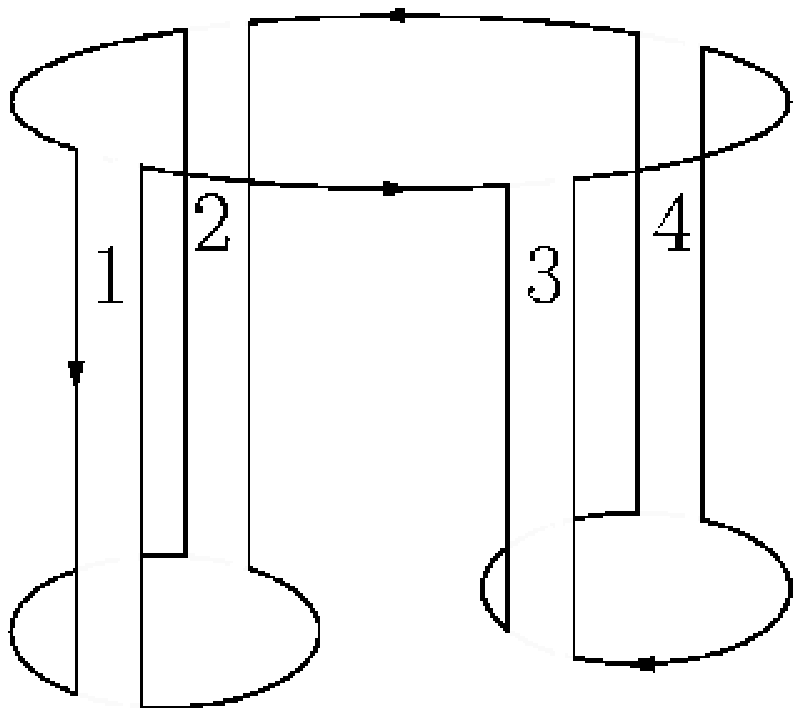}}  
    \parbox{3cm}{\center   \includegraphics[width=2cm]{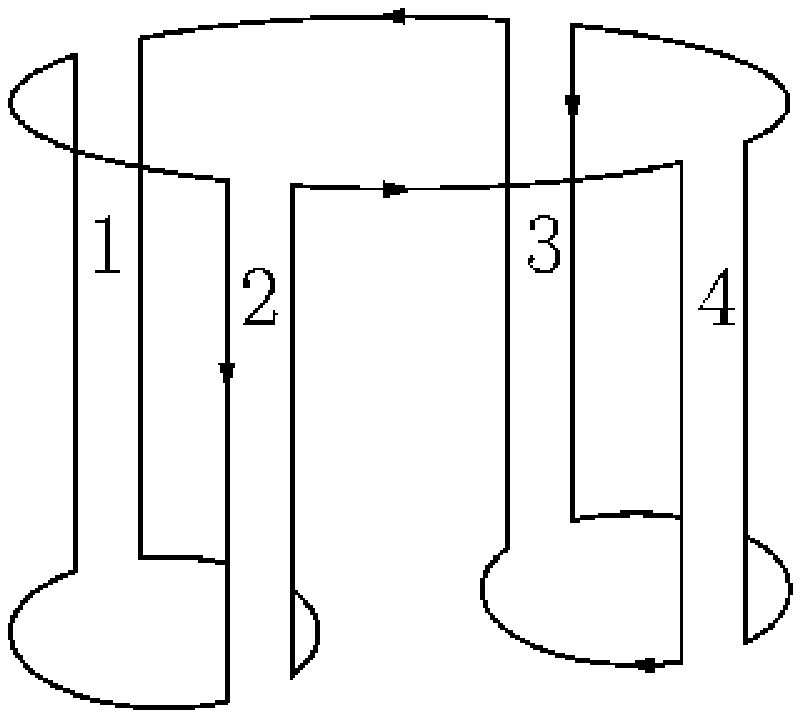}} 
    \parbox{3cm}{\center   \includegraphics[width=2cm]{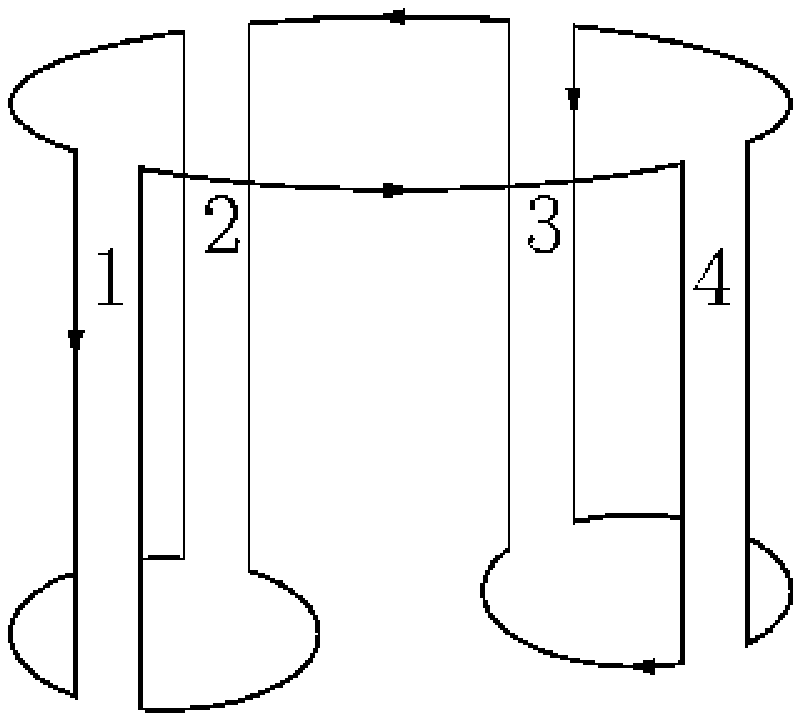}} \\
    \parbox{3cm}{\center $(1234)  $}
    \parbox{3cm}{\center ${(2134)}  $} 
     \parbox{3cm}{\center ${(1243)} $}
    \parbox{3cm}{\center ${(2143)}  $}
  \end{minipage}
  \caption{\small The four different orderings of color factors of the Born-term.}
  \label{fig:born}
\end{figure}

All diagrams that contribute to the Born approximation of the 
triple-discontinuity have the form of Fig.\ref{fig:trouser}, with $t$-channel
gluons coupling to the the quark-loop in all possible ways. 
Overall, one finds sixteen different diagrams. Because of the triple 
energy discontinuity $t$-channel gluons never 
intersect and hence can be labeled from the left to the right.
A closer look then shows that, inside the quark loop, we have 
the four different orderings of color matrices:
(1234), (2134), (1243), and (2143) (not distinguishing between cyclic permutations ). These four structures are illustrated in 
Fig.\ref{fig:born}
and are all of the order
\begin{equation} 
g^8 N_c^3 = N_c^{-1} \lambda^4 .
\label{eq:borncounting}
\end{equation}

\subsection{Gluon production on the pair-of-pants surface}
\label{sec:prod_pop}

In order to study corrections to these Born-amplitudes, 
we consider real gluon production processes on the pair-of-pants surface 
($t$-channel gluons are always reggeized). 
As to the selection of diagrams, we are searching the maximal power 
of color factors $N_c$. With (\ref{eq:borncounting}) the diagrams 
we are going to collect will come with the weight 
\begin{align}
\label{eq:BBB}
g^8 N_c^3 (g^2 N_c)^k = N_c^{-1} \lambda^{4+k}, 
\end{align}
with $k$ being some positive integer number.
This leads to the requirement that, for each gluon, the pair of color 
lines has to be planar, i.e. it never intersects.   
\begin{figure}[htbp]
  \centering
 \parbox{7cm}{\center \includegraphics[height=4.5cm]{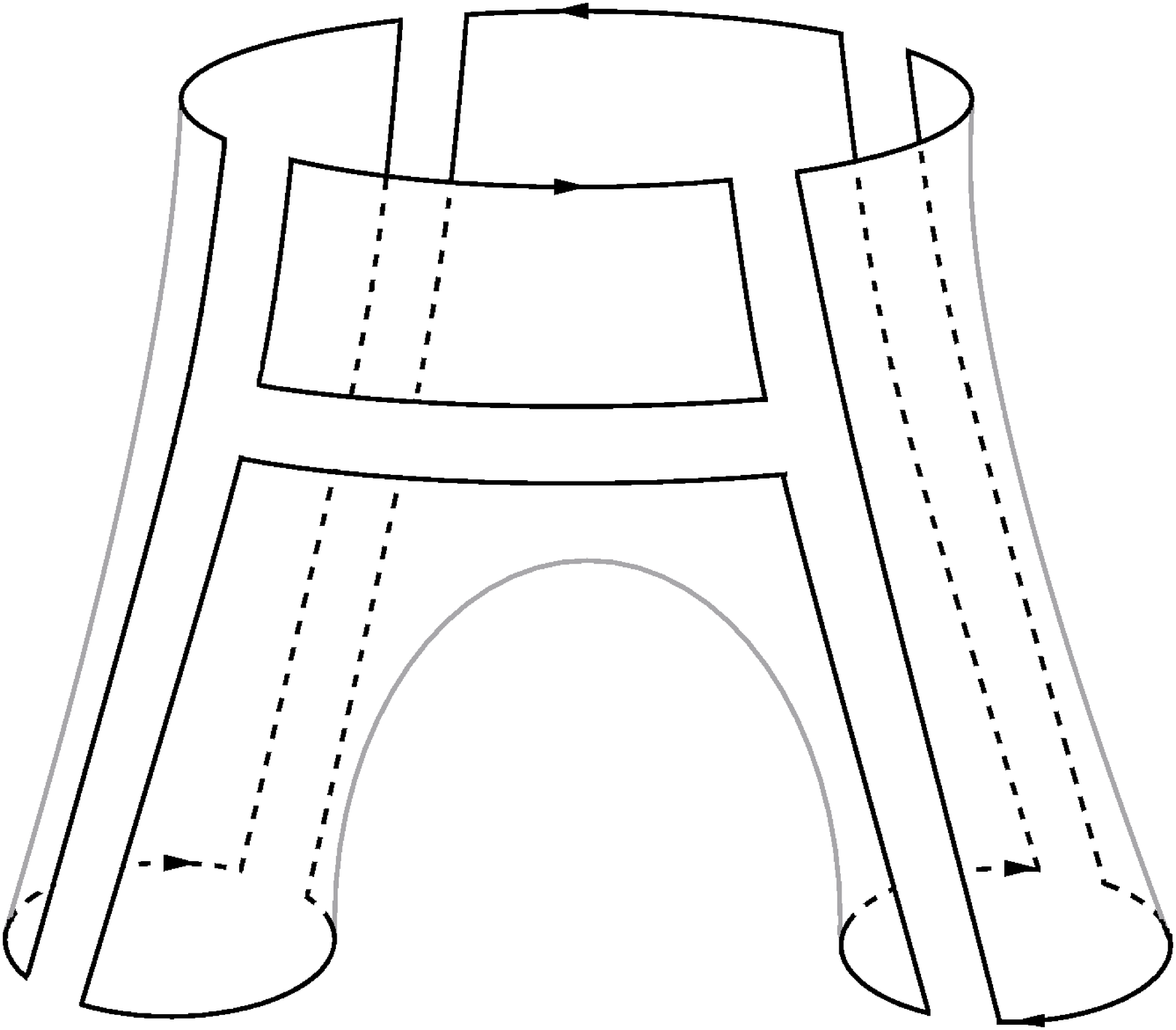}}
 \parbox{7cm}{\center \includegraphics[height=4.5cm]{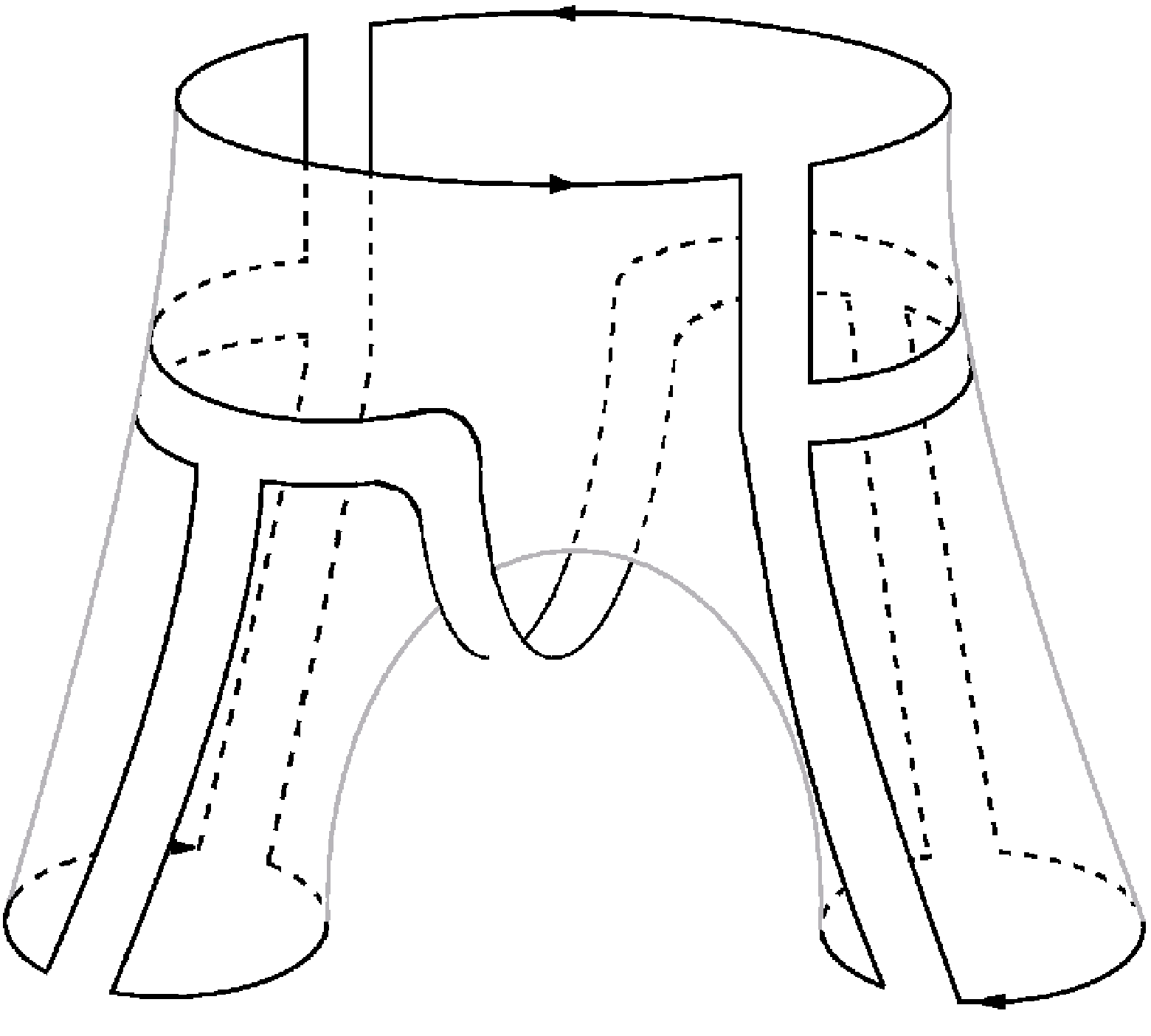}}\\
\parbox{7cm}{\center (a)}\parbox{7cm}{\center (b)}
  \caption{\small (a) A typical color factor of the planar class (A). (b) A typical color factor of the non-planar class (B) which has the interpretation of the Mandelstam diagram. }
  \label{fig:classes}
\end{figure}

As a result we find two classes of color
diagrams. The first class (A) consists of all diagrams which, by
contracting closed color loops, coincide with one of the lowest order
diagrams in Fig.\ref{fig:born}. An example for such a diagram is given
in Fig.\ref{fig:classes}.a. In the following we will refer to these
diagrams as 'planar' diagrams.  The second class (B) consists of those
diagrams where the 'last' vertex before the separation into the two
lower cylinders is of the form illustrated in Fig.\ref{fig:classes}.b:
one easily verifies that the power of color factors is $g^8 N_c^3 (g^2
N_c) = N_c^{-1} \lambda^{4+1}$, as required by condition
Eq.(\ref{eq:BBB}). Also, there is no intersection of color lines on
the pair-of-pants surface. These diagrams cannot be redrawn, by
contracting closed color loops in a straightforward way, such that they
coincide with one of the lowest order diagrams of Fig.\ref{fig:born}.
The structure shown in Fig.\ref{fig:classes}.b is reminiscent of the
Mandelstam diagram ('Mandelstam cross') which, when integrated over
the diffractive mass $M$, couples to the two Pomeron cut; we will
therefore call it 'non-planar' (although it fits onto the surface of
the pair-of-pants).  Note that this non-planar structure appears only
once, namely at the point where the upper cylinder splits into the two
lower ones.

In the remainder of this section, we describe these two sets of diagrams in 
more detail. In particular, we list the momentum space expressions which 
belong to the gluon transition kernels. In the following section, we derive 
integral equations which sum all diagrams of set (A). For set (B) we 
re-derive the triple Pomeron vertex.

\subsection{Two-to-two Reggeon transitions }
\label{sec:amplitudes-with-4}
\begin{figure}[htbp]
  \centering
  \parbox{3cm}{\includegraphics[height=1.5cm]{bfkl_1lopp_a.eps}}
  \parbox{3.5cm}{\includegraphics[height=2.5cm]{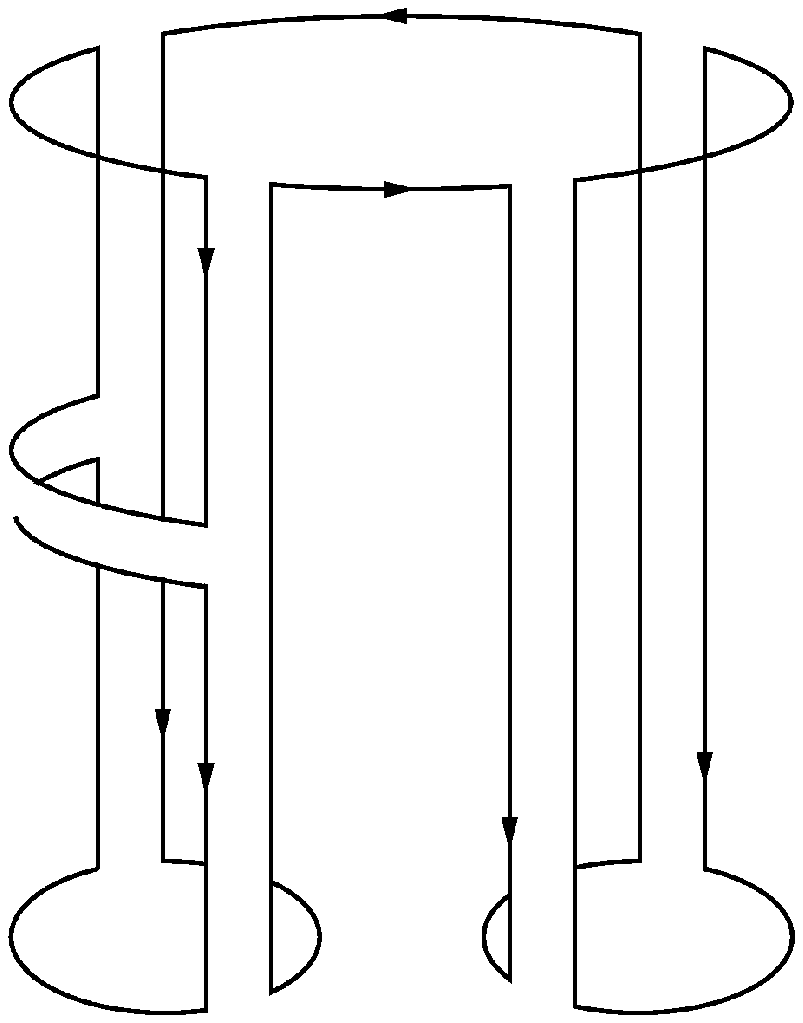}} 
  \parbox{3cm}{\includegraphics[height=1.5cm]{bfkl_1lopp_b.eps}}
  \parbox{3.5cm}{\includegraphics[height=2.5cm]{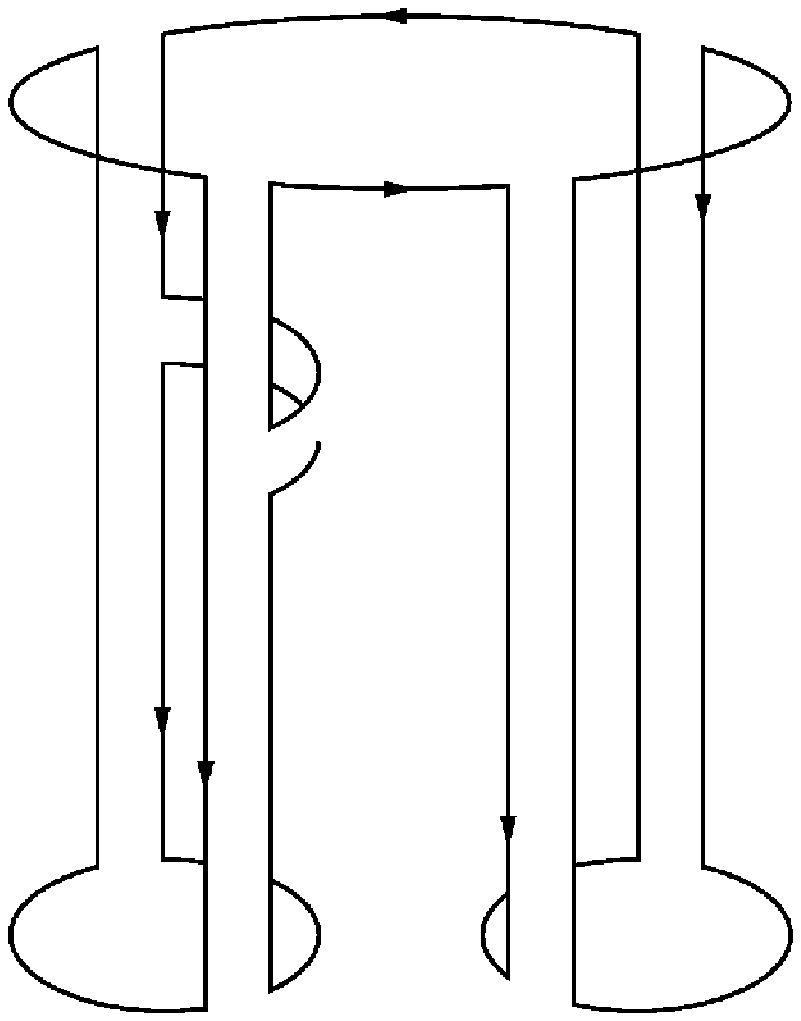}}
  \caption{\small Two combination of  color-factors for the interaction 
between gluons '1' and '2'.}
  \label{fig:same_ql}
\end{figure}

\begin{figure}[htbp]
  \centering
   \parbox{2cm}{\includegraphics[height=1.5cm]{bfkl_1lopp_a.eps}}\parbox{3.5cm}{\center \includegraphics[height=2.5cm]{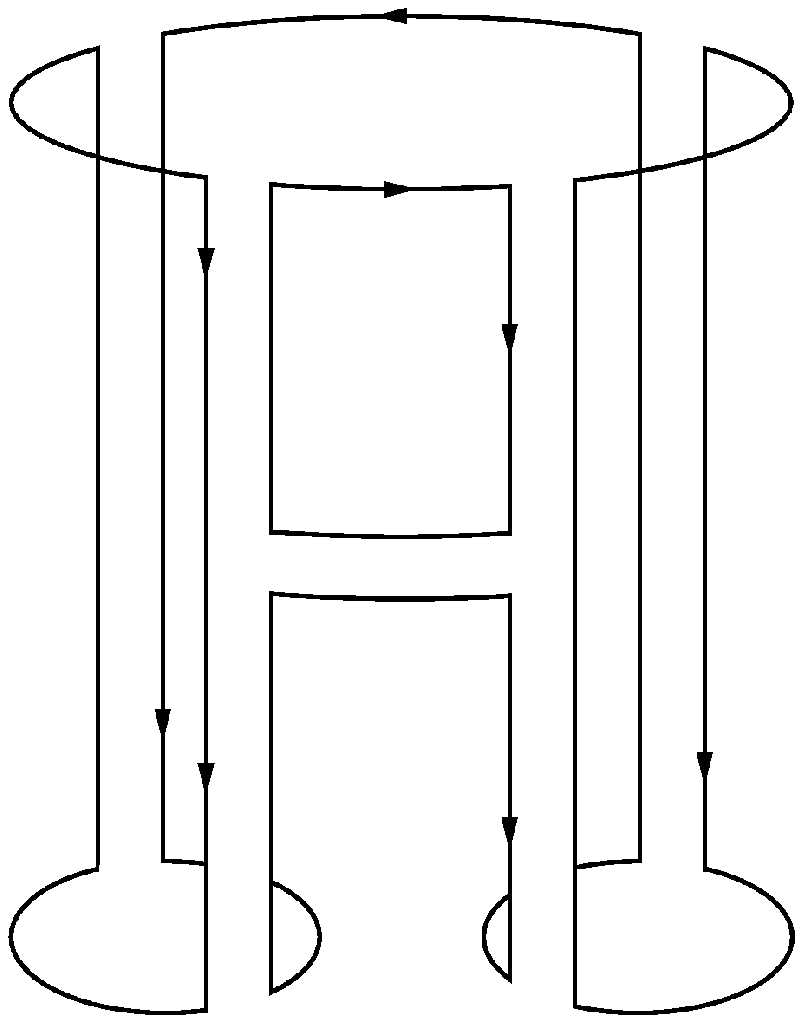}} 
  \caption{\small Color-factors of an interacting between two neighboring $t$-channel gluons.}
  \label{fig:diff_ql_nearby}
\end{figure}

\begin{figure}[htbp]
  \centering
   \parbox{3cm}{\includegraphics[height=1.5cm]{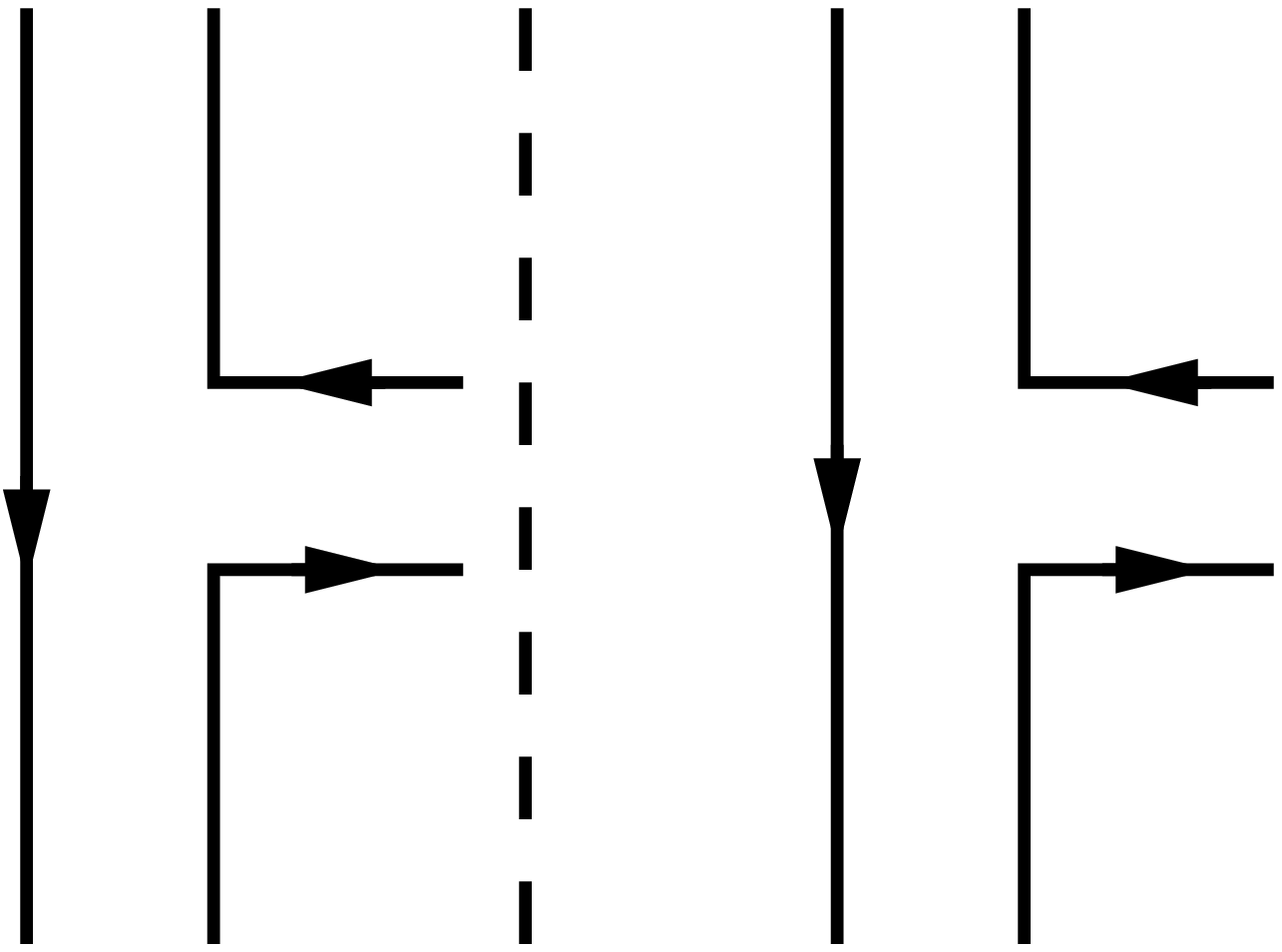}}\parbox{3.5cm}{\center \includegraphics[height=2.5cm]{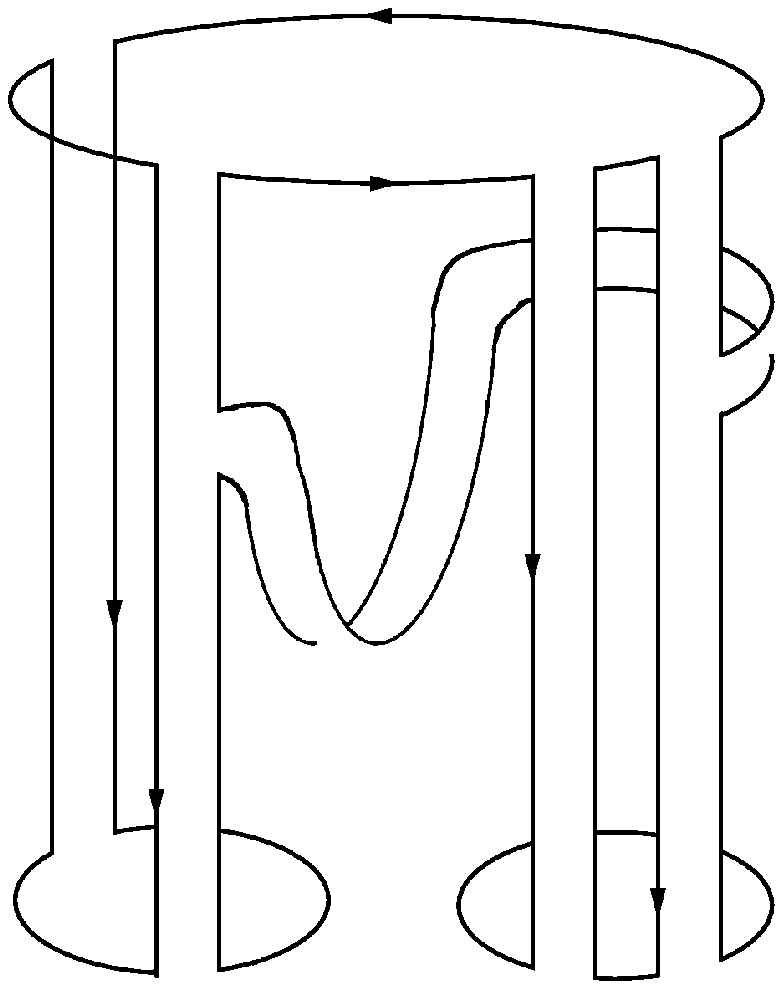}} 
  \caption{\small Color factors of an interaction between two non-neighboring $t$-channel gluons.}
  \label{fig:diff_ql_no_nearby}
\end{figure}

We start with the case, where all four $t$-channel gluons couple to
the upper quark-loop. We therefore consider only insertions of the
two-to-two transition kernel,
\begin{align}
  \label{eq:22kernel_color}
\parbox{2cm}{\includegraphics[width=1.5cm]{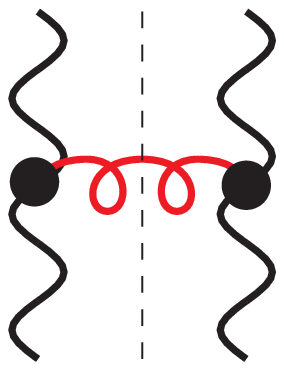}}  = 
\bar{g}^2\big(
     \parbox{.5cm}{\includegraphics[width=.5cm]{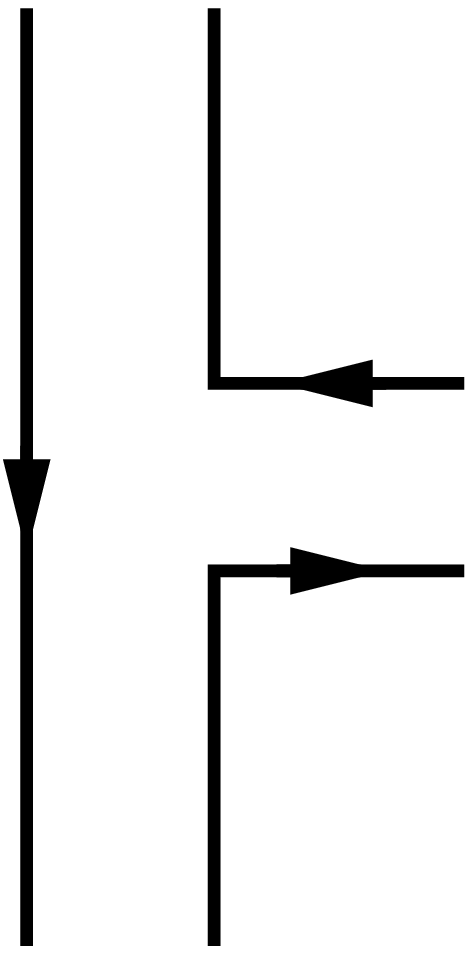}} 
     -
     \parbox{.5cm}{\includegraphics[width=.5cm]{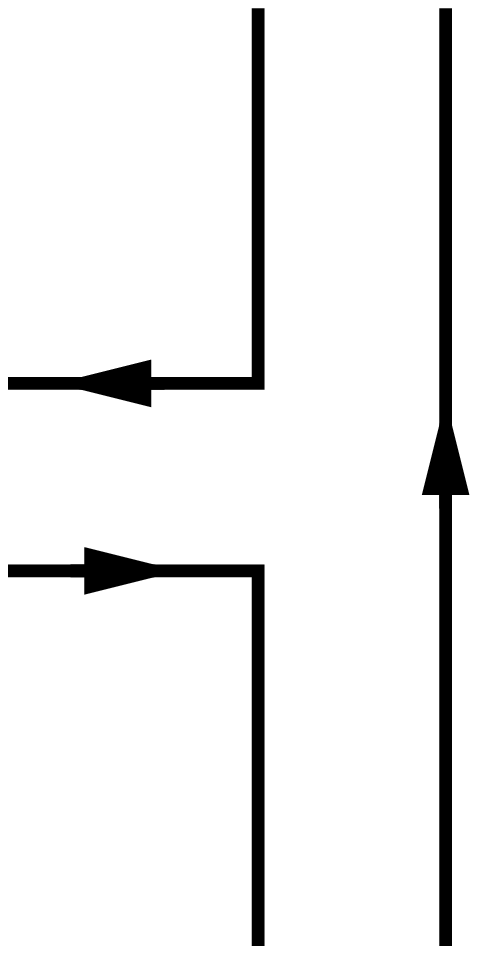}}
\big) \cdot
\big(
     \parbox{.5cm}{\includegraphics[width=.5cm]{structure2.eps}} 
     -
     \parbox{.5cm}{\includegraphics[width=.5cm]{structure1.eps}}
\big)
 K_{2 \to 2}({\bf l}_1, {\bf l }_2;{\bf k}_1, {\bf k}_2), 
\end{align}
into the Born diagrams of Fig.\ref{fig:born}, with $K_{2 \to 2}$ given
by Eq.(\ref{eq:22kernel_momis}). To be definite, we consider corrections 
to the first  diagram of Fig.\ref{fig:born}; the other Born diagrams are treated in the same way. 
We start with the case where
the interaction is between gluons which end up in the same lower
quark-loop, i.e. the interaction is inside the gluon pairs (12) or
(34). In the case of the gluon pair (12), the two combinations of
color factors of Fig.\ref{fig:same_ql} fit on the pair-of-pants,
similar to the cylinder discussed before. An analogous result holds for the 
gluon pair (34). If the
interacting $t$-channel gluons end up in different quark-loops we need
to distinguish between two different cases. In the first 
case, shown in Fig.\ref{fig:diff_ql_nearby}, the interaction is between 
gluons '2' and '3'; on the surface of the upper cylinder the two gluons are 
neighboring.
We call this interaction planar: by contracting the color loop above the rung between gluon '2' and '3',
we are back to the first Born diagram in Fig.\ref{fig:born}.  
In the second case, shown in Fig.\ref{fig:diff_ql_no_nearby}, the 
interaction is between gluon '2' and '4', These two gluons are  
not neighboring, and although the color diagram fits onto the surface of the 
pair-of-pants without any intersection of color lines, it will be referred to  
as 'non-planar'. It cannot be reduced to the Born diagram in Fig.\ref{fig:born}.
Counting closed color loops, it is of the same order as 
that of Fig.\ref{fig:diff_ql_nearby}. 
Note, however, that compared to the planar one it has a relative minus sign.  
The same discussion applies if the interaction is between gluon '1' and '3'.

\begin{figure}[htbp]
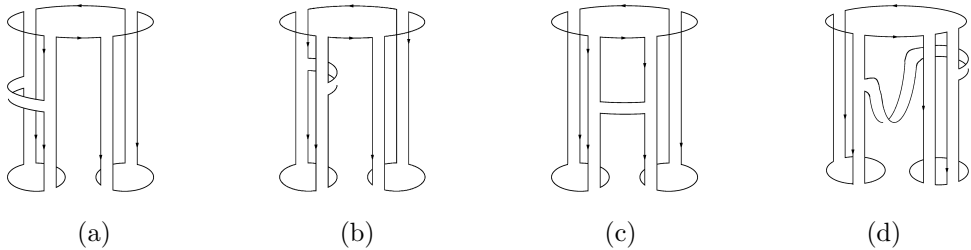

  \centering
     \parbox{3.5cm}{  \center \includegraphics[height=2.5cm]{trouser_inta12_cyl_voll.eps}} 
     \parbox{3.5cm}{\center \includegraphics[height=2.5cm]{trouser_inta.eps}} 
     \parbox{3.5cm}{\center \includegraphics[height=2.5cm]{trouser_inta_cyl_voll12.eps}} 
  \parbox{3.5cm}{\center \includegraphics[height=2.5cm]{trans22_alt2.eps}}   
\\
     \parbox{3.5cm}{\center (a)}\parbox{3.5cm}{\center (b)}\parbox{3.5cm}{\center (c)}\parbox{3.5cm}{\center (d)}
  \caption{\small  Different structures due to insertion of a two-to-two kernel on the pair-of-pants.}
  \label{fig:differentstructures}
\end{figure}

We summarize the four possibilities in Fig.\ref{fig:differentstructures}.
Apart form the last diagram, all diagrams belong to the class of
'planar' graphs.  Note, however, that while in the first graph we can
contract the closed color loop either above or below, in the second
graph we can contract only the lower loop on the lower cylinder: it is
planar w.r.t. the lower left cylinder.

When inserting more two-to-two interactions, it is useful to descend
from the top to the bottom of the diagram: we start by inserting
$s$-channel gluons between $t$-channel gluons which, on the upper
cylinder, are neighboring (Figs.\ref{fig:differentstructures}a and c).
This generates planar graphs.  Moving further down, these planar
insertions come to a stop as soon as one of the
following interactions is included:\\
(i) either an interaction of the type
Fig.\ref{fig:differentstructures}b which is still planar but belongs
to the lower left cylinder. Below such an interactions, further
interactions lie on the surface of one of the two lower cylinders,
i.e. they are inside the pairs $(12)$ or $(34)$. An interaction
between the two cylinders, e.g., between gluon '1' and '3',
loses a color factor $N_c$ and does not contribute on the pair-of-pants.\\
(ii) alternatively, a non-planar interaction of the type
Fig.\ref{fig:differentstructures}: This interaction occurs at most
once, and any further interaction below, again, lies on the surface of
one of the two lower cylinders and hence is inside the pairs $(12)$ or
$(34)$.

\subsection{Two-to-four Reggeon transition }
\label{sec:amplitudes-with-24}
\begin{figure}[htbp]
  \centering
  \parbox{6cm}{\includegraphics[width=4cm]{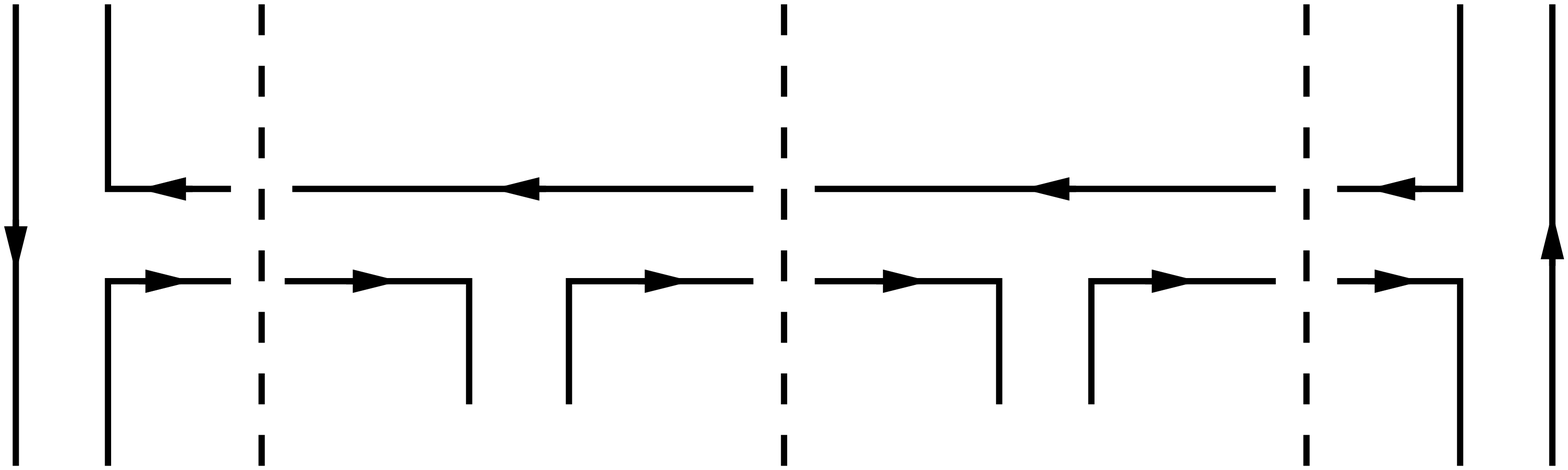}} \parbox{4cm}{    \includegraphics[height=2.2cm]{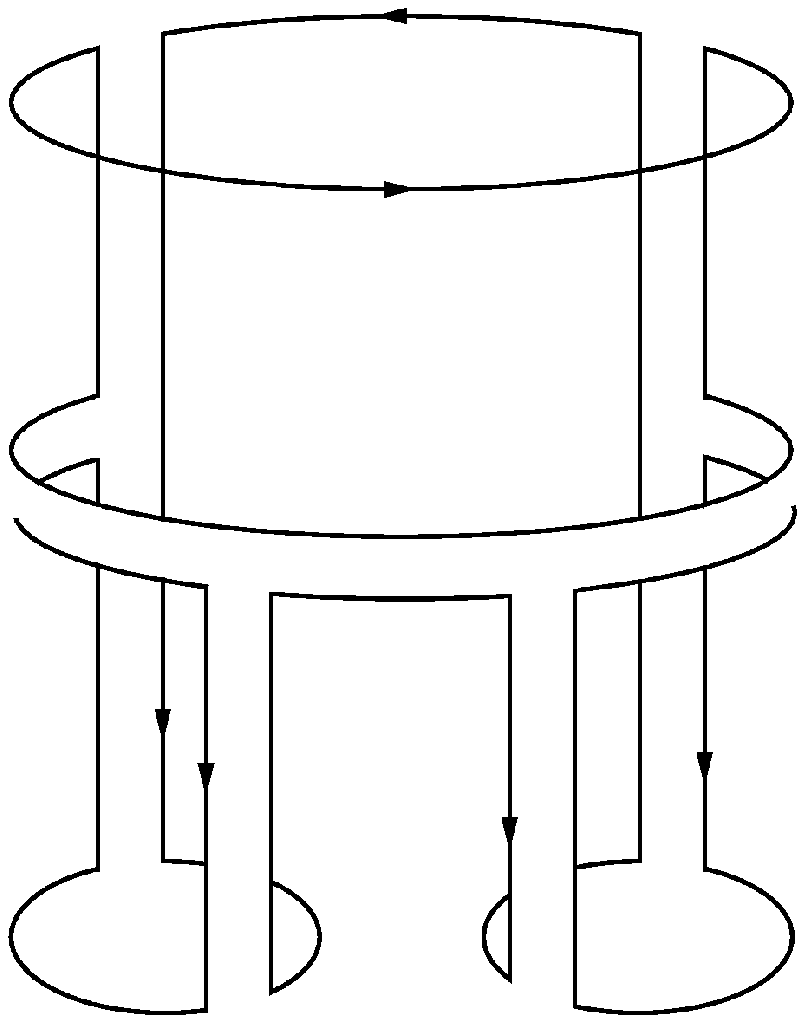}} \\
 \parbox{6cm}{\includegraphics[width=4cm]{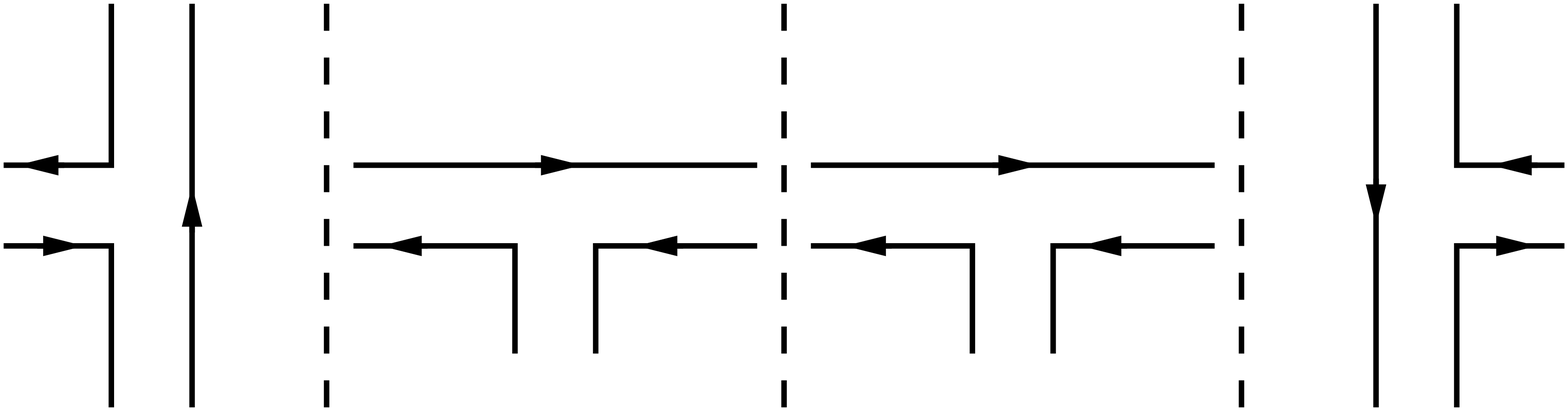}} \parbox{4cm}{    \includegraphics[height=2.2cm]{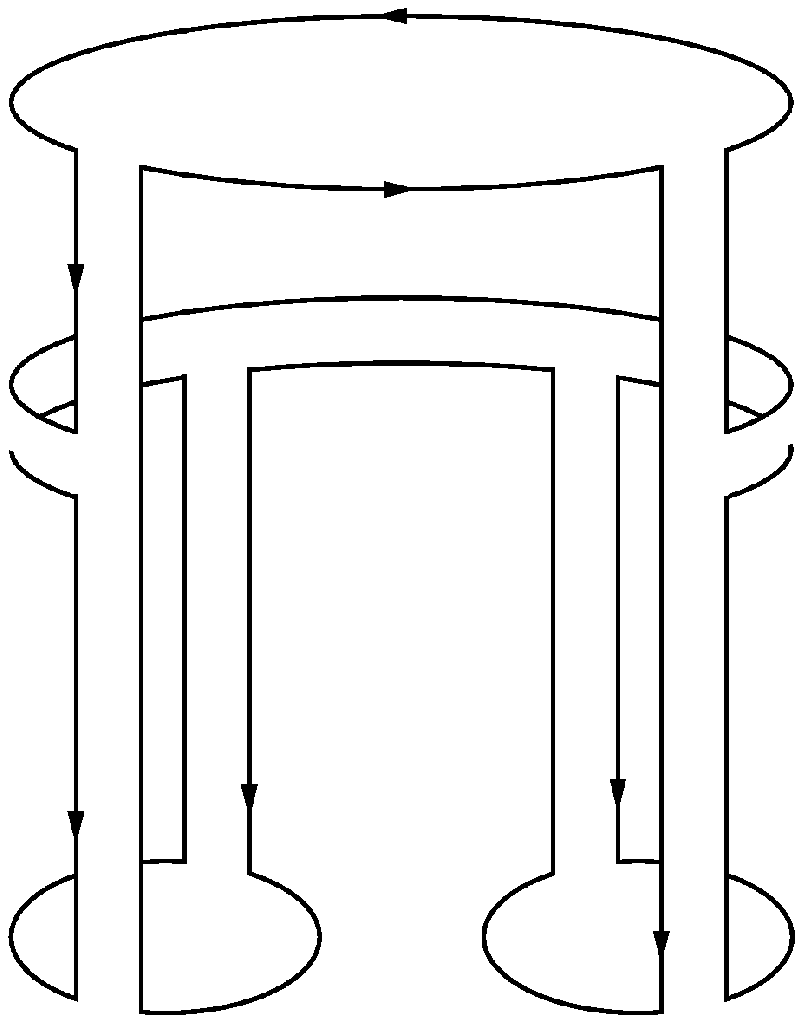}}
  \caption{\small The planar two-to-four Reggeon transition.}
  \label{fig:two4_1}
\end{figure}

\begin{figure}[htbp]
  \centering \parbox{6cm}{\includegraphics[width=4cm]{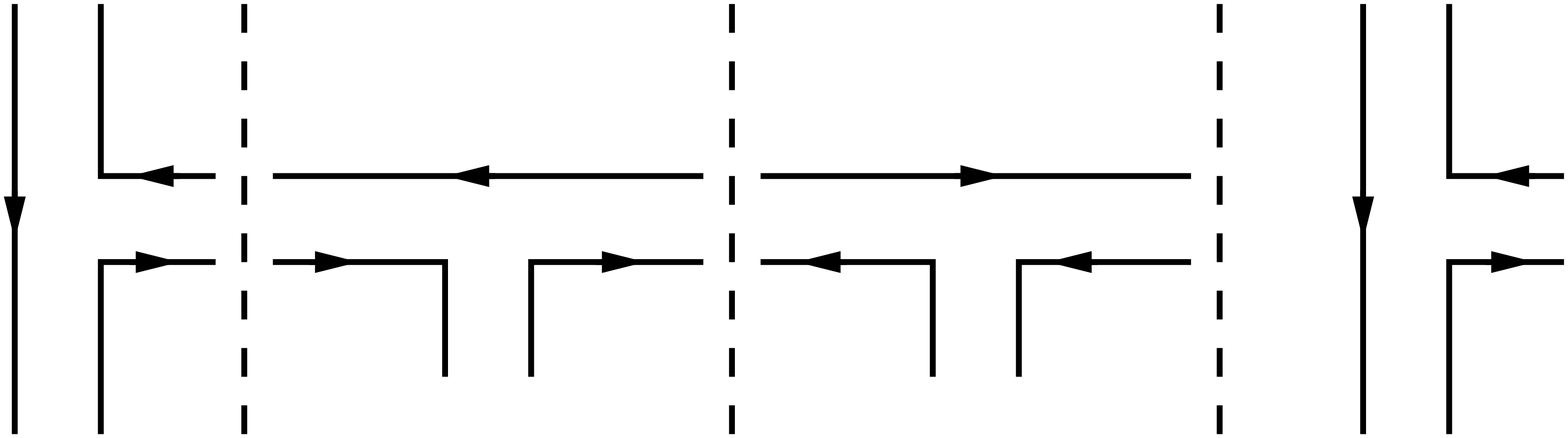}}
  \parbox{4cm}{
    \includegraphics[width=2cm]{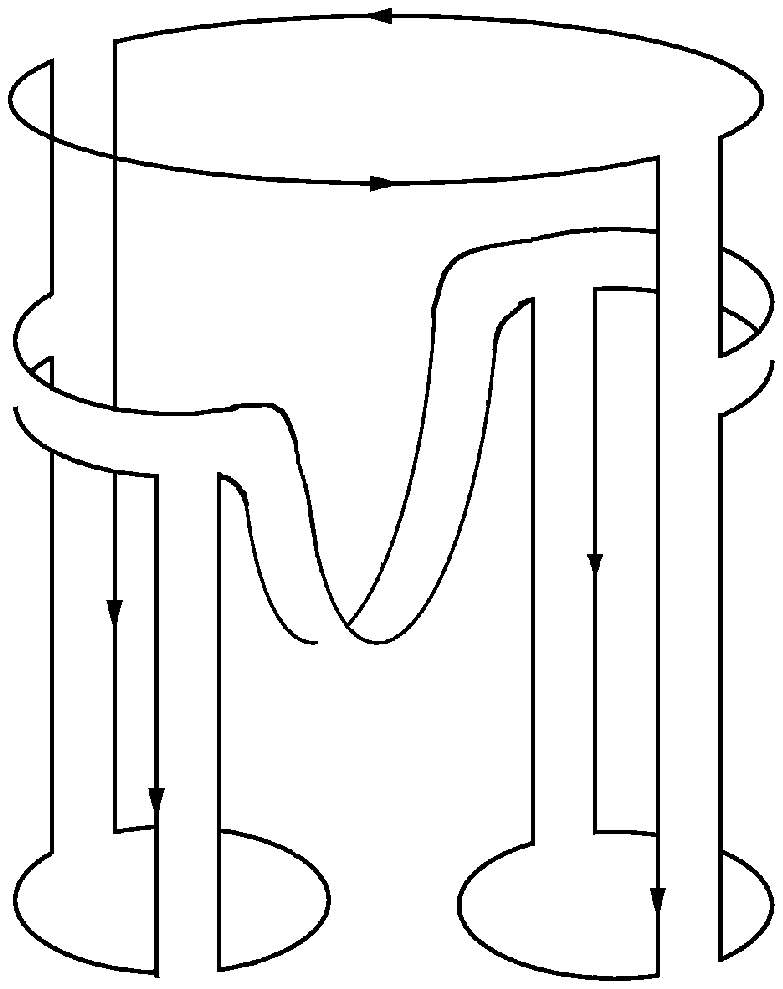}}
  \caption{\small The non-planar two-to-four Reggeon transition.}
  \label{fig:two4_2}
\end{figure}

Apart from the above examples we have further the possibility that some of
the $t$-channel gluons do not start at the upper quark loop but emerge
from produced $s$-channel gluons. We start with the transition from
two to four $t$-channel gluons. In this case we start, at the upper
quark loop, with two $t$-channel gluons which end at a two-to-four
transition. Leaving for the moment details on the momentum structure
of the transition kernel $K_{2\to 4}$ aside, the transition from two
to four reggeized gluons is given by
\begin{align}
  \label{eq:kernel24}
\parbox{2cm}{\includegraphics[height=1cm]{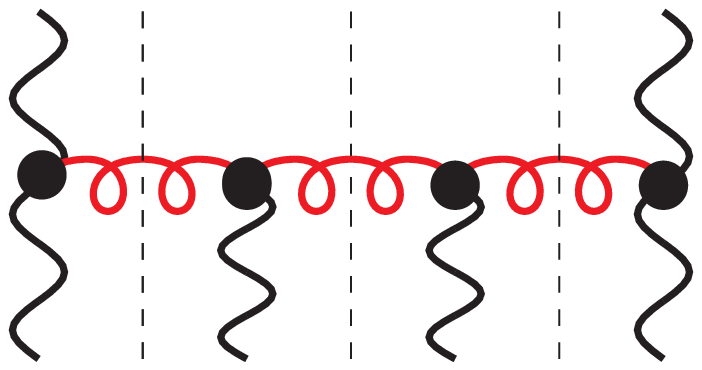}}  
 = {\bar g}^4
 \bigg(
       \parbox{.5cm}{\includegraphics[width=.5cm]{structure1.eps}}
       -
       \parbox{.5cm}{\includegraphics[width=.5cm]{structure2.eps}}
\bigg)
\bigg(
      \parbox{1.1cm}{\includegraphics[height=.5cm]{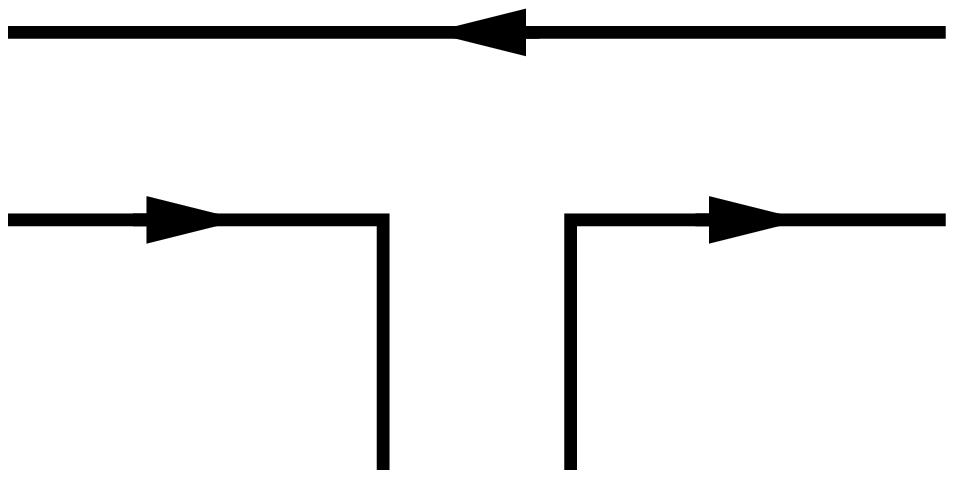}} 
       -
       \parbox{1.1cm}{\includegraphics[height=.5cm]{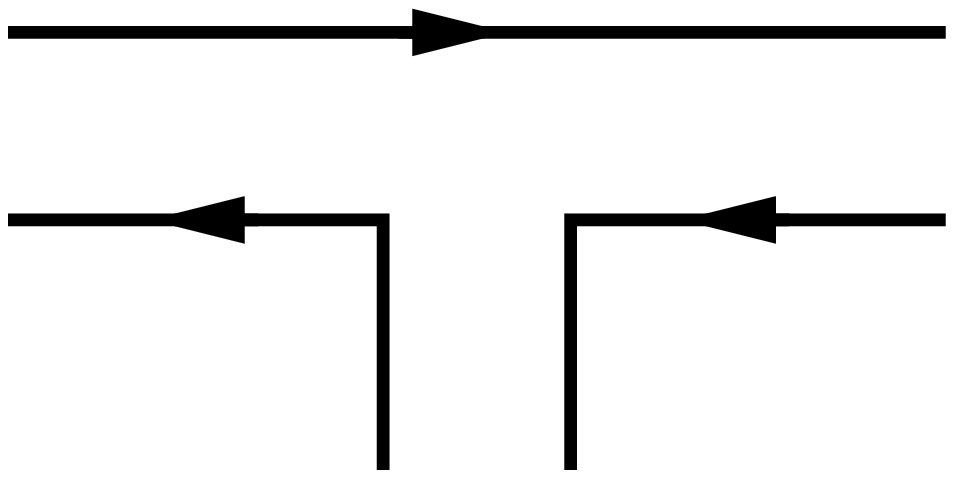}}
\bigg)
\bigg(
      \parbox{1.1cm}{\includegraphics[height=.5cm]{structure4.eps}} 
       -
       \parbox{1.1cm}{\includegraphics[height=.5cm]{structure3.eps}}
\bigg)
\bigg(
      \parbox{.5cm}{\includegraphics[width=.5cm]{structure2.eps}}
      -
      \parbox{.5cm}{\includegraphics[width=.5cm]{structure1.eps}}
\bigg)
K_{2 \to 4}.
\end{align}
Of the 16 possible combinations of color factors, only a subset fits
on the pair-of-pants, of which two examples are shown in
Figs.\ref{fig:two4_1} and \ref{fig:two4_2}.
Again, we have two distinct structures: Fig.\ref{fig:two4_1}, by
contracting the closed color loop on the upper cylinder, can be
reduced to one of the Born diagrams, and thus belongs to class of
planar diagrams. In contrast, Fig.\ref{fig:two4_2} cannot be
contracted and is non-planar.

Further rungs above the $2 \to 4$ transition vertex can always be
contracted to either Figs.\ref{fig:two4_1} or \ref{fig:two4_2}. As to
the interactions below the $2 \to 4$ transition, we have to
distinguish between the two classes.  For the planar class in
Figs.\ref{fig:two4_1}, we can continue by inserting $2\to2$
interactions as described in the previous subsection. For the
non-planar class in Fig.\ref{fig:two4_2} any further $2\to2$
interaction is either inside the pair $(12)$ or $(34)$ and can always
be reduced to Fig.\ref{fig:two4_2}.

For the momentum structure of the $2 \to 4$ transition vertex we need,
apart from the real gluon production vertex in
Eq.(\ref{eq:lipatov_factor}), the vertex which describes coupling of a
$t$-channel gluon to a real $s$-channel gluon, which is known as the
Reggeon-Particle-Particle (RPP)-vertex. Similar to the production
vertex, the RPP-vertex is an effective vertex. To lowest order, for
the scattering of a gluon on an antiquark at high center of mass
energies, this RPP-vertex is built up from the following diagrams:
\begin{align}
  \label{eq:1diag_rpr}
\parbox{1.5cm}{\includegraphics[width=1.5cm]{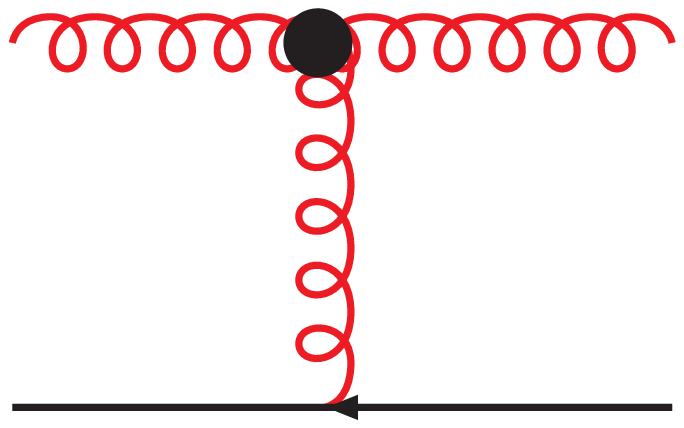}}
=
\parbox{1.5cm}{\includegraphics[width=1.5cm]{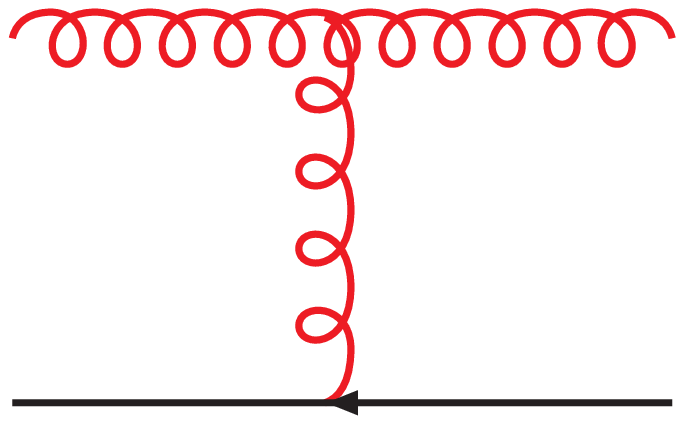}}
+
\parbox{1.5cm}{\includegraphics[width=1.5cm]{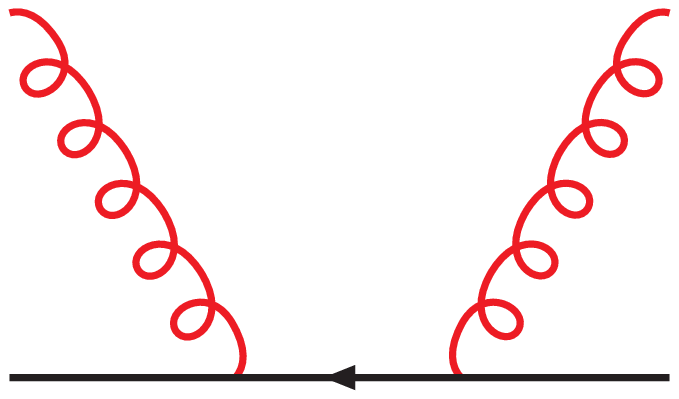}}
+
\parbox{1.5cm}{\includegraphics[width=1.5cm]{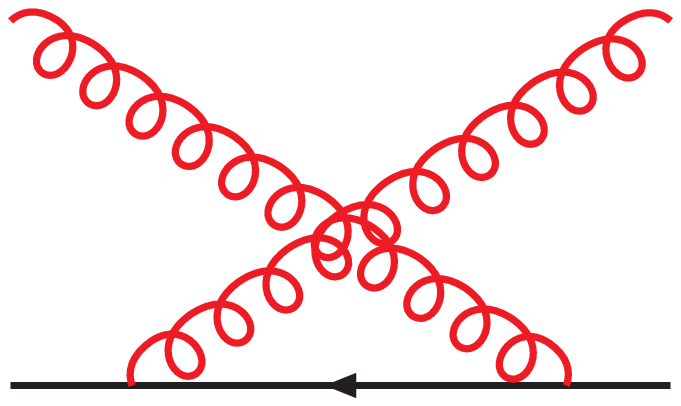}.}
\end{align}
Similar to the production vertex, at high center of mass energies the
last two diagrams coincide with each other up to a sign, and color
factor and momentum part of the RPP-vertex can be written in the 
factorized form
\begin{align}
  \label{eq:colorrpp_abstract}
-i{\bar{g}} \left[
\parbox{1.5cm}{\includegraphics[width=1.5cm]{structure4.eps}} 
-
 \parbox{1.5cm}{\includegraphics[width=1.5cm]{structure3.eps}}
\right] \epsilon^*_{(\lambda')} \!\!\cdot\!\Gamma\!\cdot\!\epsilon_{(\lambda)}.
\end{align}
Sandwiching one or two RPP vertices between two production vertices
then leads to two-to-three and two-to-four Reggeon transition kernels.
They have been constructed in \cite{bkp} and for details we refer to
this reference.  The momentum part of the the $2\to 4$ transition is
then given by
\begin{align}
  \label{eq:def_k24}
K_{2 \to 4} =  {\bf q}^2
   &-
   \frac{{\bf l}_1^2 ({\bf q} - {\bf k}_1)^2}{ ({\bf l}_1 - {\bf k}_1)^2 } 
   -
   \frac{{\bf l}_2^2 ({\bf k}_1 + {\bf k}_2 + {\bf k}_3)^2}{ ({\bf l}_2 - {\bf k}_4)^2 }
   +
   \frac{{\bf l}_1^2 ({\bf q}  - {\bf k}_1)^2 {\bf l}^2_2 }{ ({\bf l}_1 - {\bf k}_1)^2 ({\bf l}_2 - {\bf k}_4)^2 }.
\end{align}

\subsection{Two-to-three Reggeon transition}
\label{sec:amplitudes-with-234}
Last we need to consider diagrams that contain, at least, one
transition from two to three gluons. Beginning with the momentum
structure, we remind that the coupling of an additional 'new'
$t$-channel to an $s$-channel gluon takes place by the RPP-vertex in
Eq.(\ref{eq:colorrpp_abstract}). The result for two-to-three Reggeon
transition is described by
\begin{align}
  \label{eq:kernel23}
\parbox{2cm}{\includegraphics[height=1cm]{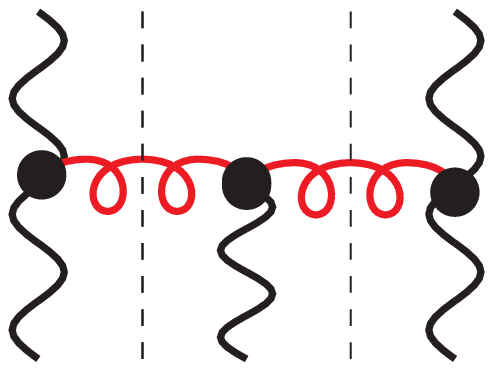}} 
 =
i\bar{g}^3 \bigg(
       \parbox{.5cm}{\includegraphics[width=.5cm]{structure1.eps}}
       -
       \parbox{.5cm}{\includegraphics[width=.5cm]{structure2.eps}}
\bigg)
\bigg(
      \parbox{1.1cm}{\includegraphics[height=.5cm]{structure4.eps}} 
       -
       \parbox{1.1cm}{\includegraphics[height=.5cm]{structure3.eps}}
\bigg)
\bigg(
      \parbox{.5cm}{\includegraphics[width=.5cm]{structure2.eps}}
      -
      \parbox{.5cm}{\includegraphics[width=.5cm]{structure1.eps}}
\bigg)
K_{2 \to 3}
\end{align}
with
\begin{align}
  \label{eq:k23_def}
 K_{2 \to 3}^{(\{12\} \to \{123\})} =   ( {\bf k}_1 +{\bf k}_2 + {\bf k}_3 )^2
         &-
         \frac{{\bf l}_1^2 ({\bf k}_2 + {\bf k}_3)^2}{ ({\bf l}_1 - {\bf k}_1)^2 } 
         -
         \frac{{\bf l}_2^2 ({\bf k}_1 + {\bf k}_2)^2}{ ({\bf l}_2 - {\bf k}_3)^2 }
         +
         \frac{{\bf l}_1^2 {\bf k}_2^2 {\bf l}^2_2 }{ ({\bf l}_1 - {\bf k}_1)^2 ({\bf l}_2 - {\bf k}_3)^2 } ,
\end{align}
where the superscripts on the kernel refer to the ingoing and outgoing
$t$-channel gluons respectively. As to the general structure of diagrams 
with $2 \to 3$ transitions, we have two possibilities: either we start, 
at the upper quark loop, 
with three gluons. In this case we need, somewhere further below, only one $2 \to 3$ transition.
Alternatively, we could start with two gluons and then need two $2 \to 3$ transition vertices.     

\begin{figure}[htbp]
  \centering
\parbox{5cm}{\includegraphics[width=3cm]{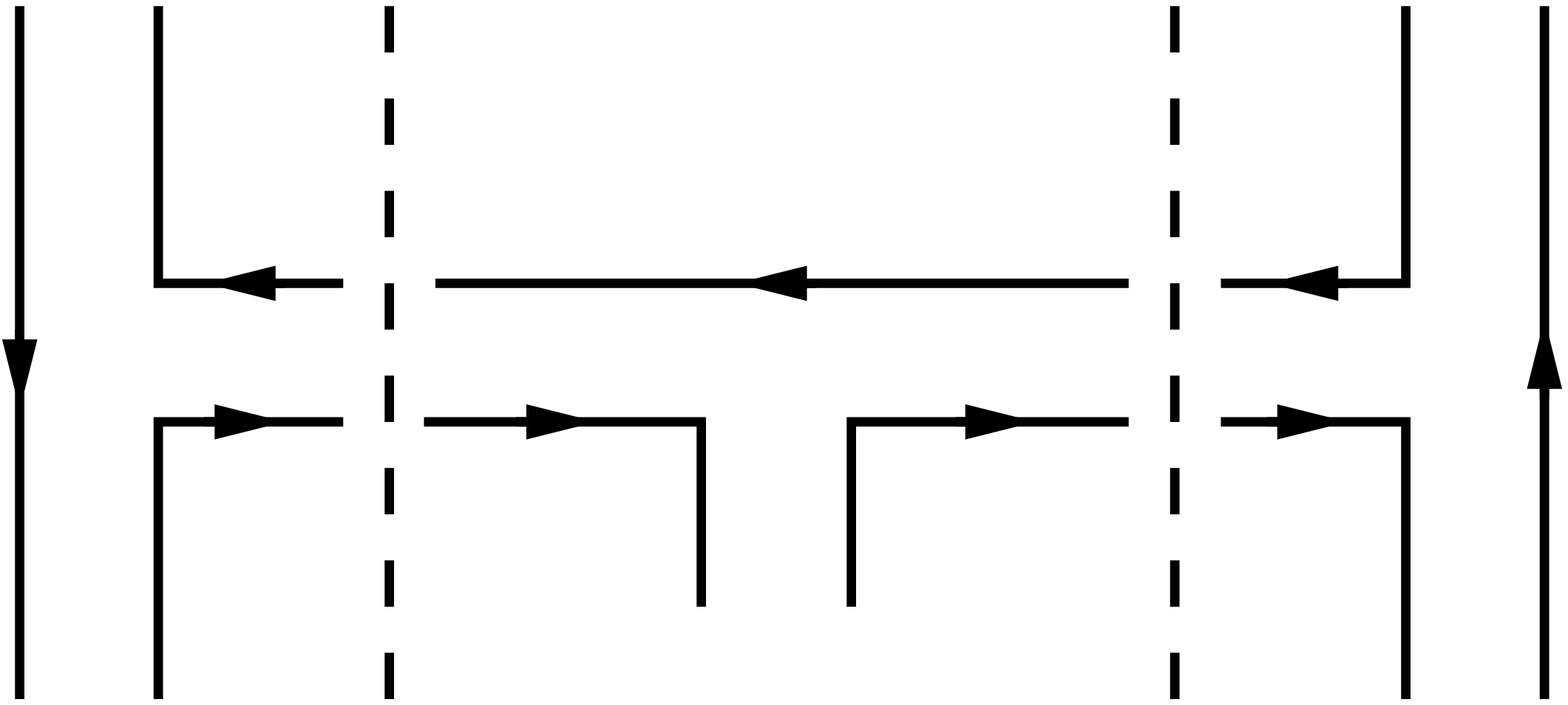}}  \parbox{4cm}{\includegraphics[height=2.5cm]{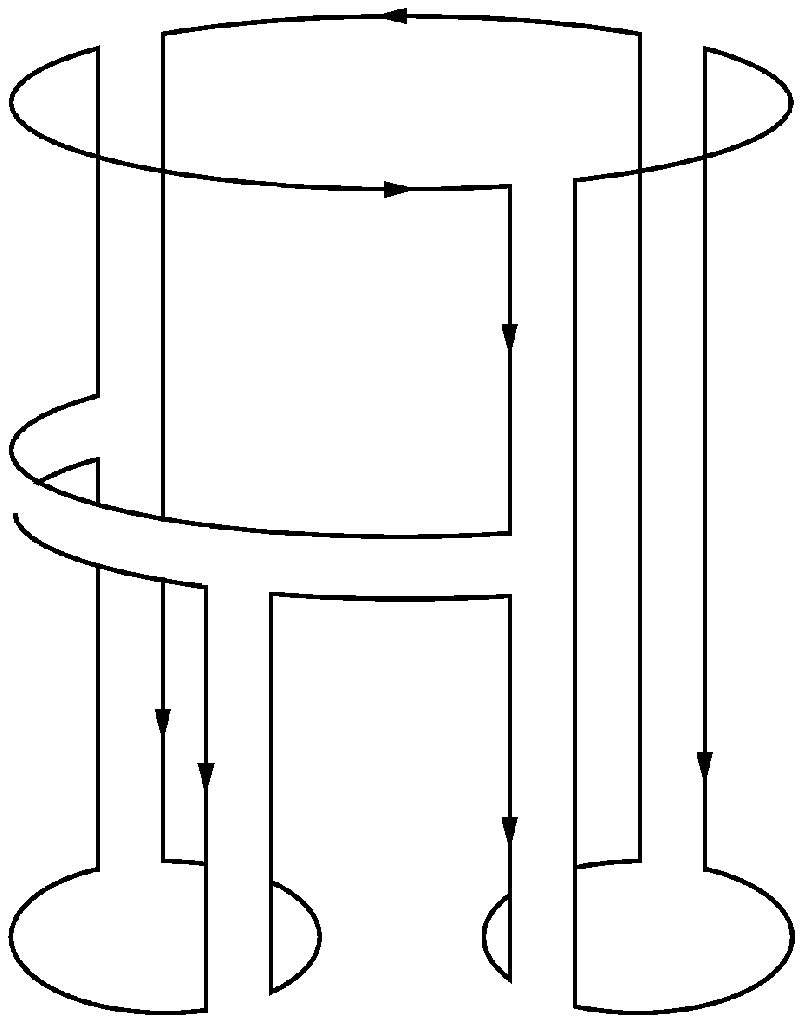}}\\
  \caption{\small Planar graph with one two-to-three transition.}
  \label{fig:two3_1}
\end{figure}

\begin{figure}[htbp]
  \centering
  \parbox{5cm}{\includegraphics[width=3cm]{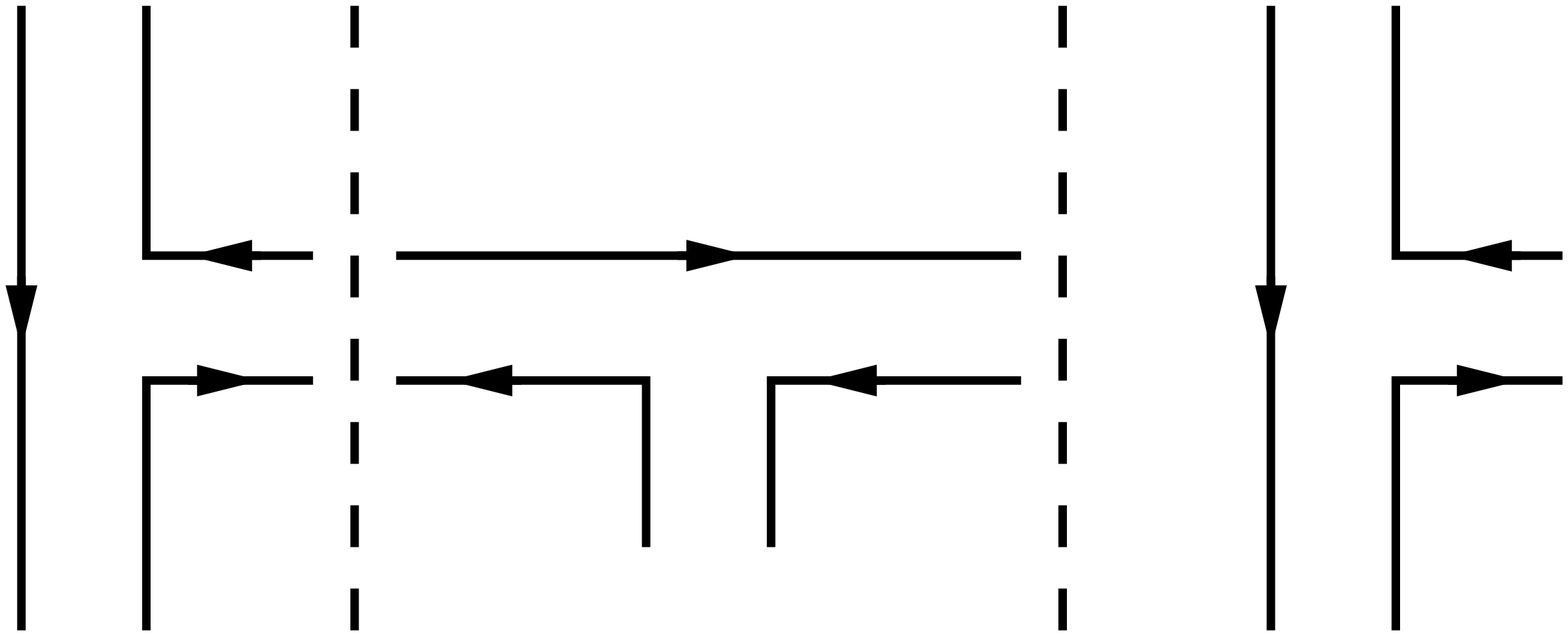}} 
  \parbox{4cm}{  \center \includegraphics[height=2.5cm]{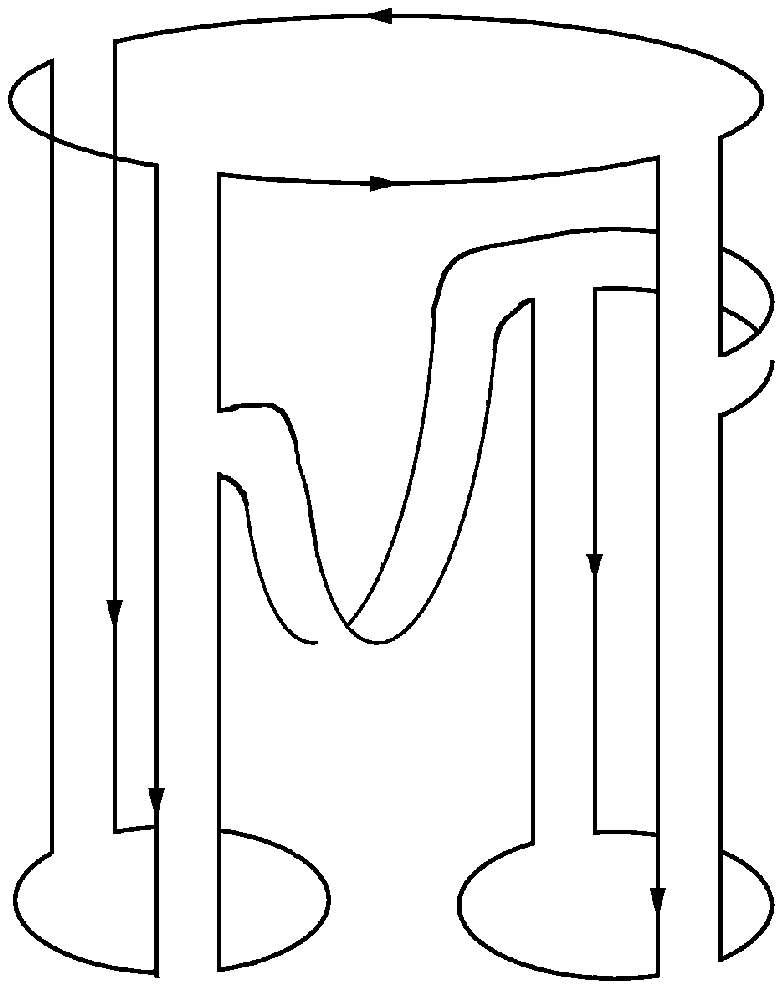}}
  \caption{\small Non-planar graph with one two-to-three transition.}
  \label{fig:two3_2}
\end{figure}

For the discussion of the color diagrams we start with the former case. 
As before we encounter planar and non-planar graphs. 
An example of a planar graph is shown in Fig.\ref{fig:two3_1}, 
a non-planar graph in Fig.\ref{fig:two3_2}. In both cases, the planar class, Fig.\ref{fig:two3_1},
 and the non-planar class, Fig.\ref{fig:two3_2},
above the vertex any further $2\to2$ interaction is between lines which
are neighbored on the surface of the upper cylinder. Below the $2\to3$ vertex 
we have the same situation as for the $2\to4$ vertex: for the planar graph 
in  Fig.\ref{fig:two3_1} we proceed as described in Sec.\ref{sec:amplitudes-with-4}, whereas for 
the non-planar graph there is no further communication between the two lower 
cylinders.     

\begin{figure}[htbp]
  \centering
  \parbox{5cm}{\includegraphics[width=4cm]{two3_a.eps}}
  \parbox{5cm}{  \center \includegraphics[height=3cm]{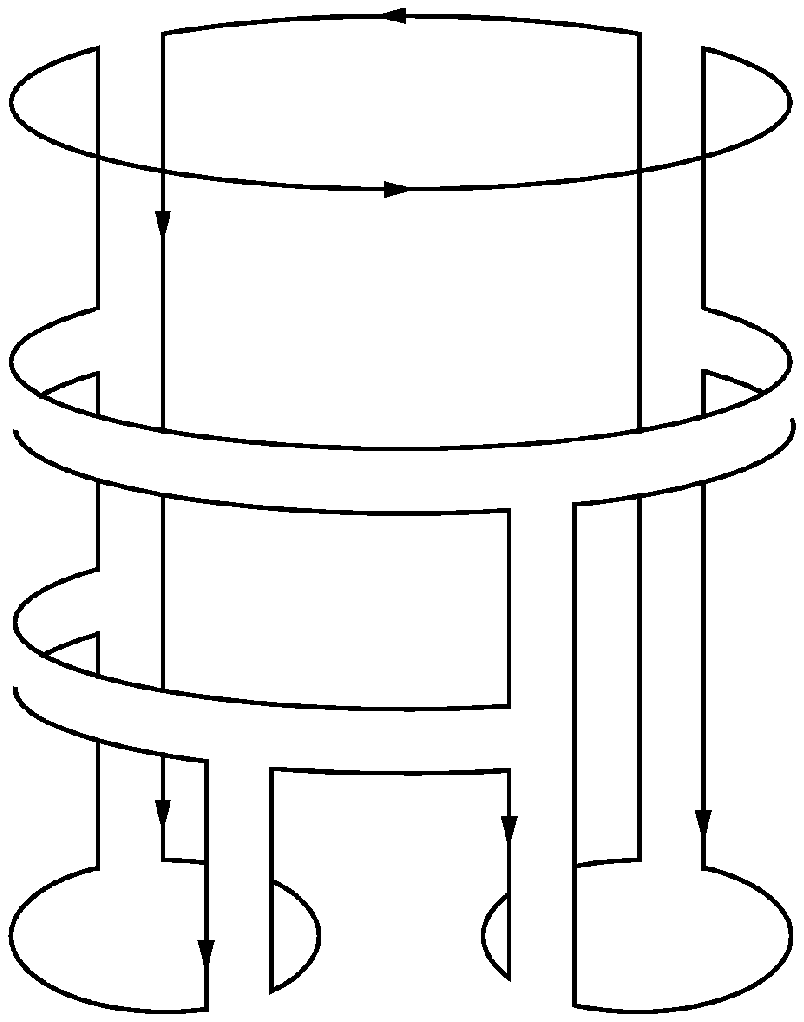}} \\
  \parbox{5cm}{\includegraphics[width=4cm]{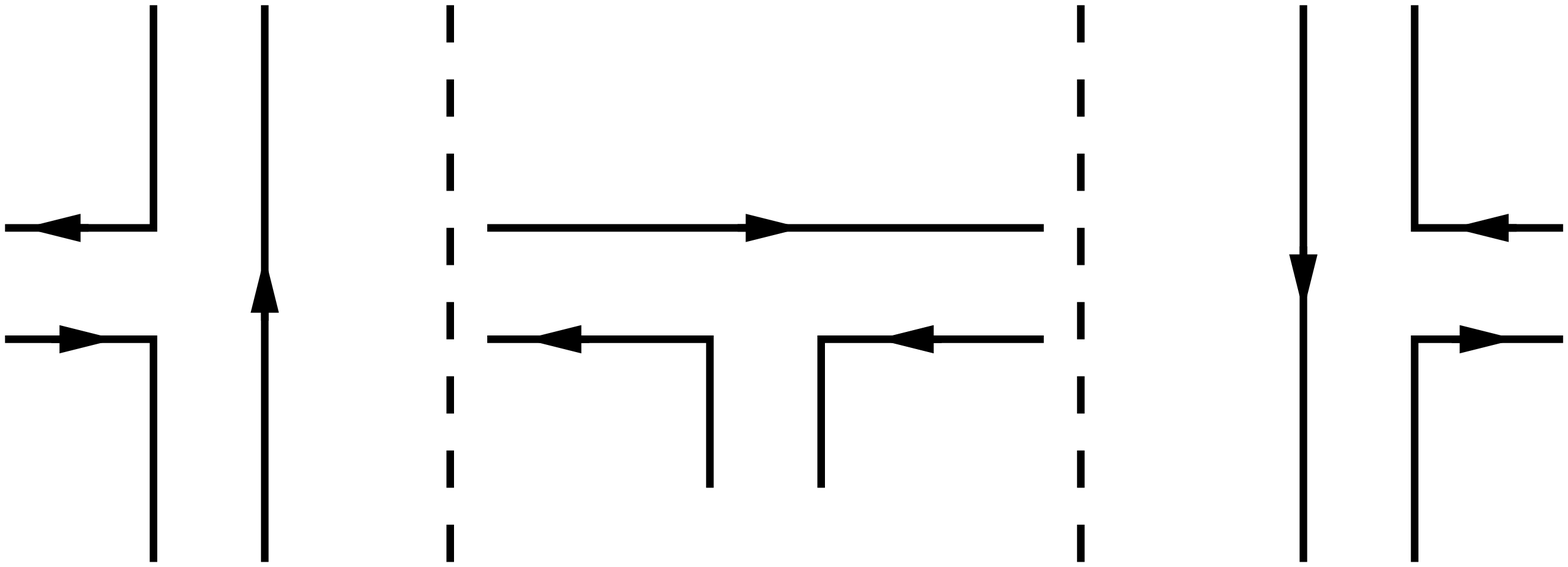}}
  \parbox{5cm}{  \center \includegraphics[height=3cm]{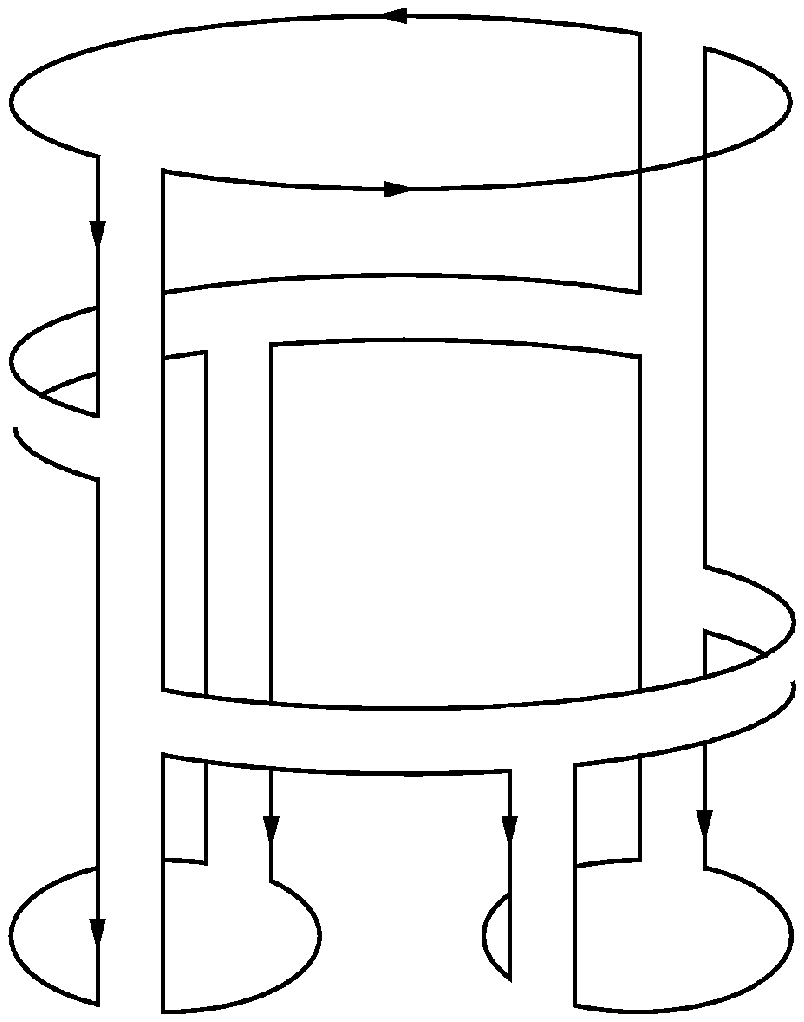}}
  \caption{\small Combination of two transitions of two to three $t$-channel gluons.     From Eq.(\ref{eq:kernel23}) one finds a relative minus sign for
   the second combination.}
  \label{fig:two3_4}
\end{figure}

For the graphs which, at the top, start with two $t$-channel gluons 
the two classes arise in the following way. Beginning at the top, the first 
$2\to3$ transition always lies on the surface of the upper cylinder.
Below this vertex we have the same situation which we have described 
a moment ago, i.e. the same as for the case where at the fermion quark loop 
we start with three gluons: the second $2\to3$ transition decides whether 
the graph is planar or non-planar. Planar examples are shown in 
Fig.\ref{fig:two3_4}.

\subsection{Planar and non-planar partial waves}
\label{sec:below}
In this final subsection we collect the results and prepare the 
summation of all diagrams on the pair-of-pants surface.
At the end of Sec.\ref{sec:sixpoint} we proposed to factorize
at the branching point, i.e. at the last interaction between the two 'legs' 
of the pair-of-pants,  the partial wave $F(\omega, \omega_1, \omega_2)$ 
into a convolution of three amplitudes $\mathcal{D}_4(\omega)$,
$\mathcal{D}_2(\omega_1)$, and $\mathcal{D}_2(\omega_2)$,
depending on $\omega$,
$\omega_1$, or  $\omega_2$, resp.. In order to find closed expressions 
for these amplitudes we shall, similar to our treatment of the diagrams in 
the plane and on the surface of the cylinder,
formulate integral equations which sum up the different classes of 
diagrams. 

The situation is easiest for the two
amplitudes that start from the two lower virtual photons,
$\mathcal{D}_2(\omega_1)$, and $\mathcal{D}_2(\omega_2)$. For these
amplitudes we need to resum, for both legs, contributions like those in
Fig.\ref{fig:same_ql}. Their resummation yields the BFKL-equation on the
cylinder Eq.(\ref{eq:bfkl-singlet}).  

The upper amplitude, $\mathcal{D}_4(\omega)$ is defined to include the
branching vertex, which by definition is the lowest interaction that
connects the gluon pair (12) with the pair (34). Below this vertex,
there are only interactions inside the pair $(12)$ or $(34)$. The
branching vertex can be either a $2\to 2$, a $2\to3$, or a $2\to4$
transition vertex.  From this definition it follows that the upper
amplitude always has four reggeized gluons at its lower end.

From now on it becomes important to distinguish between 'planar' and
'non-planar' graphs.
\begin{figure}[htbp]
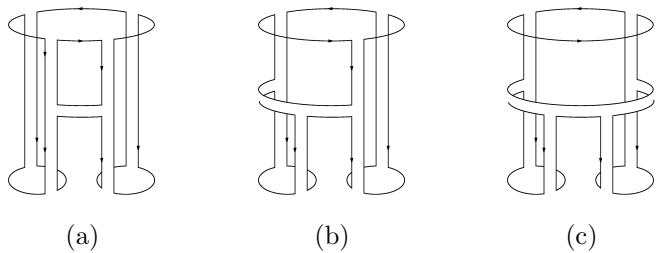

  \centering
  \parbox{3.2cm}{\center \includegraphics[height=2.5cm]{trouser_inta_cyl_voll12.eps}} 
  \parbox{3.2cm}{\center \includegraphics[height=2.5cm]{trouser_inta_cyl323.eps}}
  \parbox{3.2cm}{ \center   \includegraphics[height=2.5cm]{trouser_inta_cyl24.eps}} \\
\parbox{3.2cm}{\center (a)}
\parbox{3.2cm}{\center (b)}
\parbox{3.2cm}{\center (c)}
  \caption{\small 'Planar' color factors on the pair-of-pants which reduce in an apparent way to the color structure of the Born-term by extracting a closed color loop with the \emph{upper} quark-loop.}
  \label{fig:planar_omega}
\end{figure}
\begin{figure}[htbp]
  \centering
  \parbox{2.9cm}{\center \includegraphics[height=2.5cm]{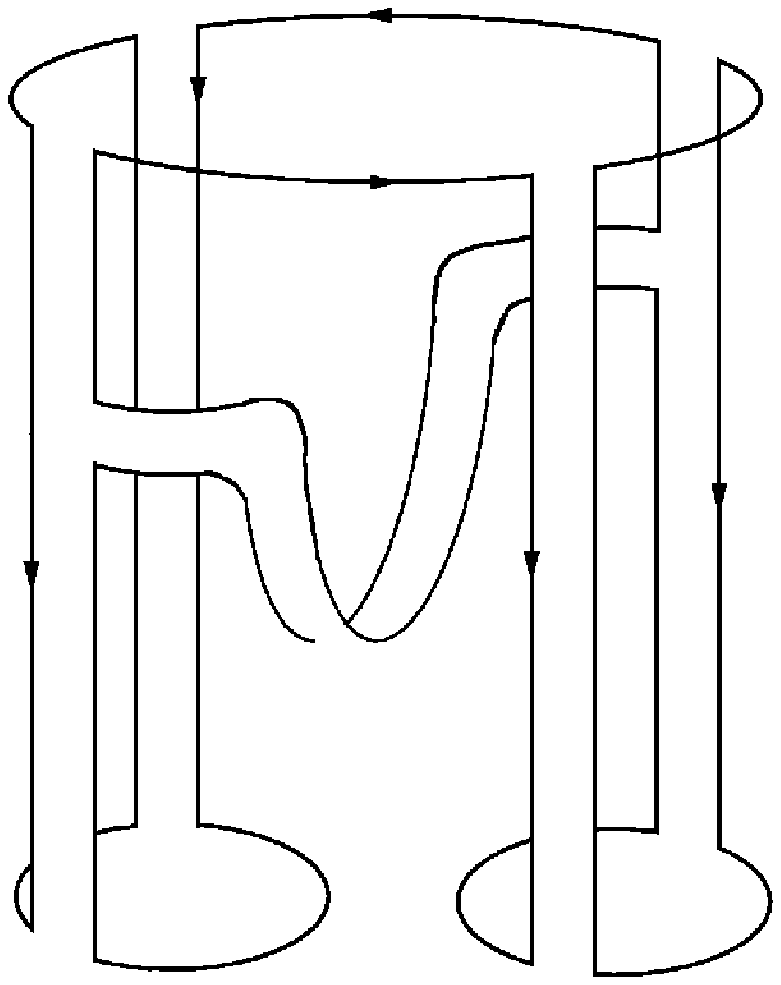}}
  \parbox{2.9cm}{\center \includegraphics[height=2.5cm]{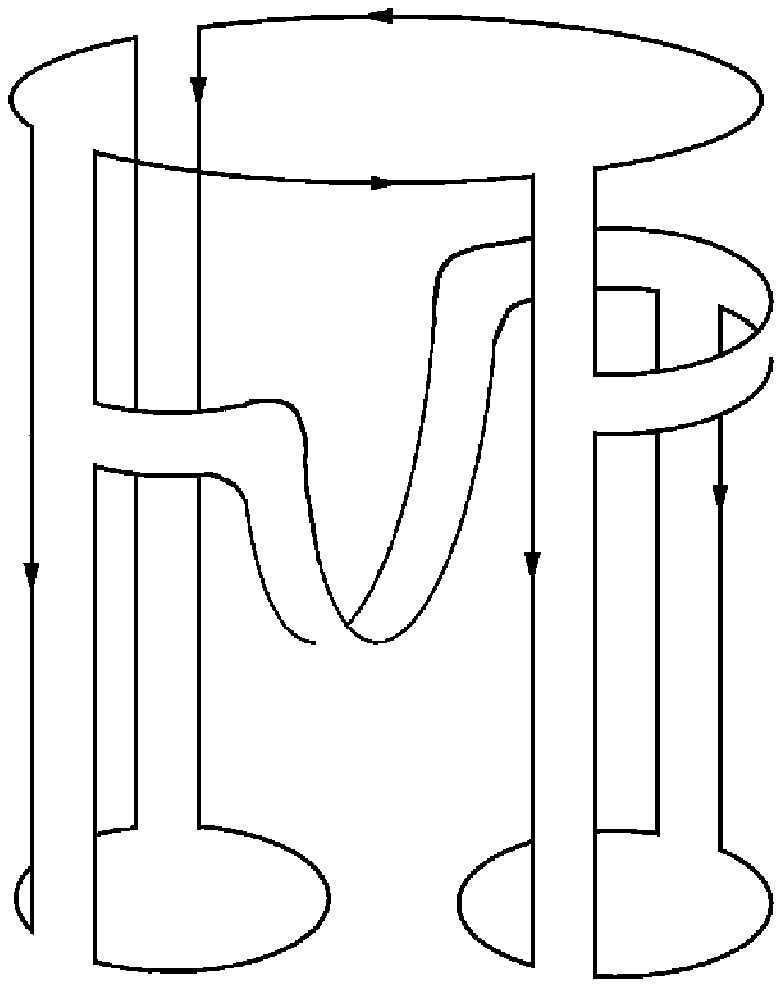}}
  \parbox{2.9cm}{\center \includegraphics[height=2.5cm]{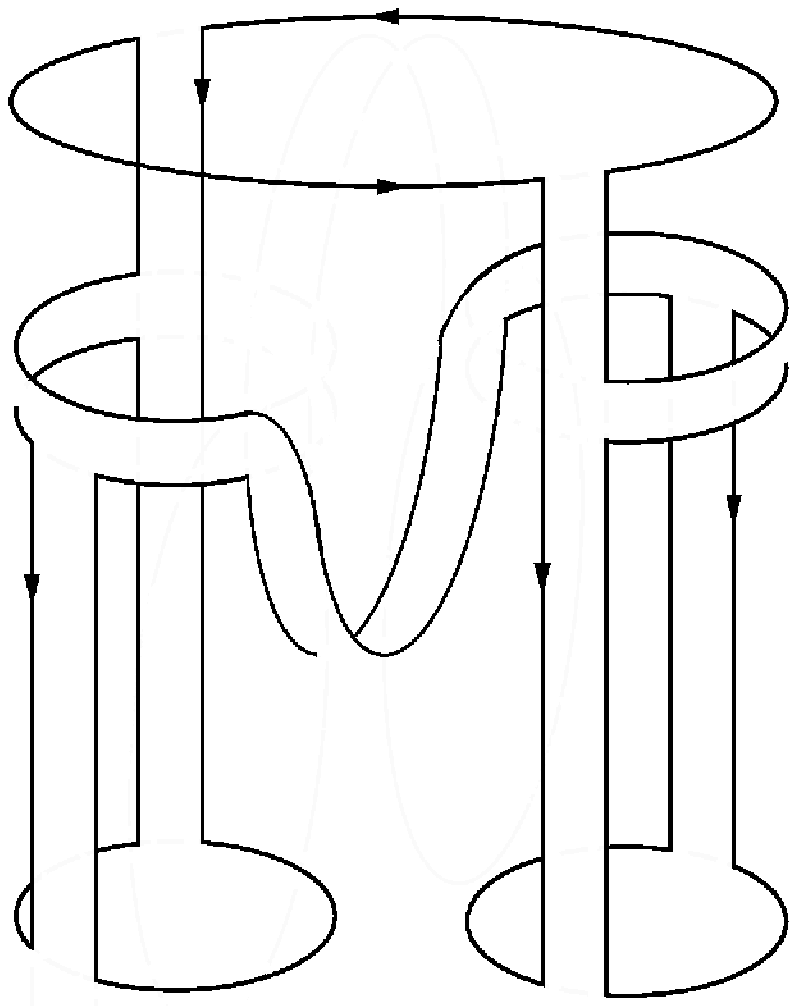}}\\
\parbox{2.9cm}{\center (a)}\parbox{2.9cm}{\center (b)}\parbox{2.9cm}{\center (c)}
  \caption{\small 'Non-planar' color factors that cannot be reduced trivially to the color structure of the Born-term by extracting a closed color loop with the \emph{upper} quark-loop.}
  \label{fig:non_planar_omega}
\end{figure}
For convenience we list, once more, three examples of planar and non-planar 
graphs: Fig.\ref{fig:planar_omega} contains the planar example, and 
Fig.\ref{fig:non_planar_omega} the non-planar ones. 
They are of the order $g^8 N_c^3 (g^2 N_c)$.

Beginning with the planar diagrams in Fig.\ref{fig:planar_omega} 
one easily verifies that we always 
can contract one closed color loop to arrive at the first diagram of Fig.\ref{fig:born}. 
For the other diagrams in  Fig.\ref{fig:born} we have analogous sets of graphs.  
Next, the figures Fig.\ref{fig:planar_omega}a-c differ from each other in that, at the fermion loop at the top, 
in (a) we start with four gluons, in (b) with only three gluons, 
and in (c) with two gluons. 
Correspondingly, in (a) the branching vertex consists of a two-to-to kernel, 
in (b) a two-to-three kernel, and in (c)  a two-to-four kernel. 
At the branching vertex, all diagrams have four $t$-channel gluon 
lines. Higher order diagrams are obtained by inserting, above the
branching vertex, pairwise interactions between neighboring $t$-channel gluons. 
Also, in Fig.\ref{fig:planar_omega}b and d we could insert a $2\to2$ 
interaction between gluons '2' and '3' or '1' and '4': in this case, the branching vertex 
consists of a two-to-to kernel. In all cases,   
by drawing a horizontal cutting line just below the branching  
vertex, we arrive at amplitudes with four $t$-channel gluons.  
We will denote the sum of all graphs by $\mathcal{D}_4^{(1234)}(\omega)$,  
 $\mathcal{D}_4^{(2134)}(\omega)$, $\mathcal{D}_4^{(2143)}(\omega)$, 
and  $\mathcal{D}_4^{(2134)}(\omega)$ where 
the upper label refers to the four terms in Fig.\ref{fig:born}, 
i.e. it indicates to which of the four Born terms the diagrams can be 
contracted (a more precise definition will be given further below in section
\ref{sec:regg_ql}). 

For the task of summing of all diagrams it will be convenient to define also 
auxiliary amplitudes with two and three gluons in the $t$-channel. 
In Fig.\ref{fig:planar_omega}c
one sees that, starting at the upper quark loop, we begin with two 
gluons which, when higher order corrections are included, interact through 
two-to-two kernels. The sum of these 
diagrams leads to the BFKL cylinder and therefore coincides with 
$\mathcal{D}_2(\omega)$ of Sec.\ref{sec:cylinder}. 
\begin{figure}[htbp]
  \centering
  \parbox{3cm}{\center    \includegraphics[width=2cm]{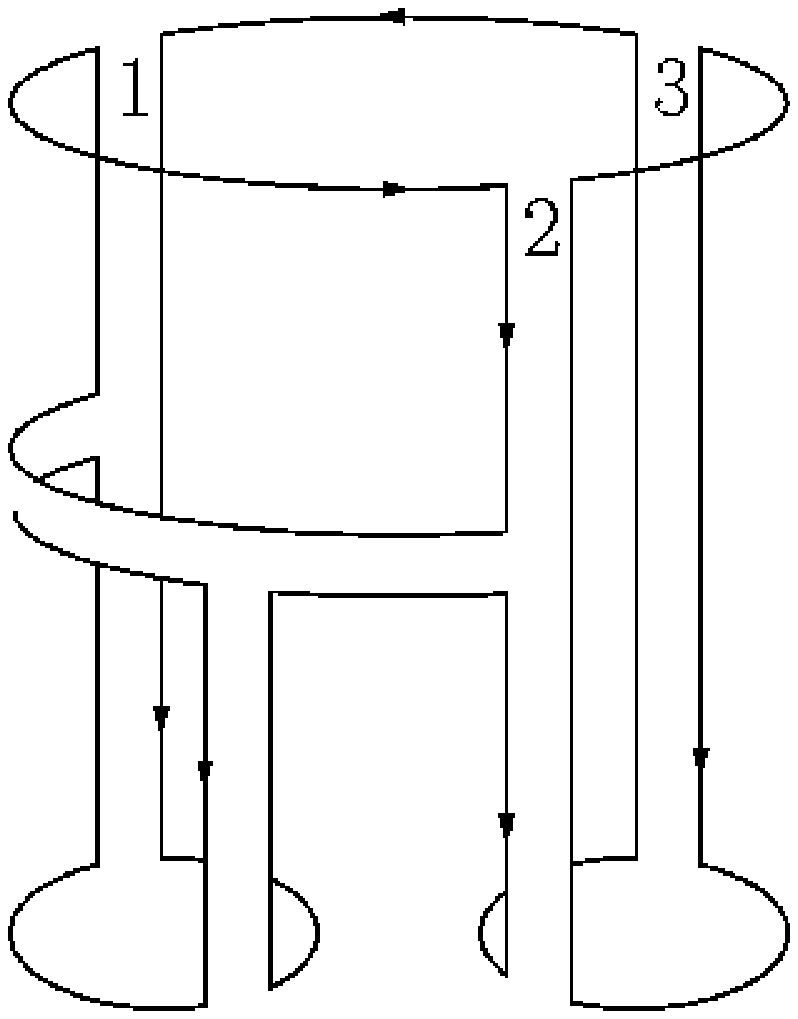} }
  \parbox{3cm}{\center    \includegraphics[width=2cm]{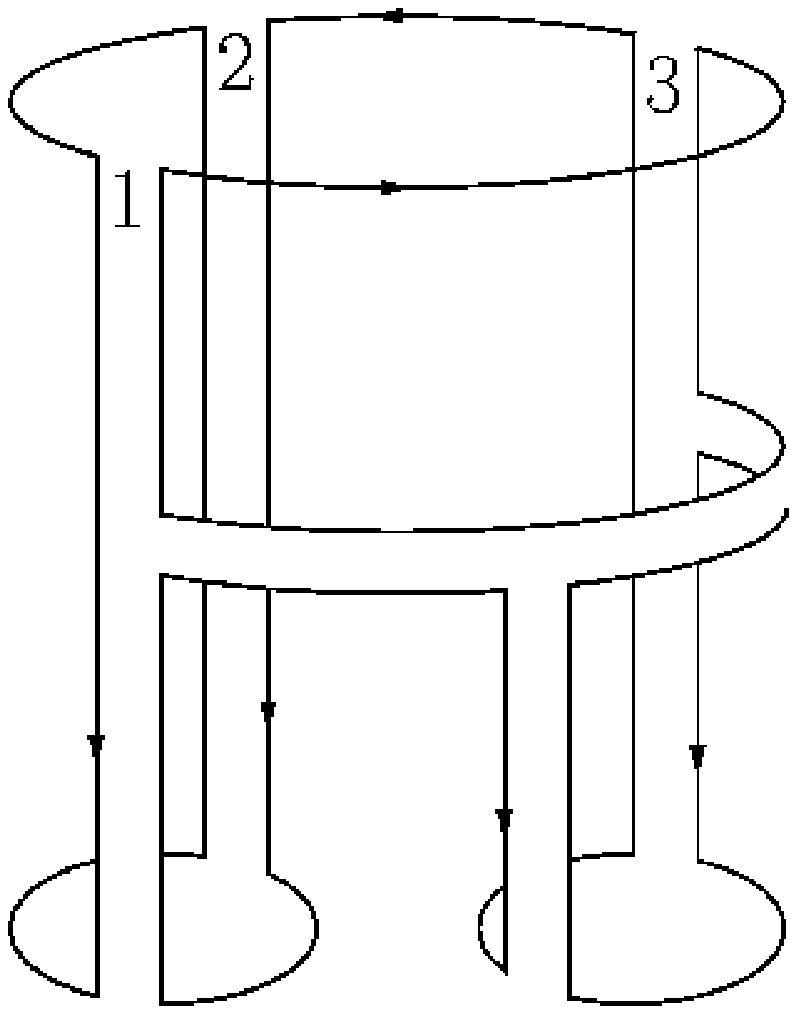} }\\
  \parbox{3cm}{\center ${(123)}$ } 
  \parbox{3cm}{\center  ${(321)}$} 
  \caption{ \small The two possible orderings of three gluons along the quark-loop  which leads to the definition of $\mathcal{D}_{3}^{(123)}(\omega)$ and  $\mathcal{D}_{3}^{(321)}(\omega)$. }
  \label{fig:born_d3}
\end{figure}
For the three gluon 
amplitude above the branching vertex in Fig.\ref{fig:planar_omega}b 
we have two inequivalent couplings to the quark loops, as illustrated 
in Fig.\ref{fig:born_d3}.  
We denote them by $\mathcal{D}_3^{(123)}(\omega)$ and 
$\mathcal{D}_3^{(132)}(\omega)$.

When writing down integral equations for the amplitudes $\mathcal{D}_3$ and 
$\mathcal{D}_4$, we observe that they will be coupled. The three gluon state 
at the lower end of $\mathcal{D}_3$ can start as a two gluons at the quark 
loop, then undergo a two-to-three transition. Similarly, the four gluon state  
can start from two or three gluons and then make transitions to the final four 
gluon state. The formulation of these coupled integral equations 
will be carried out in the subsequent section. 

Next we turn to the non-planar diagrams of
Fig.\ref{fig:non_planar_omega}.  They cannot be reduced to the color
structure of the Born diagrams, and we therefore define an additional
partial wave that sums up these terms. They all have in common that the 
non-planar structure at the lower end of the upper cylinder is always 
the branching vertex: below this vertex, there is no further communication 
between the two lower cylinders. As an example, an interaction between 
gluon '2' and '3' would be subleading in powers $N_c$. Similar to the 
planar diagrams
in Fig.\ref{fig:planar_omega}, we see three different structures:
above the branching vertex we have four (Fig.\ref{fig:planar_omega}a),
three (Fig.\ref{fig:planar_omega}b), or two
(Fig.\ref{fig:planar_omega}c) $t$-channel gluons. This suggests that
above the branching vertex the structure is the same as for the planar
graphs, and we simply have to convolute the non-planar branching
vertex with the planar amplitudes $\mathcal{D}_2$, $\mathcal{D}_3$,
and $\mathcal{D}_4$.  The resulting expression will be denoted by
$\mathcal{D}_4^{(\text{NP})}(\omega)$.  Its derivation will be the
content of Sec.\ref{sec:trip_pom}.

\section{Integral equations: gluon amplitudes with planar color structure}
\label{sec:regg_ql}

In the following we will formulate integral equations for amplitudes
of the planar class with two, three and four $t$-channel gluons. The
amplitude with four $t$-channel gluons that belongs to the non-planar
class will be addressed in Sec.\ref{sec:trip_pom}.

\subsection{Integral equations for the three gluon amplitude}
\label{sec:inteq_d3}

The evolution of the three gluon state is described by the sum of
pairwise interactions, i.e. the two-to-two transition kernels acting
on the three gluon state
\begin{align}
  \label{eq:d3_22}
         \parbox{2cm}{\includegraphics[width=2cm]{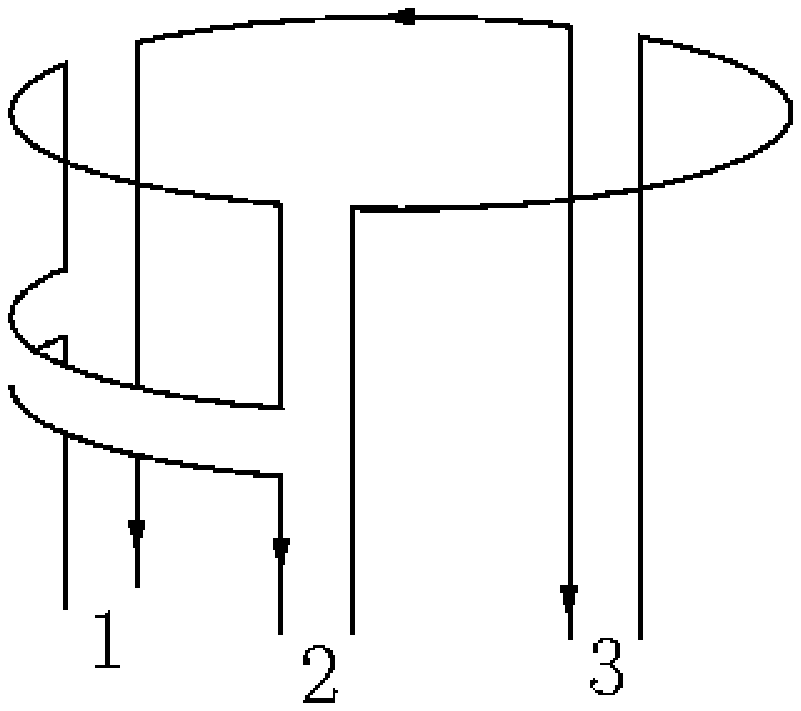}} 
         + 
         \parbox{2.3cm}{\includegraphics[width=2cm]{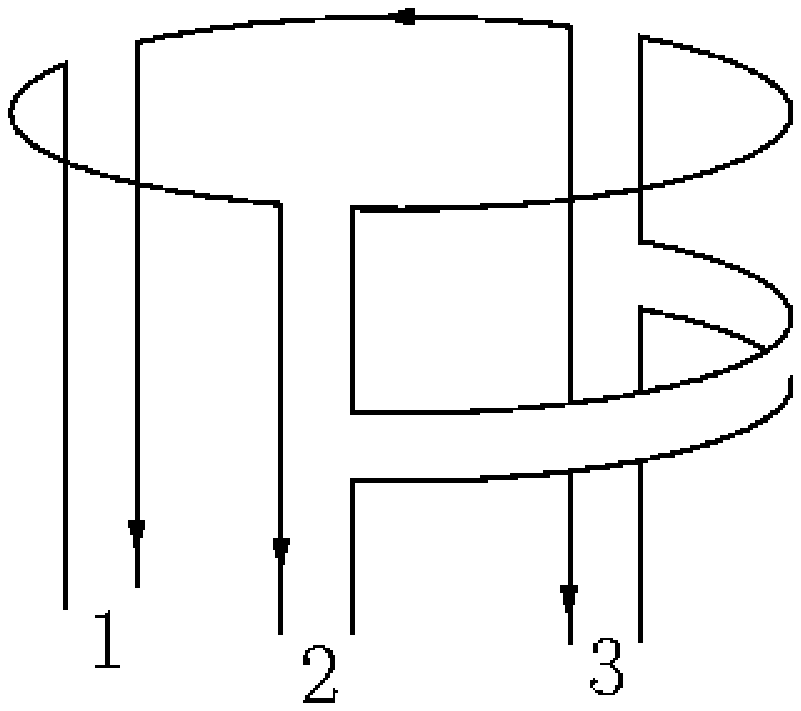}} 
         +\,
         \parbox{2cm}{\includegraphics[width=2cm]{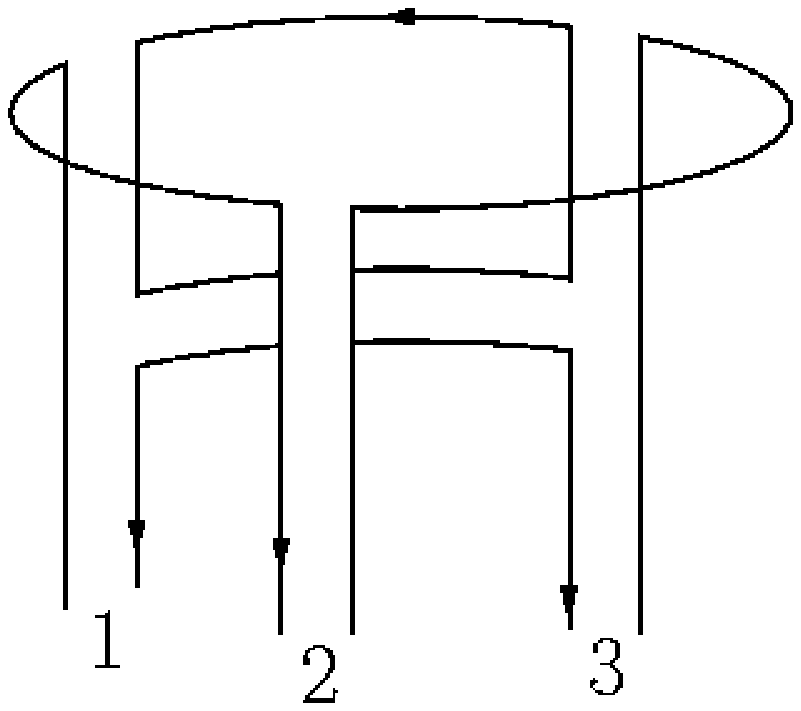}} 
         \to 
          \bar{g}^2N_c \sum_{\substack{(12), (23) \\ (31)}} K_{2 \to 2}\otimes \mathcal{D}_3^{(123)}(\omega).
\end{align}
The transition from the two gluon state to the three gluon state is 
mediated by the two-to-three transition kernels acting on the two-gluon state,
\begin{align}
  \label{eq:d2_23}
 \parbox{2.5cm}{\includegraphics[width=2.5cm]{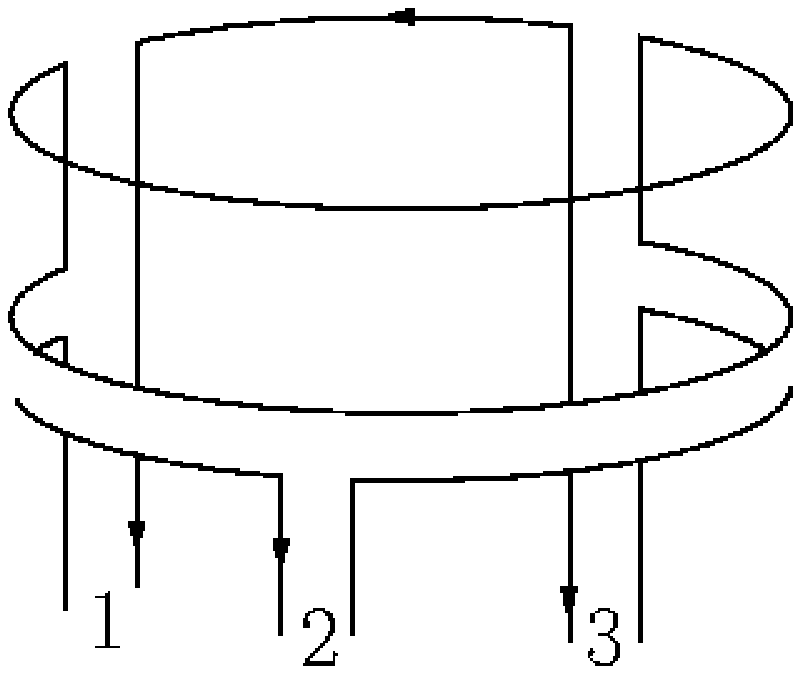}} \to  {\bar{g}^3}(-N_c) K_{2 \to 3}^{( \{12\} \to \{123\})} \otimes \mathcal{D}_2(\omega).
\end{align}
 Similar to $\mathcal{D}_2(\omega)$, also the
amplitude $\mathcal{D}_3(\omega)$ is defined to contain the Reggeon
propagator of the external reggeized gluons, $1/(\omega - \sum_i^3
\beta({\bf k}_i))$. With these ingredients, the integral equation for
$\mathcal{D}_{3}^{(123)}(\omega)$ is given by
\begin{align}
  \label{eq:bfkl-singlet3sincolor}
  \big(\omega - \sum_i^3 \beta({\bf{k}}_i)\big)
  \mathcal{D}_{3}^{(123)} (\omega| {\bf k}_1,{\bf k}_2,{\bf k}_3) =&
  \mathcal{D}^{(123)}_{(3,0)} + {\bar g}^3(-N_c) K^{( \{12\} \to
    \{123\})}_{2 \to 3}\otimes \mathcal{D}_2(\omega)
  \notag \\
  &+ \bar{g}^2N_c\sum K_{2 \to 2}\otimes
  \mathcal{D}^{(123)}_3(\omega).
\end{align}
In complete analogy the equation for
$\mathcal{D}_3^{(321)}(\omega)$ contains the two-to-three transition kernel
\begin{align}
  \label{eq:d2_23_backwards}
 \parbox{2.5cm}{\includegraphics[width=2.5cm]{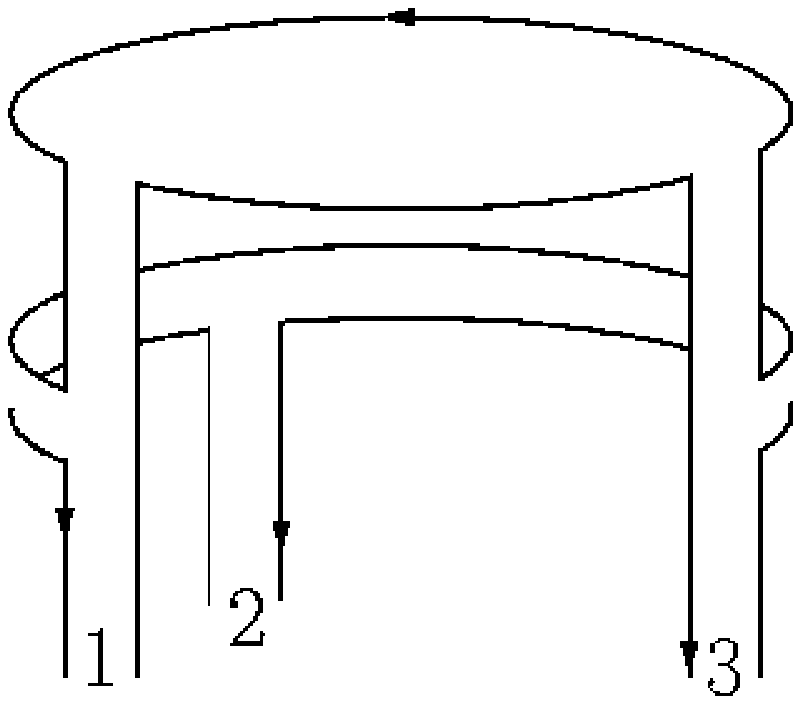}} \to {\bar g}^3N_c K_{2 \to 3}^{( \{12\} \to \{123\})} \otimes \mathcal{D}_2(\omega).
\end{align}
The integral equation is given by:
\begin{align}
  \label{eq:bfkl-singlet3sincolor_321}
\big((\omega - \sum_i^3 \beta({\bf{k}}_i)\big) \mathcal{D}_{3}^{(321)} (\omega| {\bf k}_1,{\bf k}_2,{\bf k}_3) =& 
 \mathcal{D}^{(321)}_{(3,0)}
+
 \bar{g}^3N_c K^{( \{12\} \to \{123\})}_{2 \to 3}\otimes \mathcal{D}_2(\omega)
\notag \\
&+
\bar{g}^2N_c \sum_{\substack{(13),(32) \\ (21)}}  K_{2 \to 2}\otimes \mathcal{D}^{(321)}_3(\omega)
.
\end{align}

\subsection{Reggeization of the three-gluon amplitude}
\label{sec:regg_d3}

To study the integral equations Eq.(\ref{eq:bfkl-singlet3sincolor})
and Eq.(\ref{eq:bfkl-singlet3sincolor_321}) we need the impact factors
$\mathcal{D}_{(3;0)}^{(123)}$ and $\mathcal{D}_{(3;0)}^{(321)}$.  Both
impact factors have the same properties as in the $N_c$ finite case
\cite{Bartels:1994jj}, and they can be written as a superposition of
two-gluon impact factors $\mathcal{D}_{(2;0)}$:
\begin{align}
  \label{eq:d30d20123}
\mathcal{D}_{(3;0)}^{(123)} ( {\bf k}_1,{\bf k}_2,{\bf k}_3) &= + \frac{{\bar g}}{2}  \big[\mathcal{D}_{(2;0)}(12,3) - \mathcal{D}_{(2;0)}(13,2) + \mathcal{D}_{(2;0)} (1,23) \big] \\
\label{eq:d30d20321}
D_{(3;0)}^{(321)} ( {\bf k}_1,{\bf k}_2,{\bf k}_3) &= -\frac{{\bar g}}{2}  \big[D_{(2;0)}(12,3) - D_{(2;0)}(13,2) + D_{(2;0)} (1,23) \big].
\end{align}
  On the right
hand side we introduced a short-hand notation: a string of numbers
stands for the sum of the corresponding transverse momenta, for
instance
\begin{align}
  \label{eq:shorthand1}
\mathcal{D}_{(2;0)}(12,3) \equiv \mathcal{D}_{(2;0)}( {\bf k}_1 +{\bf k}_2,{\bf k}_3).
\end{align}
We will make use of this notation in the following, whenever it proves
useful to clarify the underlying structure of complex expressions.  It
can be now demonstrated that the structure on the rhs of
Eqs.(\ref{eq:d30d20123}] and Eqs.(\ref{eq:d30d20321} also holds for
the solutions of the integral equations
Eq.(\ref{eq:bfkl-singlet3sincolor}) and
Eq.(\ref{eq:bfkl-singlet3sincolor_321}):
\begin{align}
  \label{eq:d3regge123}
  \mathcal{D}_{3}^{(123)} (\omega | {\bf k}_1,{\bf k}_2,{\bf k}_3) = 
+ \frac{{\bar g}}{2}   \big[\mathcal{D}_{2}(\omega|12,3) - \mathcal{D}_{2}(\omega|13,2) + \mathcal{D}_{2} (\omega| 1,23) \big] \\
\mathcal{D}_{3}^{(321)} (\omega| {\bf k}_1,{\bf k}_2,{\bf k}_3) =
 - \frac{{\bar g}}{2}  
\big[\mathcal{D}_{2}(\omega|12,3) - \mathcal{D}_{2}(\omega|13,2) + \mathcal{D}_{2} (\omega|1,23) \big].
\end{align}
Obviously this solution shares important properties with the
reggeization of the planar gluon discussed in
Sec.\ref{sec:reggeization}: in each term on the rhs, two gluons
'collapse' into a single gluon which, at the end, decays into two
gluons. This allows also to deform the color factor as shown in
Fig.\ref{fig:deform_D3}:
\begin{figure}[htbp]
  \centering
     \parbox{2.5cm}{ \includegraphics[height=1.7cm]{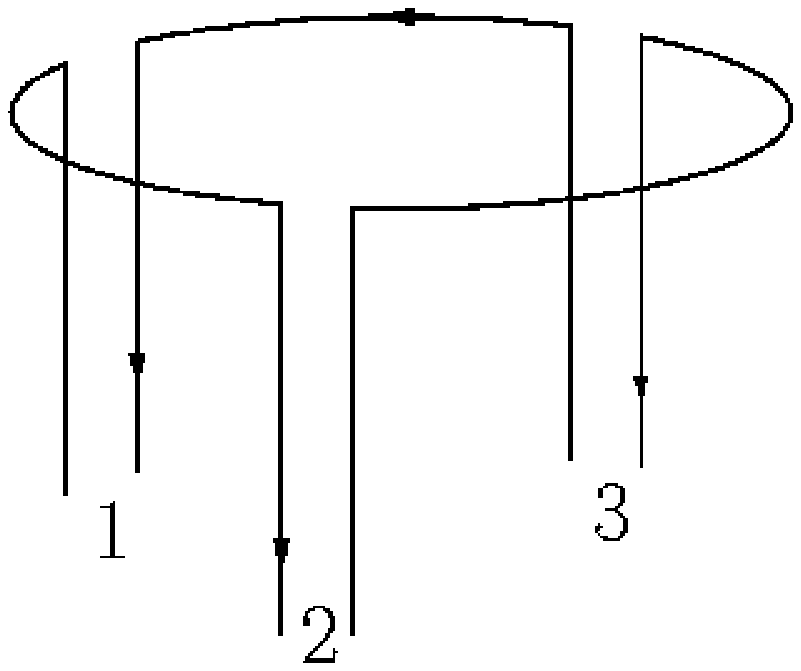}} = 
     \parbox{2.5cm}{ \includegraphics[height=1.7cm]{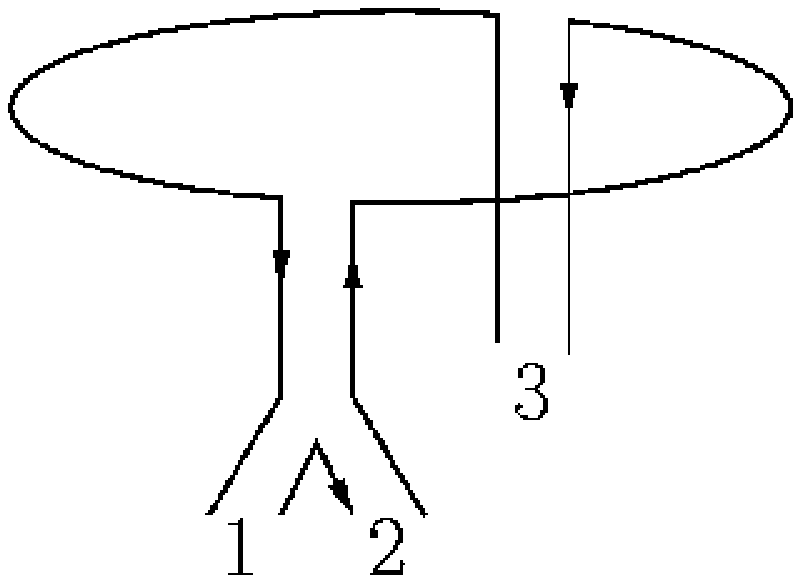}} =
     \parbox{2.5cm}{ \includegraphics[height=1.7cm]{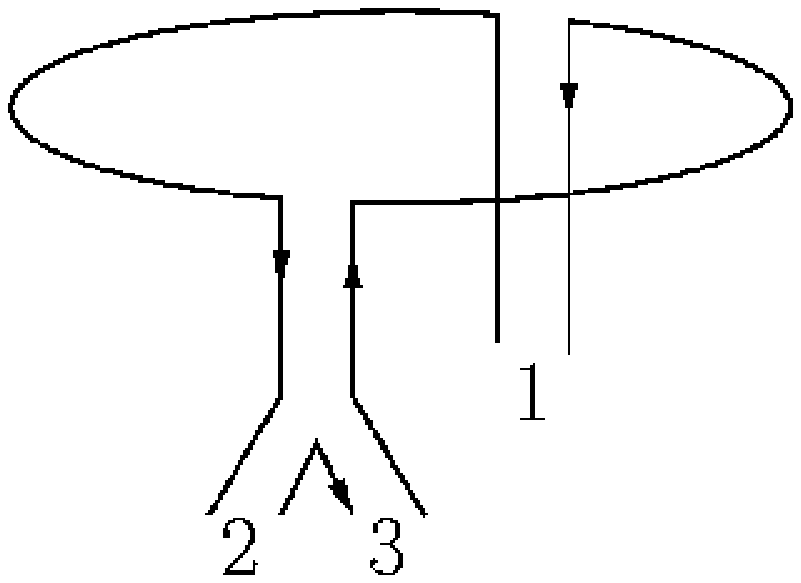}} =
     \parbox{2.5cm}{ \includegraphics[height=1.7cm]{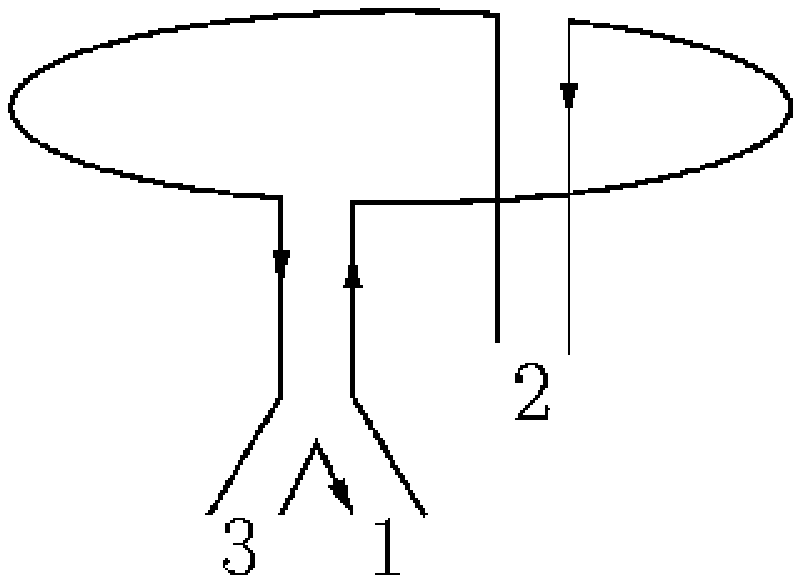}} 
  \caption{ \small The color factor  associated with $\mathcal{D}_3^{(123)}(\omega)$  can be deformed such that the form of the color factor  coincides with the  momentum structure of the solution in terms of $\mathcal{D}_2$. A similar observation holds for  $\mathcal{D}_3^{(321)}(\omega)$. }
  \label{fig:deform_D3}
\end{figure}

\subsection{The integral equations for the four gluon amplitude with planar  color structure}
\label{sec:inteq_d4}

For the four gluon amplitude, $\mathcal{D}_4^{(1234)}(\omega)$, the
evolution of the four gluon state is describes by the sum of pairwise
interactions between neighboring gluons. For the amplitude
$\mathcal{D}_4^{(1234)}(\omega)$ the sum of the two-to-two transition
is given by
\begin{align}
  \label{eq:22_d4_1234}
\parbox{2cm}{ \includegraphics[width=2cm]{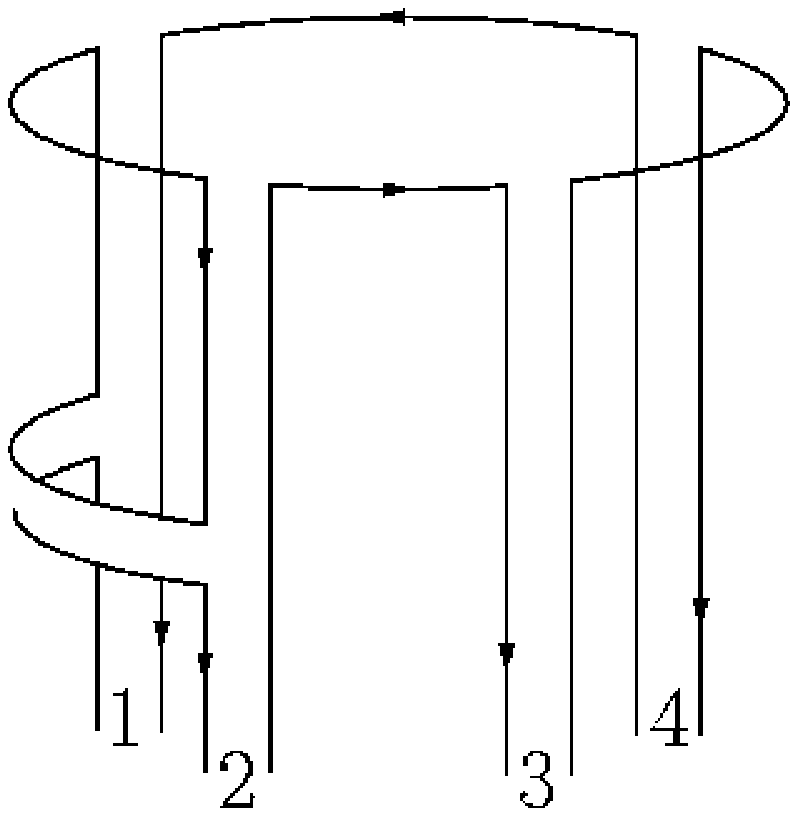}}
+
\parbox{2cm}{ \includegraphics[width=2cm]{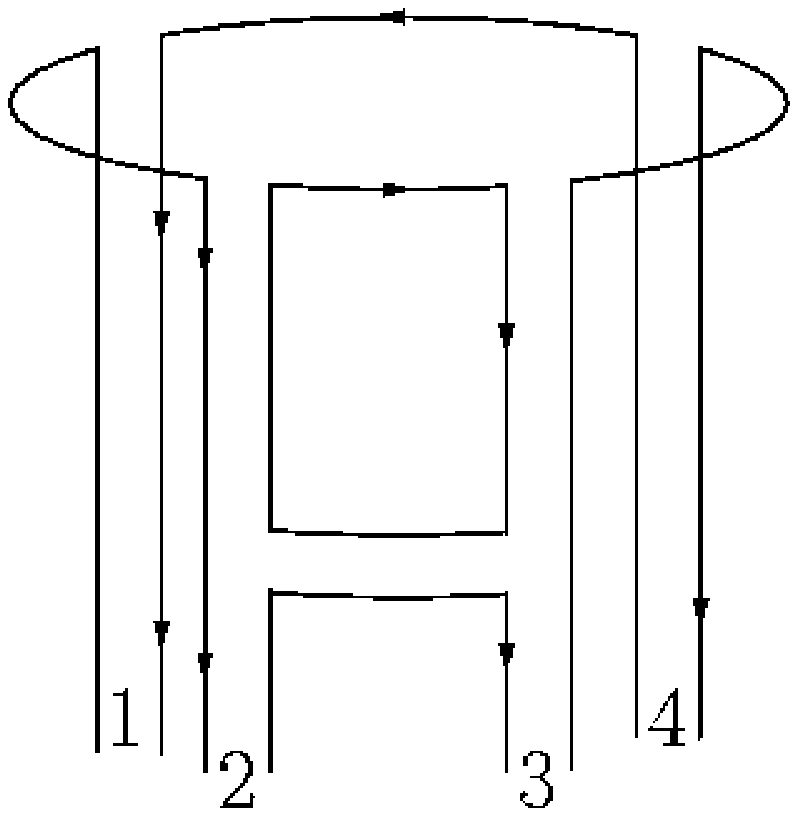}}
+
\parbox{2cm}{ \includegraphics[width=2cm]{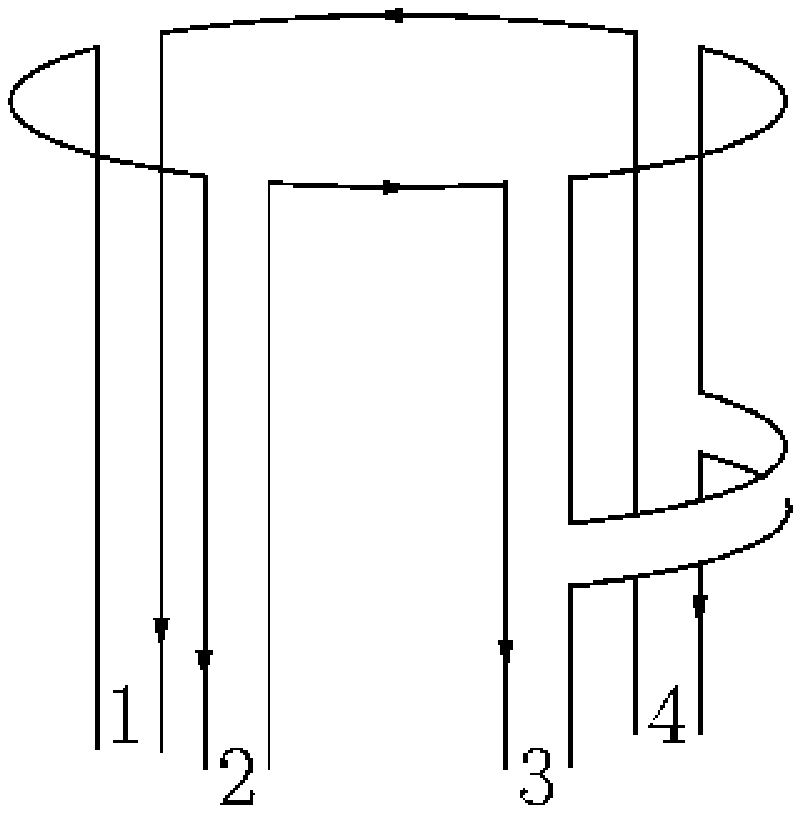}}
+
\parbox{2cm}{ \includegraphics[width=2cm]{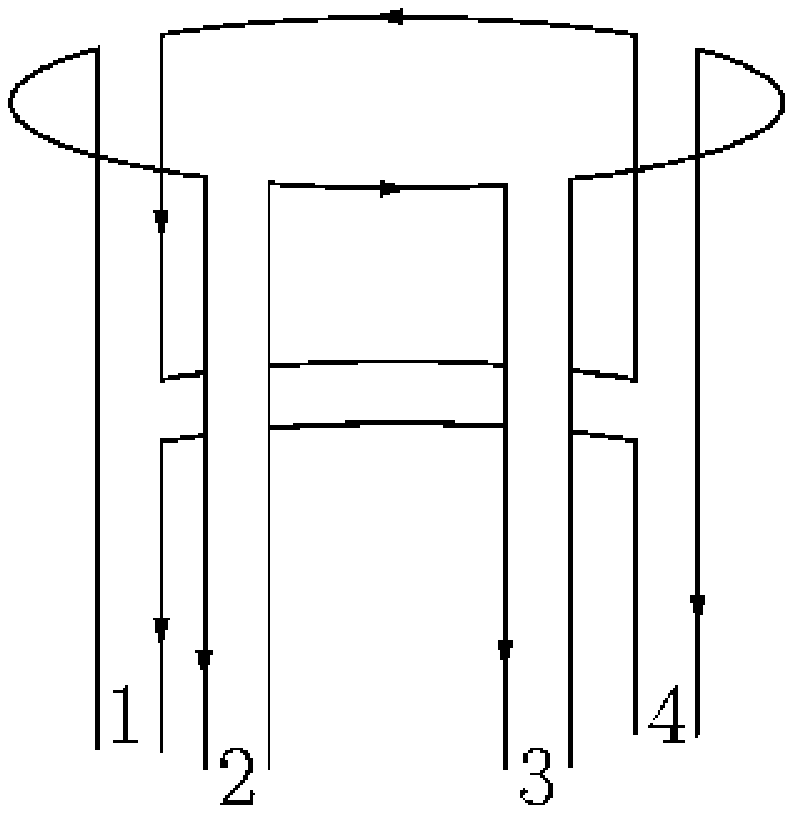}}
\to  \bar{g}^2N_c \sum_{\substack{(12), (23) \\ (34), (41)}} K_{2 \to 2} 
\otimes \mathcal{D}_4^{(1234)}(\omega).
\end{align}
The corresponding expressions for $\mathcal{D}_4^{(2134)}(\omega)$,
$\mathcal{D}_4^{(2143)}(\omega)$ and $\mathcal{D}_4^{(1243)}(\omega)$
are easily obtained by simply exchanging the corresponding labels on
the $t$-channel gluons. For $\mathcal{D}_4^{(1234)}(\omega)$, the
two-to-three transitions are
\begin{align}
  \label{eq:23_d4_1234}
\parbox{2cm}{  \includegraphics[width=2cm]{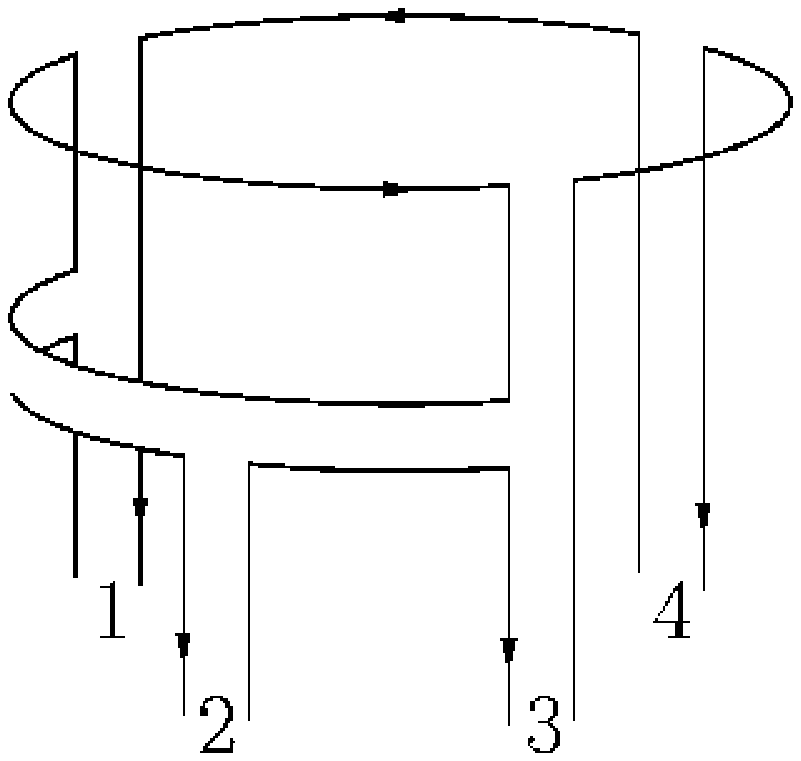}}  
+
\parbox{2cm}{  \includegraphics[width=2cm]{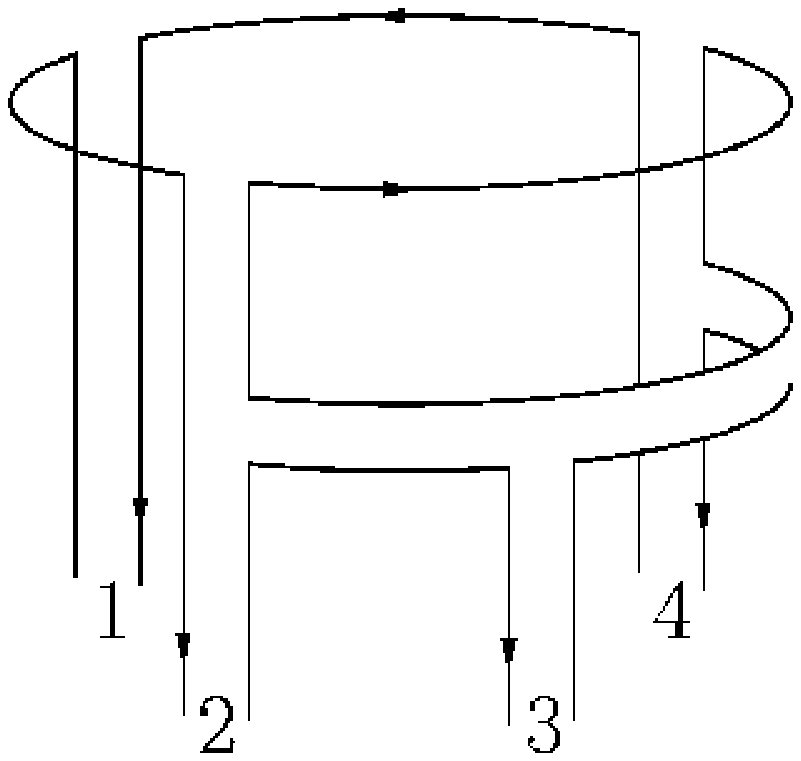}}
    \quad \to
 \begin{array}{l}
    \bar{g}^3N_c  K_{2 \to 3}^{( \{12\} \to \{123\})} \otimes D_3^{(123)}(\omega) \\ \\
    + \quad \bar{g}^3 N_c 
    K_{2 \to 3}^{( \{23\} \to \{234\})} \otimes D_3^{(123)}(\omega),
\end{array}
\end{align}
while the two-to-four  transition is
\begin{align}
  \label{eq:24_d4_1234}
\parbox{2cm}{ \includegraphics[width=2cm]{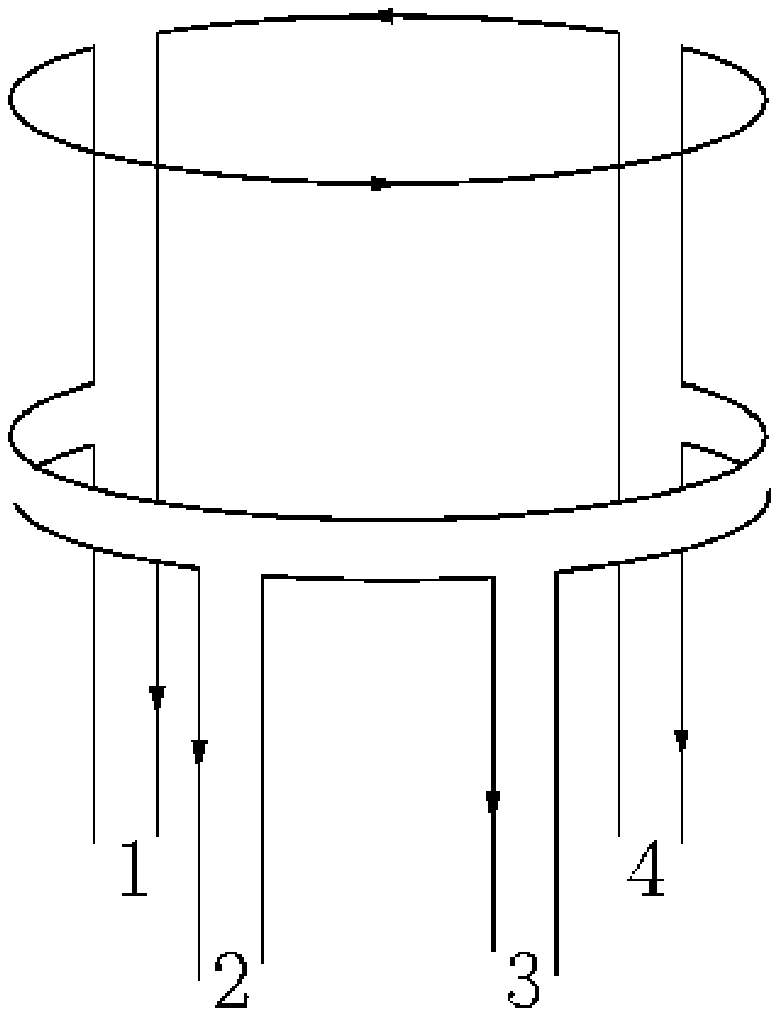}}
\quad \to
\bar{g}^4{N_c} K_{2 \to 4}^{( \{12\} \to \{1234\})} \otimes \mathcal{D}_2(\omega).
\end{align}
Also $\mathcal{D}_4^{(1234)}(\omega)$ is defined to contain the
Reggeon-propagator of the external reggeized gluons. Including this
contribution together with the four gluon impact factor,
$\mathcal{D}_{(4;0)}^{(1234)}$, the integral equation for
$\mathcal{D}_4^{(1234)}(\omega)$ is given by
\begin{align}
  \label{eq:int_eq_d4_1234}
 (\omega - \sum_i^4 \beta({\bf k}_i))  \mathcal{D}_4^{(1234)}(\omega) &= 
          \mathcal{D}^{(1234)}_{(4;0)}  
          +
          {\bar g}^3N_c\bigg[ K_{2 \to 3}^{( \{12\} \to \{123\})} \!\!\otimes\! \mathcal{D}_3^{(123)}(\omega) 
          +
          K_{2 \to 3}^{( \{23\} \to \{234\})}\!\! \otimes\! \mathcal{D}_3^{(123)}(\omega)\bigg]
          \notag \\
          &
          +            
          {\bar{g}^4}{N_c}^2  K_{2 \to 4}^{( \{12\} \to \{1234\})} \otimes \mathcal{D}_2(\omega)
          +
          \bar{g}^2N_c\sum_{\substack{(12), (23) \\ (34), (41)}} K_{2 \to 2} \otimes \mathcal{D}_4^{(1234)}(\omega).
\end{align}
For $\mathcal{D}_4^{(2143)}(\omega)$ the two-to-three transition arises from
\begin{align}
  \label{eq:23_d4_4321}
\parbox{2cm}{  \includegraphics[width=2cm]{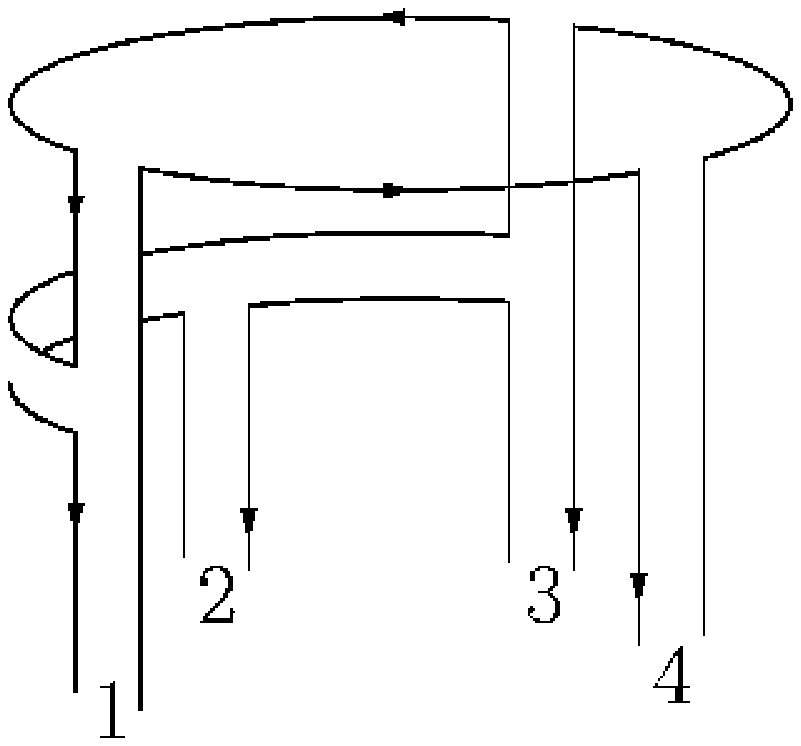}}  
+
\parbox{2cm}{  \includegraphics[width=2cm]{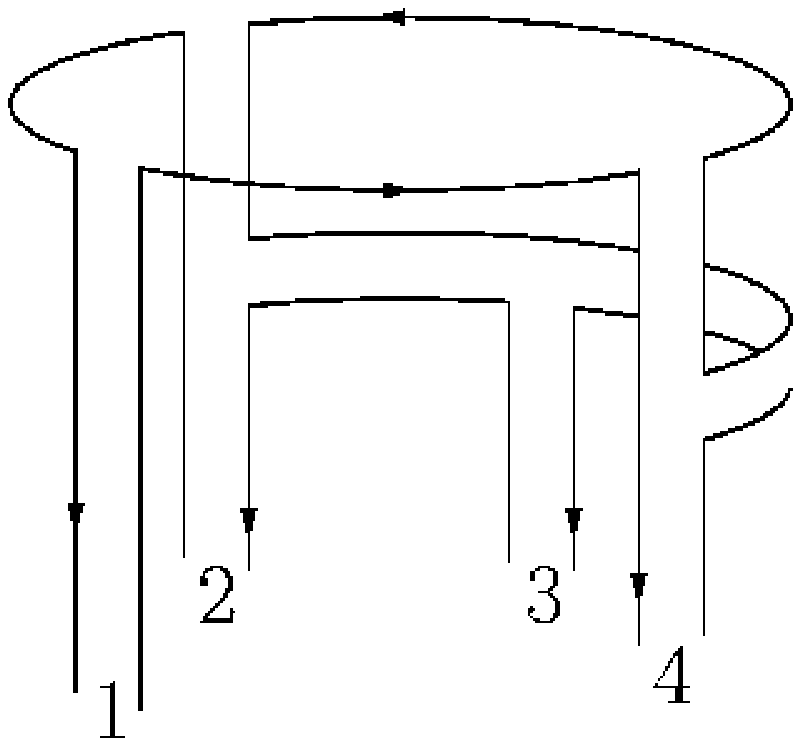}}
  \quad  \to
    \begin{array}{l}
\\ {\bar g}^3 (-N_c)  K_{2 \to 3}^{( \{12\} \to \{123\})} \otimes \mathcal{D}_3^{(321)}(\omega) 
   \\ \\   + \quad 
    {\bar g}^3 (-N_c)      K_{2 \to 3}^{( \{23\} \to \{234\})} \otimes \mathcal{D}_3^{(321)}(\omega),
    \end{array}
   \end{align}
and the two-to-four transition comes from
\begin{align}
  \label{eq:24_d4_4321}
\parbox{2cm}{ \includegraphics[width=2cm]{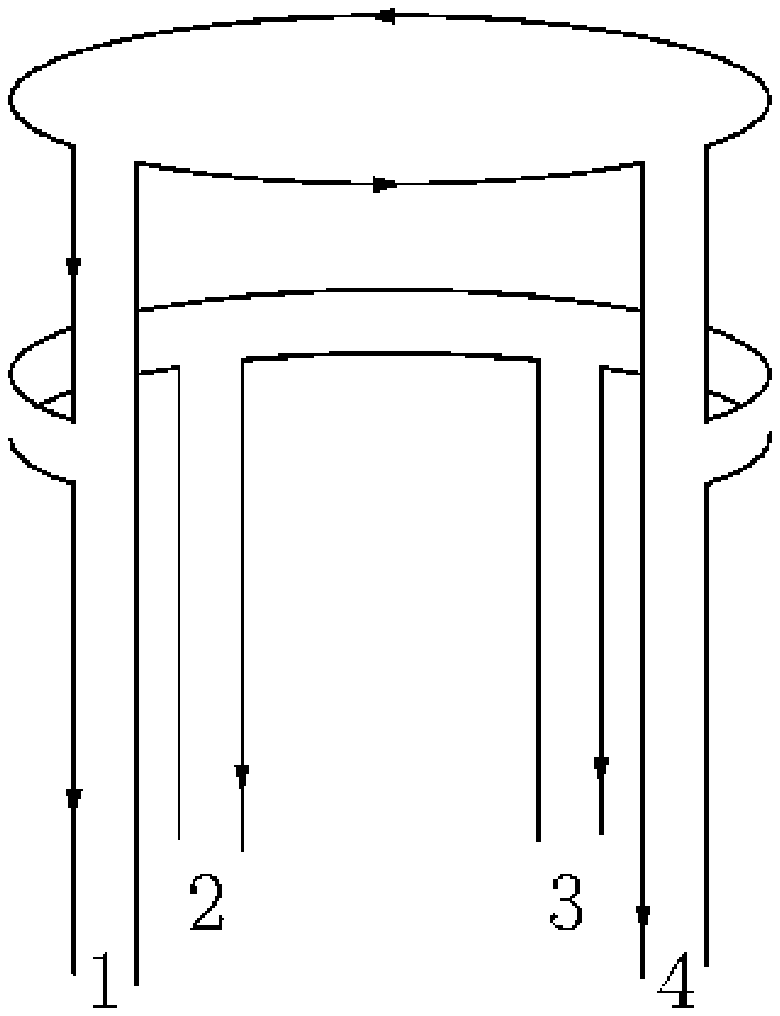}}
\quad \to
{ \bar{g}^4}{N_c} K_{2 \to 4}^{( \{12\} \to \{1234\})} \otimes \mathcal{D}_2(\omega).
\end{align}
We obtain the integral equation
\begin{align}
  \label{eq:int_eq_d4_4321}
(\omega - \sum_i^4 \beta({\bf k}_i))  \mathcal{D}_4^{(2143)}(\omega) &= 
          \mathcal{D}^{(2143)}_{(4;0)}  
          +
          {\bar g}^3 (\!-\!N_c) \bigg[ K_{2 \to 3}^{( \{12\}\! \to\! \{123\})} \!\!\otimes\! \mathcal{D}_3^{(321)}(\omega) 
          +
          K_{2 \to 3}^{( \{23\}\! \to\! \{234\})}\!\! \otimes\! \mathcal{D}_3^{(321)}(\omega)\bigg]
          \notag \\
          &
          +            
          {\bar{g}^4}{N_c}^2 K_{2 \to 4}^{( \{12\} \to \{1234\})} \otimes \mathcal{D}_2(\omega)
          +
          \bar{g}^2\sum_{\substack{(21), (14) \\ (43) , (32)}} K_{2 \to 2} \otimes \mathcal{D}_4^{(2143)}(\omega)
\end{align}
Finally, for $\mathcal{D}_4^{(2134)}(\omega)$ and
$\mathcal{D}_4^{(1243)}(\omega)$ the two-to-three transition are
\begin{align}
  \label{eq:23_d4_2134}
\parbox{2cm}{  \includegraphics[width=2cm]{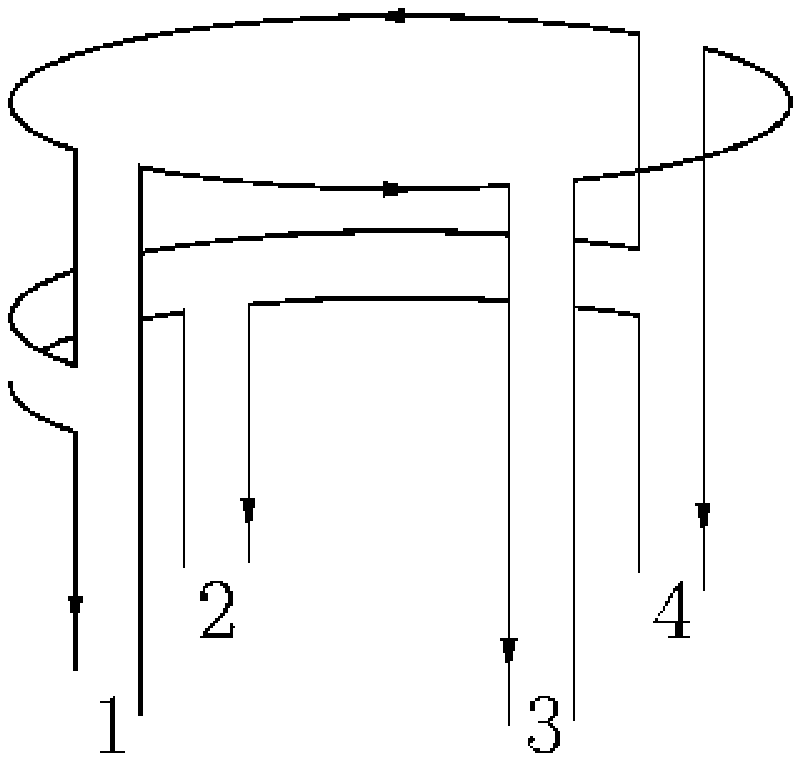}}  
+
\parbox{2cm}{  \includegraphics[width=2cm]{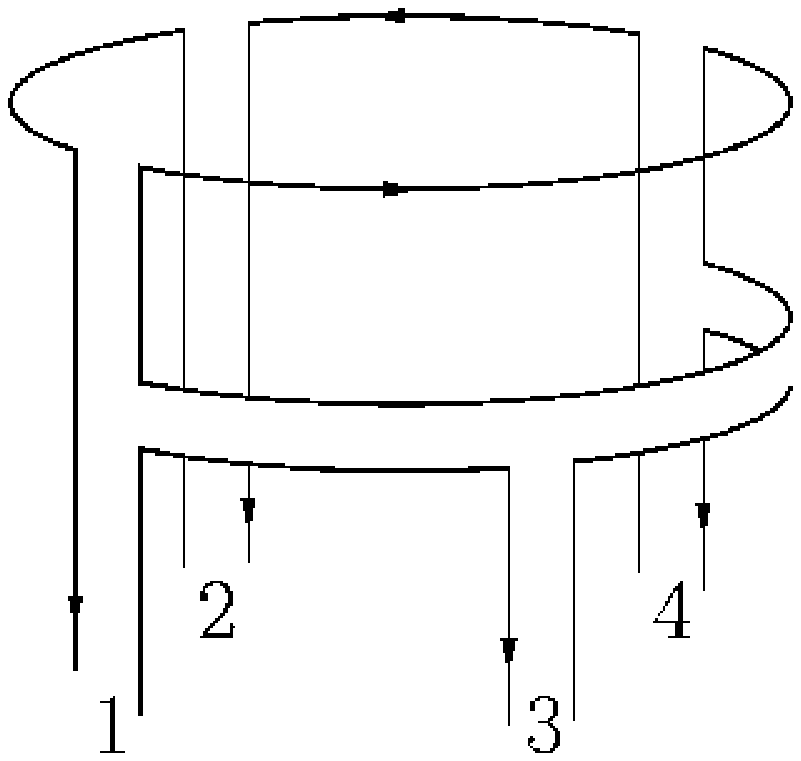}}
    \quad  \to
\begin{array}{l} 
     \bar{g}^3(-N_c) 
      K_{2 \to 3}^{( \{13\} \to \{124\})} \otimes 
\mathcal{D}_3^{(123)}(\omega)  \\ \\    
+ \quad
   {\bar g}^3N_c
     K_{2 \to 3}^{( \{13\} \to \{134\})} \otimes \mathcal{D}_3^{(321)}(\omega)
\end{array}
\end{align}
and
\begin{align}
  \label{eq:23_d4_4312}
\parbox{2cm}{  \includegraphics[width=2cm]{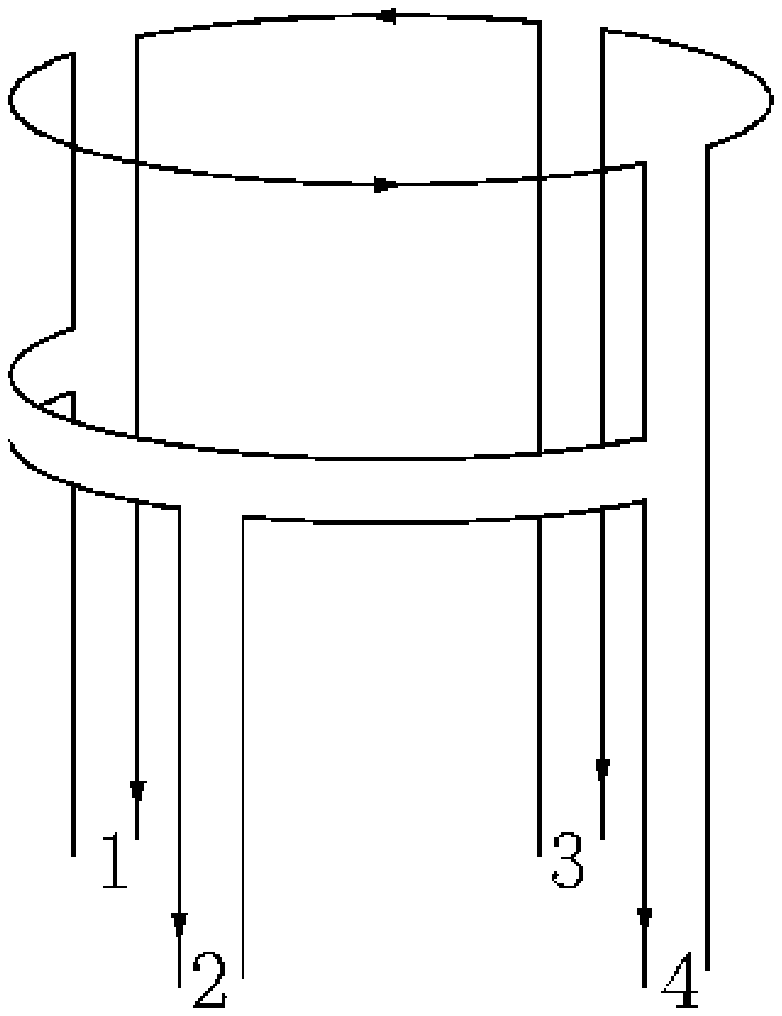}}  
+
\parbox{2cm}{  \includegraphics[width=2cm]{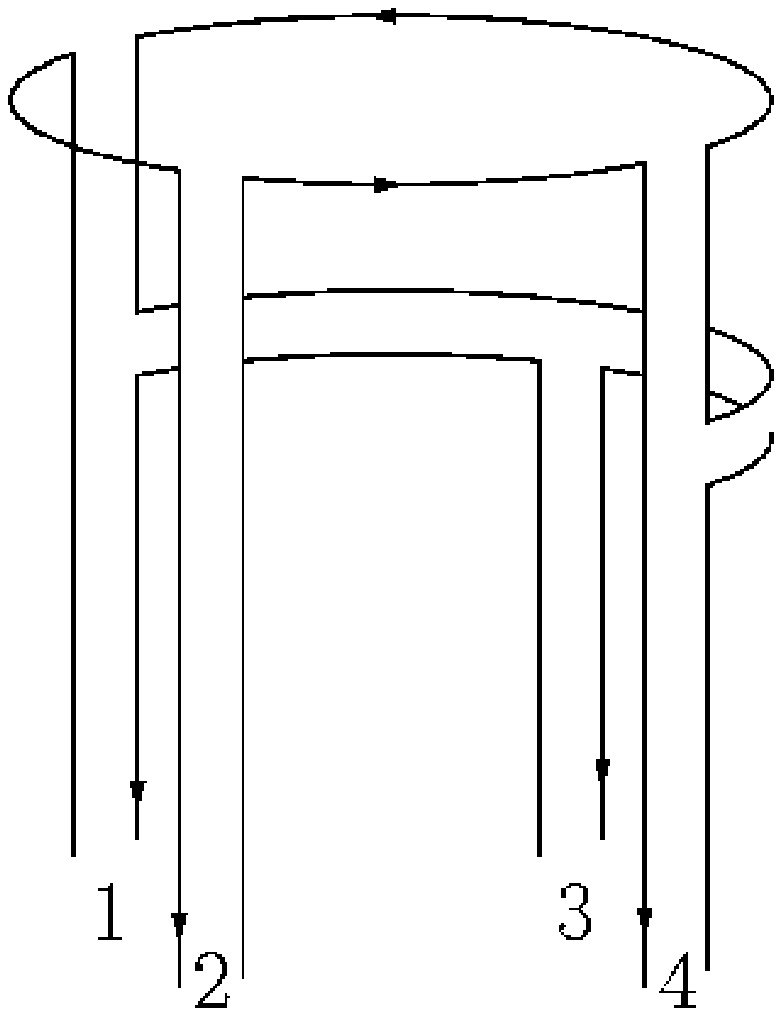}}
   \quad \to
\begin{array}{l}
    {\bar g}^3N_c 
    K_{2 \to 3}^{( \{13\} \to \{124\})} \otimes \mathcal{D}_3^{(321)}(\omega) 
     \\ \\
     + \quad
    {\bar g}^3 (-N_c) 
    K_{2 \to 3}^{( \{13\} \to \{134\})} \otimes \mathcal{D}_3^{(123)}(\omega)
\end{array}
\end{align}
respectively, while two-to-four transition are absent in
that case. The integral equations are given by
\begin{align}
 (\omega - \sum_i^4 \beta({\bf k}_i))  \mathcal{D}_4^{(2134)}(\omega) &= 
          \mathcal{D}^{(2134)}_{(4;0)}  
          + {\bar g}^3 
         (-N_c) K_{2 \to 3}^{( \{13\} \to \{124\})} \!\!\otimes\! \mathcal{D}_3^{(123)}(\omega) 
         \notag \\
          & +{\bar g}^3 N_c
          K_{2 \to 3}^{( \{13\} \to \{134\})}\!\! \otimes\! \mathcal{D}_3^{(321)}(\omega)
          +
           \bar{g}^2N_c\sum_{\substack{(21),(13)\\(34),(42)}} K_{2 \to 2} \otimes \mathcal{D}_4^{(2134)}(\omega) 
  \label{eq:int_eq_d4_2134} \\
(\omega - \sum_i^4 \beta({\bf k}_i))  \mathcal{D}_4^{(1243)}(\omega) &= 
          \mathcal{D}^{(1243)}_{(4;0)}  
          +
          {\bar g}^3 
                         (-N_c)  K_{2 \to 3}^{( \{13\} \to \{134\})}\!\! \otimes\! \mathcal{D}_3^{(123)}(\omega)
          \notag \\
          &                
 +
                {\bar g}^3  N_c K_{2 \to 3}^{( \{13\} \to \{124\})} \!\!\otimes\! \mathcal{D}_3^{(321)}(\omega) 
                     +
           \bar{g}^2N_c\sum_{\substack{(12),(24) \\(43), (31)}} K_{2 \to 2} \otimes \mathcal{D}_4^{(1243)}(\omega) .  \label{eq:int_eq_d4_4312}
\end{align}

\subsection{Reggeization of four-gluon-amplitudes with planar color  structure}
\label{sec:regg_d4}
\begin{figure}[htbp]
\centering   
  \parbox{3cm}{   \includegraphics[height=2.2cm]{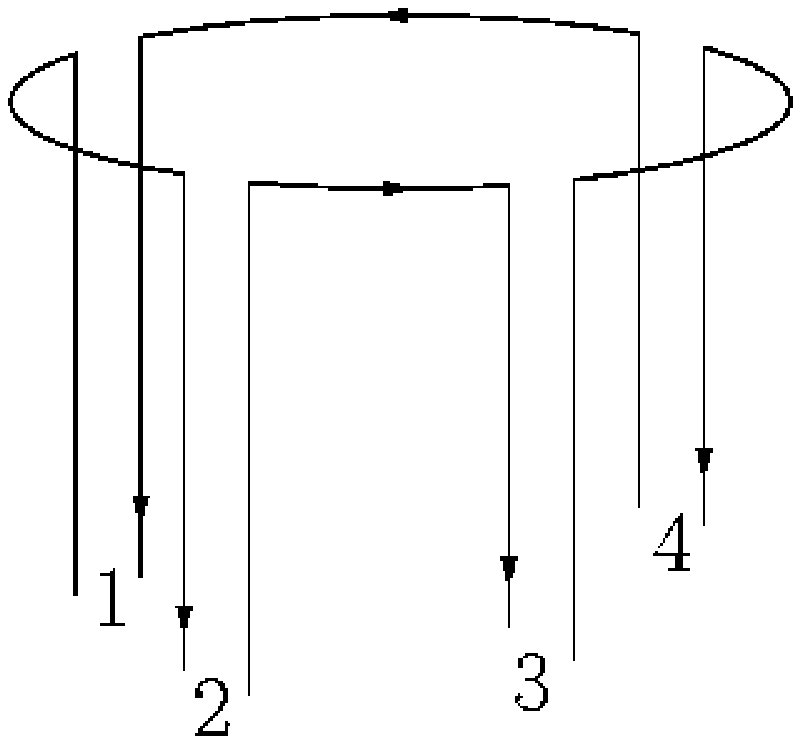}}
=    \parbox{3cm}{ \includegraphics[height=1.7cm]{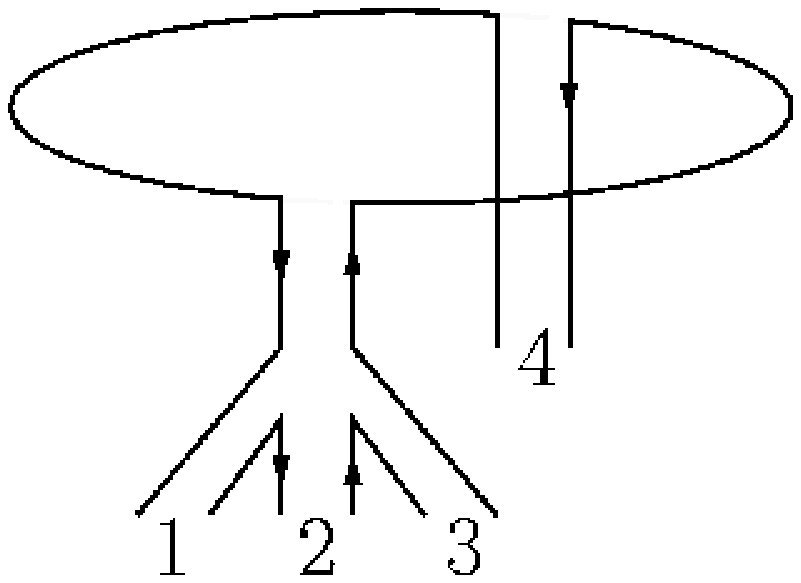}}
  =\parbox{3cm}{ \includegraphics[height=1.7cm]{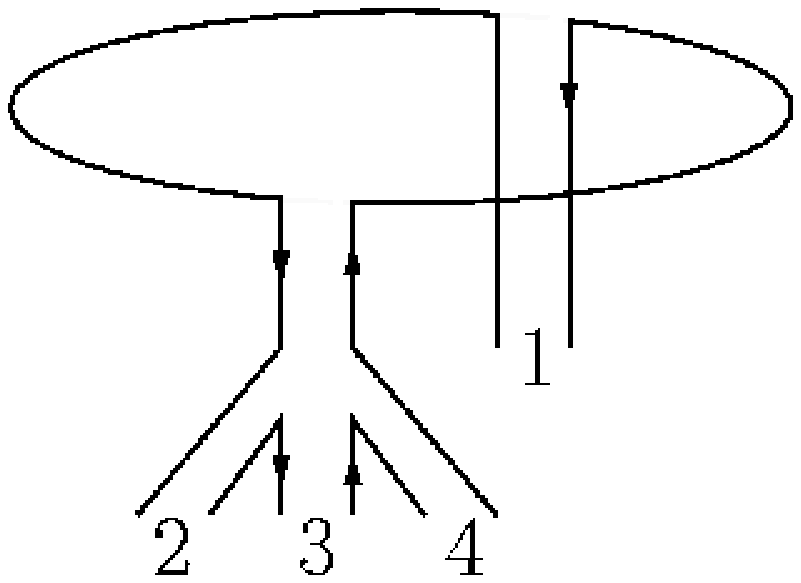}}
= \parbox{3cm}{ \includegraphics[height=1.7cm]{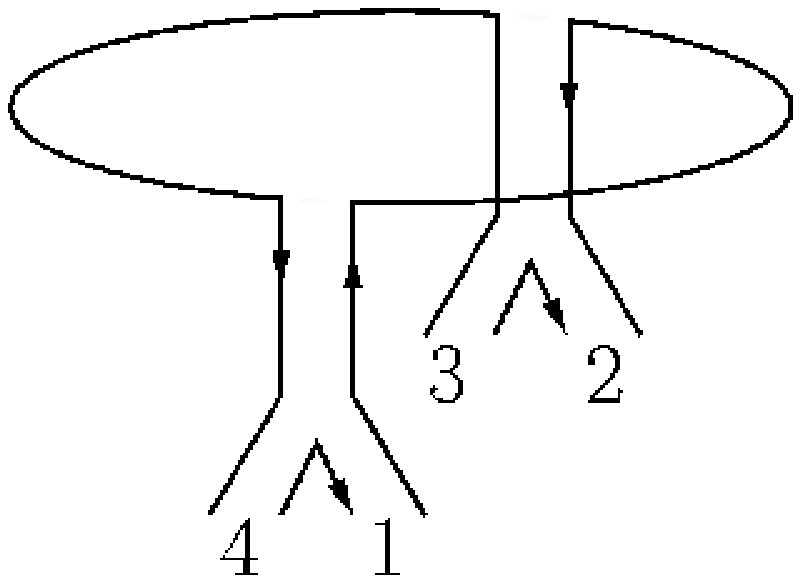}}  
   \caption{ \small Color factors associated with the four gluon amplitude with planar color structure can be deformed to coincide with the momentum structure of the solution in terms of $\mathcal{D}_2$. Above we show the ordering (1234). }
  \label{fig:d4_color}
\end{figure}

Similar to the three-gluon amplitude, the four independent integral
equations for the four-gluon-amplitudes with planar color structure
are solved by a reggeization ansatz. Again the starting point is given by 
the virtual photon impact factors which, similar to the three-gluon
case, can be expressed in terms of the two-gluon impact factor:
\begin{align}
  \label{eq:d40_1234}
\mathcal{D}_{(4;0)}^{(1234)}({\bf k}_1,{\bf k}_2,{\bf k}_3,{\bf k}_4) 
    &=
    -\frac{\bar{\lambda}}{2N_c} \big[
\mathcal{D}_{(2;0)}(123,4) + \mathcal{D}_{(2;0)}(1,234) - \mathcal{D}_{(2;0)}(14,23)
\big] \\
  \label{eq:d40_4321}
\mathcal{D}_{(4;0)}^{(2143)}({\bf k}_1,{\bf k}_2,{\bf k}_3,{\bf k}_4) 
    &=
    - \frac{\bar{\lambda}}{2N_c} \big[
\mathcal{D}_{(2;0)}(123,4) + \mathcal{D}_{(2;0)}(1,234) - \mathcal{D}_{(2;0)}(14,23)
\big] \\
\label{eq:d40_2134}
\mathcal{D}_{(4;0)}^{(2134)}({\bf k}_1,{\bf k}_2,{\bf k}_3,{\bf k}_4) 
    &=
    -\frac{\bar{\lambda}}{2N_c} \big[
\mathcal{D}_{(2;0)}(134,2) + \mathcal{D}_{(2;0)}(124,3) - \mathcal{D}_{(2;0)}(12,34) - \mathcal{D}_{(2;0)}(13,24)
\big] \\
\label{eq:d40_4312}
\mathcal{D}_{(4;0)}^{(1243)}({\bf k}_1,{\bf k}_2,{\bf k}_3,{\bf k}_4) 
    &=
    - \frac{\bar{\lambda}}{2N_c} \big[
\mathcal{D}_{(2;0)}(134,2) + \mathcal{D}_{(2;0)}(124,3) - \mathcal{D}_{(2;0)}(12,34) - \mathcal{D}_{(2;0)}(13,24)
\big] .
\end{align}
This decomposition also holds for the solutions of the integral 
equations, Eq.(\ref{eq:int_eq_d4_1234}),
Eq.(\ref{eq:int_eq_d4_4321}), Eq.(\ref{eq:int_eq_d4_2134}) and
Eq.(\ref{eq:int_eq_d4_4312}):
\begin{align}
    \label{eq:d4_1234}
\mathcal{D}_{4}^{(1234)}(\omega|{\bf k}_1,{\bf k}_2,{\bf k}_3,{\bf k}_4) 
    &=
     -\frac{\bar{\lambda}}{2N_c} \big[
\mathcal{D}_{2}(\omega|123,4) + \mathcal{D}_{2}(\omega|1,234) - \mathcal{D}_{2}(\omega|14,23)
\big] \\
  \label{eq:d4_4321}
\mathcal{D}_{4}^{(2143)}(\omega|{\bf k}_1,{\bf k}_2,{\bf k}_3,{\bf k}_4) 
    &=
     -\frac{\bar{\lambda}}{2N_c} \big[
\mathcal{D}_{2}(\omega|123,4) + \mathcal{D}_{2}(\omega|1,234) - \mathcal{D}_{2}(\omega|14,23)
\big]\\ 
\label{eq:d4_2134}
\mathcal{D}_{4}^{(2134)}(\omega|{\bf k}_1,{\bf k}_2,{\bf k}_3,{\bf k}_4) 
    &=
     -\frac{\bar{\lambda}}{2N_c} \big[
\mathcal{D}_{2}(\omega|134,2) + \mathcal{D}_{2}(\omega|124,3) - \mathcal{D}_{2}(\omega|12,34) - \mathcal{D}_{2}(\omega|13,24)
\big] \\
\label{eq:d4_4312}
\mathcal{D}_{4}^{(1243)}(\omega|{\bf k}_1,{\bf k}_2,{\bf k}_3,{\bf k}_4) 
    &=
     -\frac{\bar{\lambda}}{2N_c} \big[
\mathcal{D}_{2}(\omega|134,2) + \mathcal{D}_{2}(\omega|124,3) - \mathcal{D}_{2}(\omega|12,34) - \mathcal{D}_{2}(\omega|13,24)
\big] .
\end{align}
Similar to the the three gluon case, each term on the rhs consists of
a two gluon state where at the end one (or both) gluons decay: this
again generalizes the reggeization discussed in the context of planar
amplitudes. It allows to redraw the color factors as shown in
Fig.\ref{fig:d4_color}.

For the sum of these planar four gluon amplitudes, it is convenient to
define, following \cite{Bartels:1994jj}, an effective $2 \to 4$ vertex
$V^{\text{R}}$, which sums the various decay-channels of the two
reggeized gluons,
\begin{align}
  \label{eq:V_R_def}
V^{\text{R}}({\bf l}_1, {\bf l}_2; {\bf k}_1,{\bf k}_2,{\bf k}_3, {\bf k}_4 ) = -{\bf l}_1^2 {\bf l}_2^2  
\bigg[&
     \delta^{(2)}({\bf l}_1 - {\bf k}_1- {\bf k}_2- {\bf k}_3 ) 
     +
     \delta^{(2)}({\bf l}_1 - {\bf k}_1- {\bf k}_2- {\bf k}_4 ) 
     \notag \\     
     &+
     \delta^{(2)}({\bf l}_1 - {\bf k}_1- {\bf k}_3- {\bf k}_4 ) 
     +  
     \delta^{(2)}({\bf l}_1 - {\bf k}_2- {\bf k}_3- {\bf k}_4 ) 
     \notag \\     
     -&
     \delta^{(2)}({\bf l}_1 - {\bf k}_1- {\bf k}_2 ) 
     -
     \delta^{(2)}({\bf l}_1 - {\bf k}_1- {\bf k}_3  )
     -
     \delta^{(2)}({\bf l}_1 - {\bf k}_1- {\bf k}_4)  \bigg],
\end{align}
such that 
\begin{align}
  \label{eq:d4_planar_vertex}
\sum_{ (\text{ijkl}) =\substack{(1234), (2143), \\(2134), (1243) }}  \!\!\!\!\!\!  \mathcal{D}_{4}^{(\text{ijkl})}(\omega|{\bf k}_1,{\bf k}_2,{\bf k}_3,{\bf k}_4)  = \frac{\bar{\lambda}}{N_c} V^{(\text{R})} \otimes \mathcal{D}_{2}(\omega).
\end{align}

It is instructive to compare the results of this section with the
finite $N_c$ results of \cite{Bartels:1994jj}. For finite $N_c$ it was
found that the four gluon amplitude $D_4$ can be written as a sum of
two terms, the reggeizing term and the triple Pomeron vertex:
\begin{equation}
D_4= D_4^R + D_4^I
\end{equation}
This decomposition was motivated by $t$-channel requirements: the
reggeizing pieces contained in $D_4^R$ are antisymmetric under the
exchange of the pair of gluons which forms the reggeized gluon,
whereas the remainder $D_4^I$ (which contains the triple Pomeron
vertex), is completely symmetric. In the present approach which is
based upon the expansion in topologies, this decomposition into
reggeizing and non-reggeizing pieces comes automatically, as a
consequence of different classes of color structures. This suggests
that the reggeization is deeply linked to the planar structure of
scattering amplitudes.

\section{The triple Pomeron vertex on the pair-of-pants}
\label{sec:trip_pom}

In this section we turn now to the partial wave amplitude associated
with the non-planar class, $\mathcal{D}_4^{(\text{NP})} (\omega)$.
As we have outlined at the end of Sec.\ref{sec:below},
we have to convolute, in Fig.\ref{fig:non_planar_omega}a - c,
the branching vertices with $\mathcal{D}_4$, $\mathcal{D}_3$, 
and $\mathcal{D}_2$. Let us go through these  
convolutions in more detail.  

We begin with those contributions where the branching vertex consists 
of a two-to-two transition (Fig.\ref{fig:non_planar_omega}a).
They are to be combined with of one of the four planar amplitudes 
discussed before. Contributions involving $\mathcal{D}_4^{(1234)}(\omega)$ are
\begin{align}
  \label{eq:d4pop_22_1234}
  \parbox{2cm}{ \includegraphics[width=2cm]{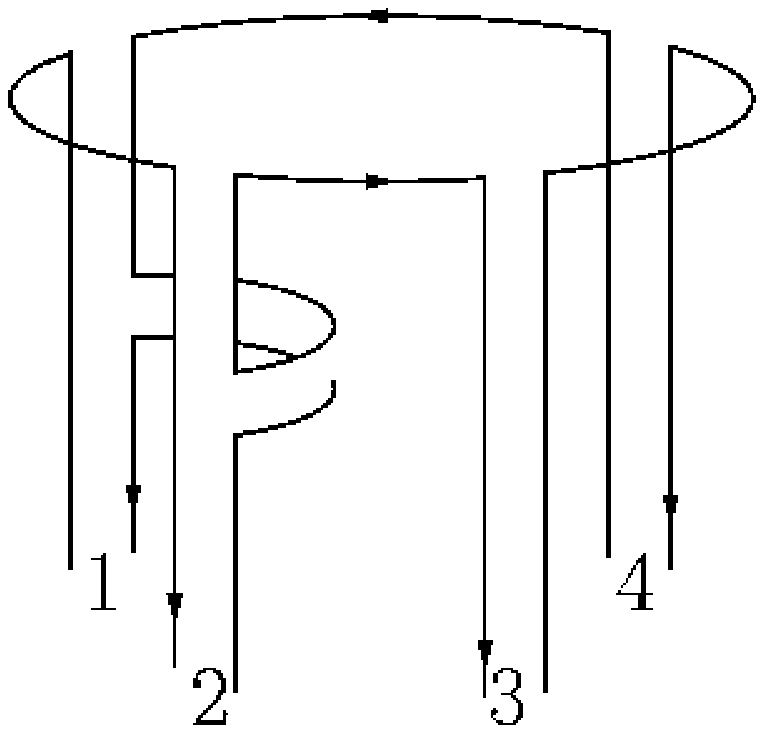}} 
  +
  \parbox{2cm}{\includegraphics[width=2cm]{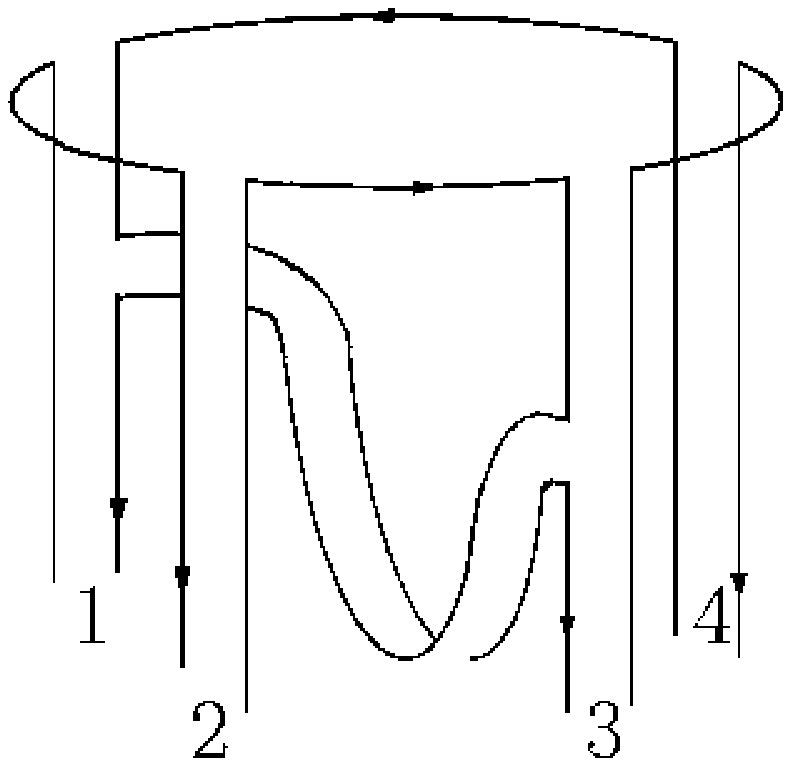}}
  + 
  \parbox{2cm}{\includegraphics[width=2cm]{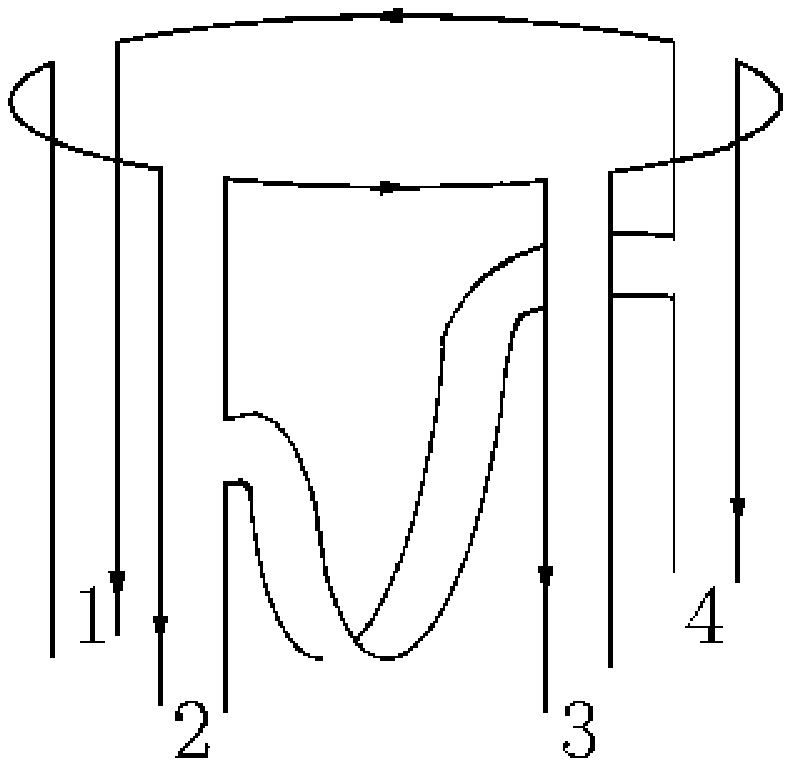}}
  +
  \parbox{2cm}{ \includegraphics[width=2cm]{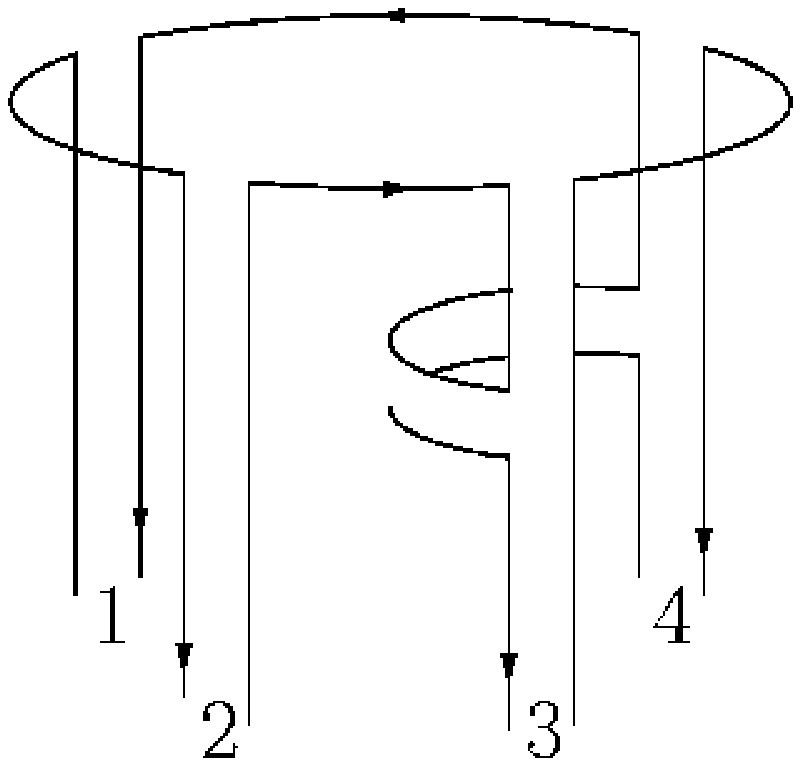}} 
\to
 \bar{g}^2N_c\sum_{\substack{(12), (13), \\(24), (34) }}  K_{2 \to 2} \otimes \mathcal{D}_4^{(1234)}(\omega),
\end{align}
where, on the rhs, the subscripts under the sum indicate which pairings of 
gluons should be included. A comment on the first and the last 
terms is in place. These interactions are inside the 
gluon pairs (12) or (34) and thus do not belong to the production     
of an $s$-channel gluon inside the $M^2$ discontinuity. Hence, these 
interactions are not part of the branching vertex. It is only for our 
convenience that we, nevertheless, include them into our definition of 
$\mathcal{D}_4^{(\text{NP})} (\omega)$ (later on, when defining the partial 
waves, we will have to subtract them again).      
In (\ref{eq:d4pop_22_1234}), the second and third term 
(the interactions (13) and (24)) carry an additional minus sign, due  
to the color structure (see Sec.\ref{sec:amplitudes-with-4}).
Similarly one has 
\begin{align}
  \label{eq:d4pop_22_2134}
  \parbox{2cm}{ \includegraphics[width=2cm]{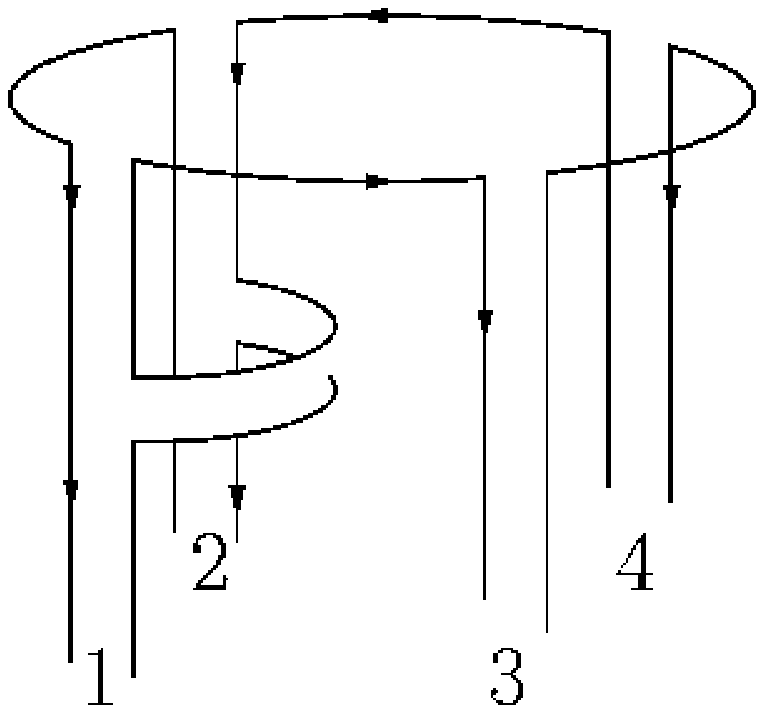}} 
  + 
  \parbox{2cm}{\includegraphics[width=2cm]{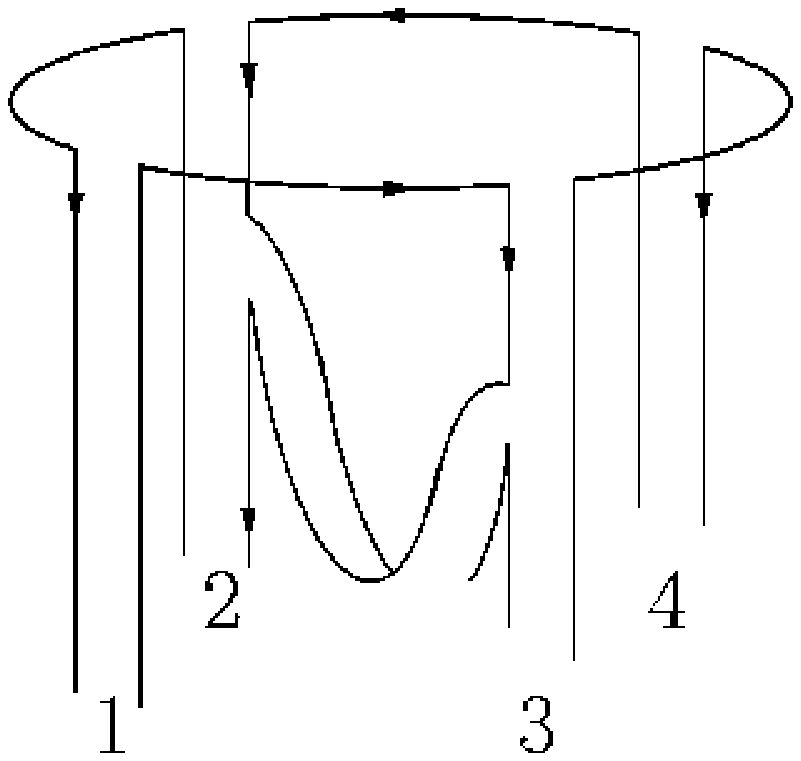}}
  + 
\parbox{2cm}{\includegraphics[width=2cm]{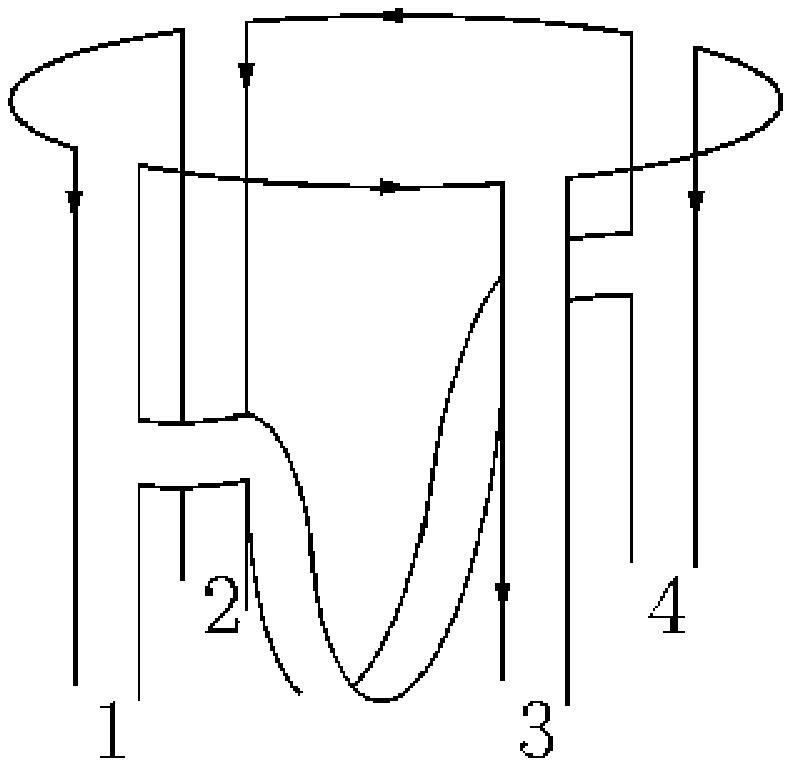}}
  +
  \parbox{2cm}{ \includegraphics[width=2cm]{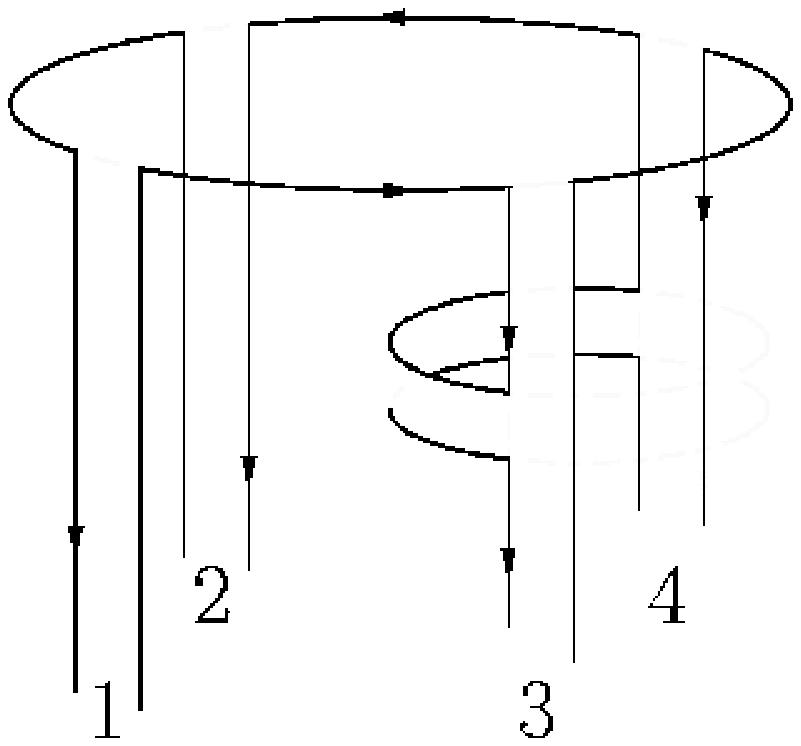}} 
\to \bar{g}^2N_c\sum_{\substack{(12), (23), \\(14), (34) }}  K_{2 \to 2} \otimes \mathcal{D}_4^{(2134)}(\omega),
\end{align}
while the contributions containing $\mathcal{D}_4^{(2143)}(\omega) $ and $\mathcal{D}_4^{(1243)}(\omega)$ are 
\begin{align}
  \label{eq:d4pop_22_4312}
  \bar{g^2}N_c\sum_{\substack{(12), (13), \\(24), (34) }}  K_{2 \to 2} \otimes \mathcal{D}_4^{(2143)}(\omega)
+
\bar{g}^2N_c\sum_{\substack{(12), (23), \\(14), (34) }}  K_{2 \to 2} \otimes \mathcal{D}_4^{(1243)}(\omega).
\end{align}
In all  three sums, the second and third terms carry minus signs,
and the first and last terms are kept for convenience and 
will have to be removed later on. 

For the two-to-three transition (Fig.\ref{fig:non_planar_omega}b) 
we have two contributions: terms that originate from $\mathcal{D}_3^{(123)}(\omega) $ are given by
\begin{align}
  \label{eq:d4pop_23_123}
  \parbox{2cm}{ \includegraphics[width=2cm]{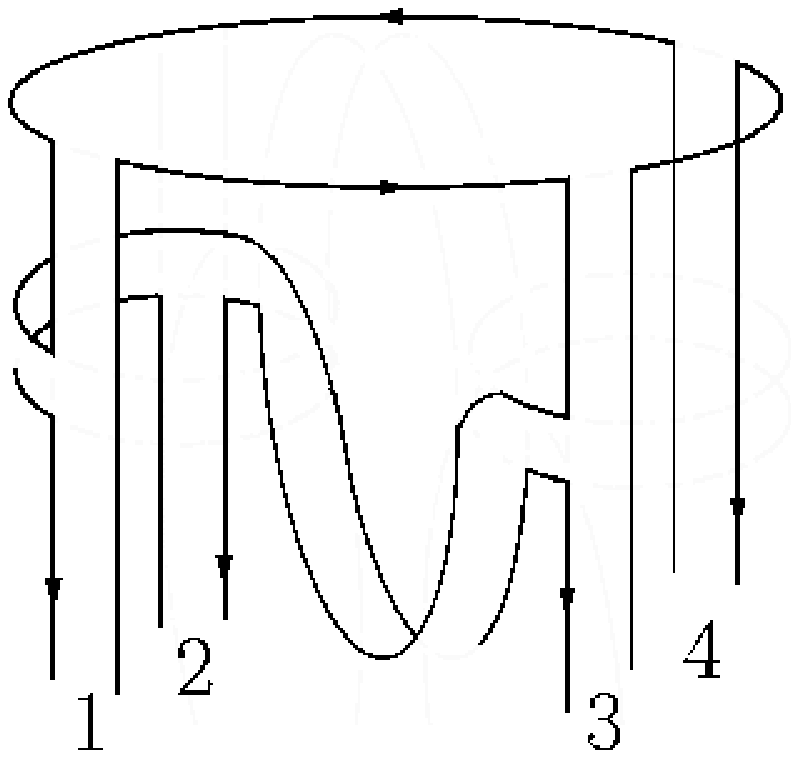}}
  +
  \parbox{2cm}{ \includegraphics[width=2cm]{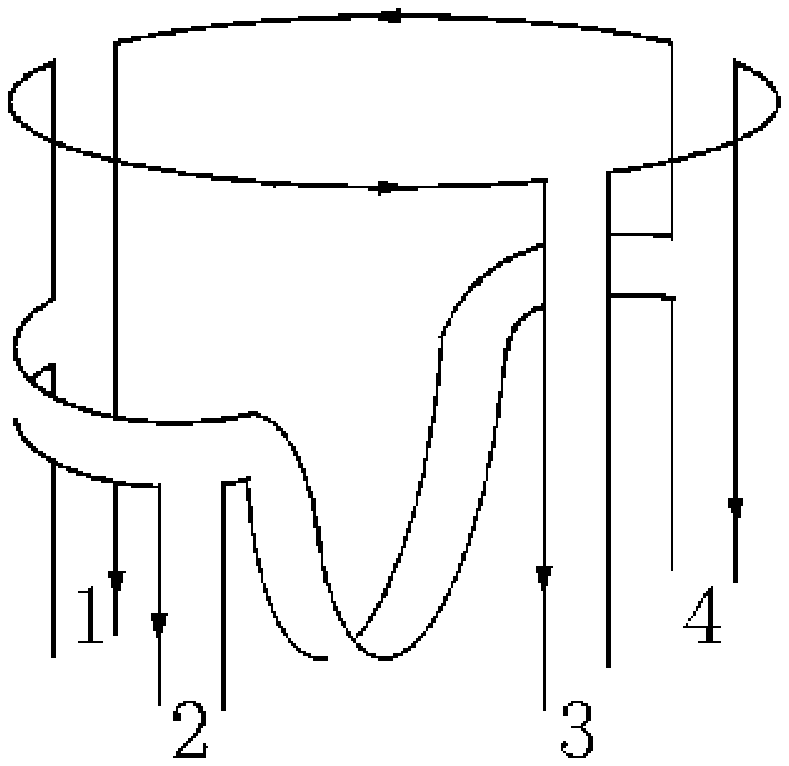}}
  + 
  \parbox{2cm}{ \includegraphics[width=2cm]{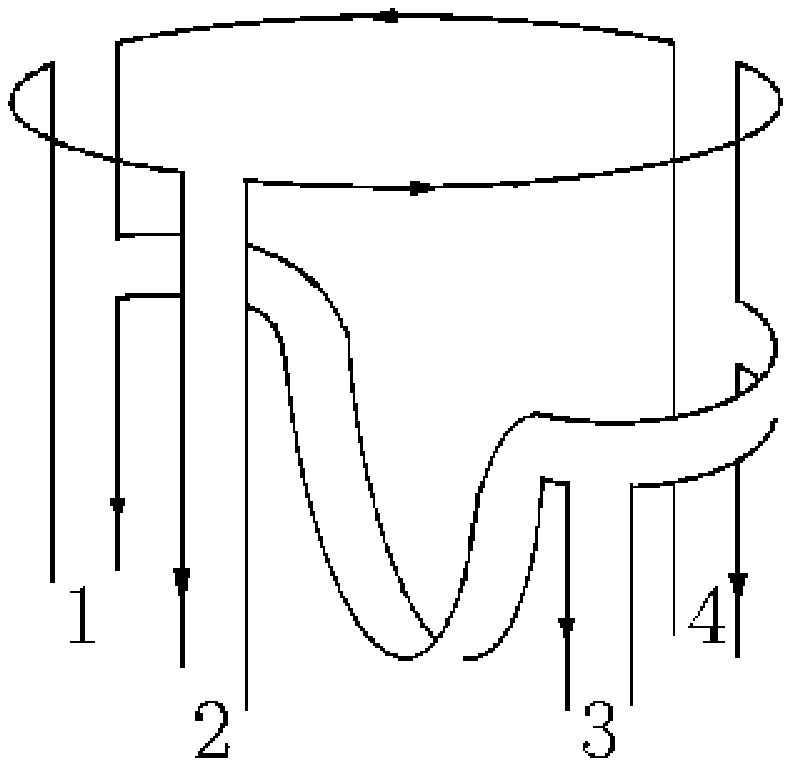}}
  + 
  \parbox{2cm}{ \includegraphics[width=2cm]{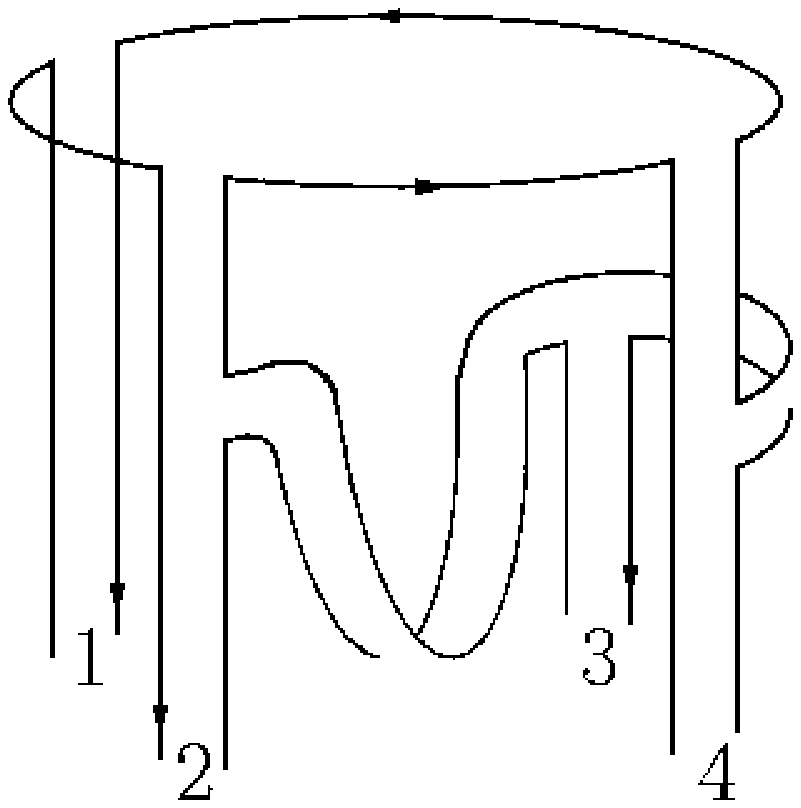}}
\to  {\bar g}^3 N_c \sum K_{2 \to 3}  \otimes \mathcal{D}_3^{(123)}(\omega),
 \end{align}
where, on the rhs, the sum is over the four possible two-to-three 
transitions, allowed by the triple discontinuity. 
Again, in the second and the third terms  
a minus sign originating from the color factor has to be included.
 Contributions with $\mathcal{D}_3^{(321)}(\omega) $ are
\begin{align}
  \label{eq:d4pop_23_321}
  \parbox{2cm}{ \includegraphics[width=2cm]{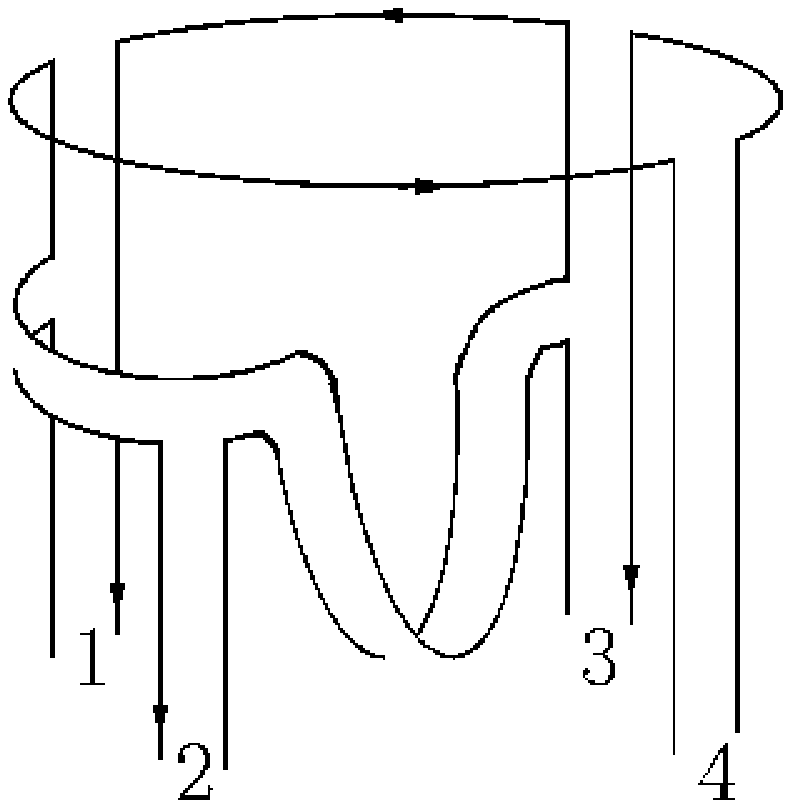}}
  +
  \parbox{2cm}{ \includegraphics[width=2cm]{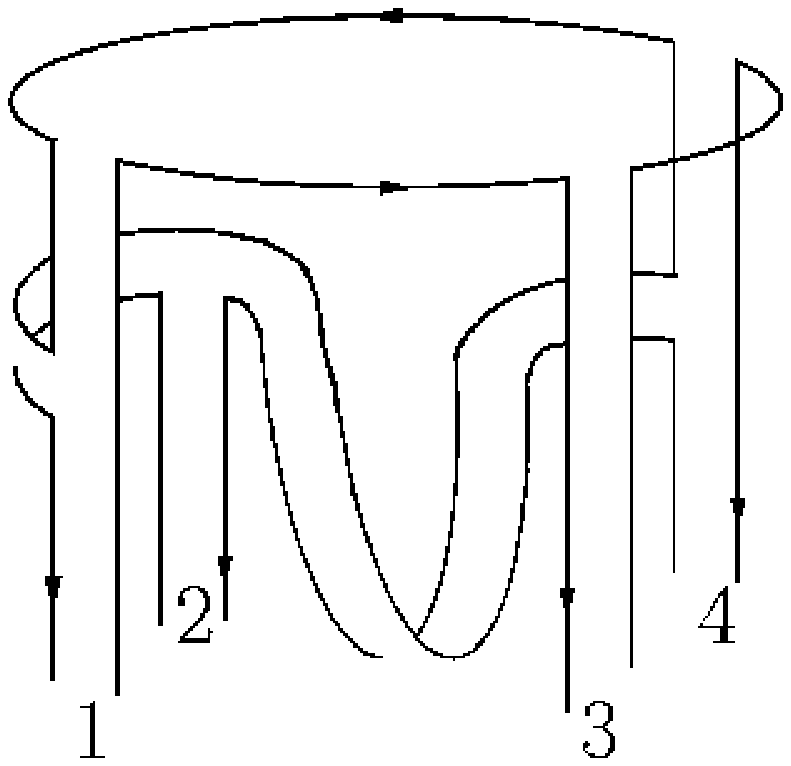}}
  + 
  \parbox{2cm}{ \includegraphics[width=2cm]{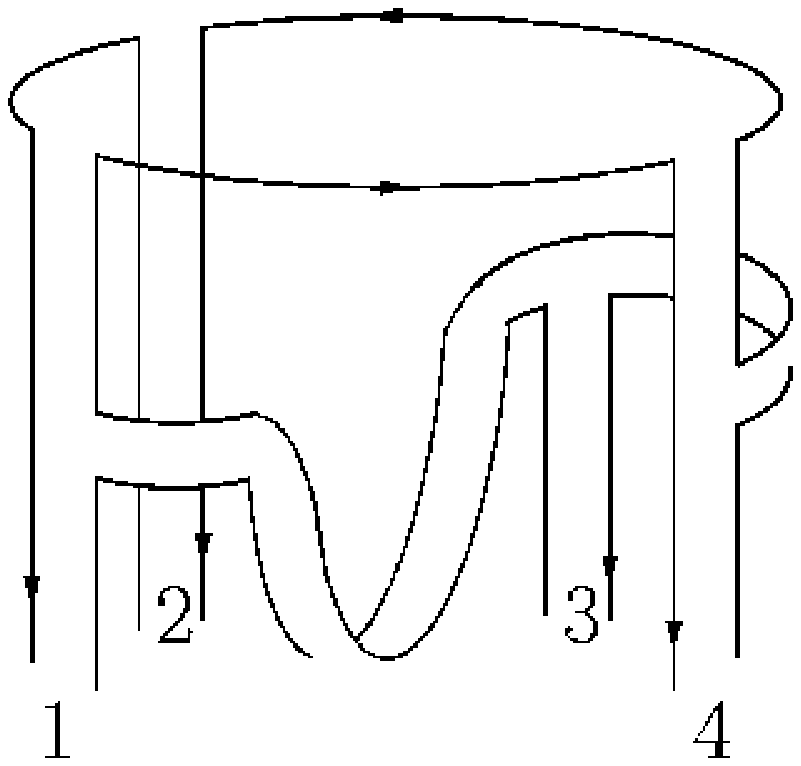}}
  + 
  \parbox{2cm}{ \includegraphics[width=2cm]{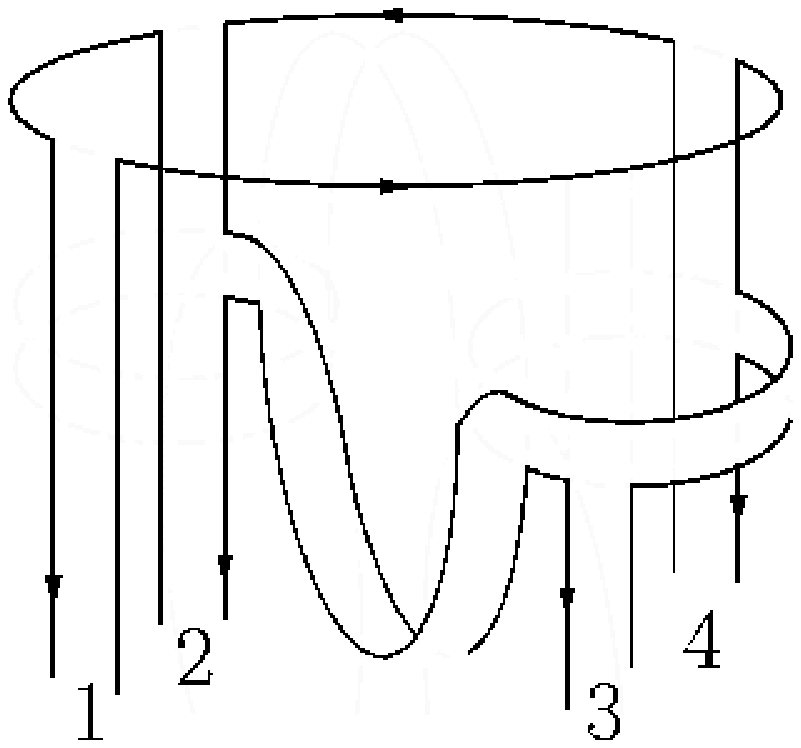}}
\to
 {\bar g}^3(-N_c)  \sum K_{2 \to 3}  \otimes \mathcal{D}_3^{(321)}(\omega),
 \end{align}
 where the sum is again over all four possible two-to-three
 transitions and for the second and third term a relative minus sign has
 to be included. Finally, from the two-to-four transition 
 (Fig.\ref{fig:non_planar_omega}c) we have two contributions.  They
 are given by
\begin{align}
  \label{eq:d4pop_24}
  \parbox{2cm}{\includegraphics[width=2cm]{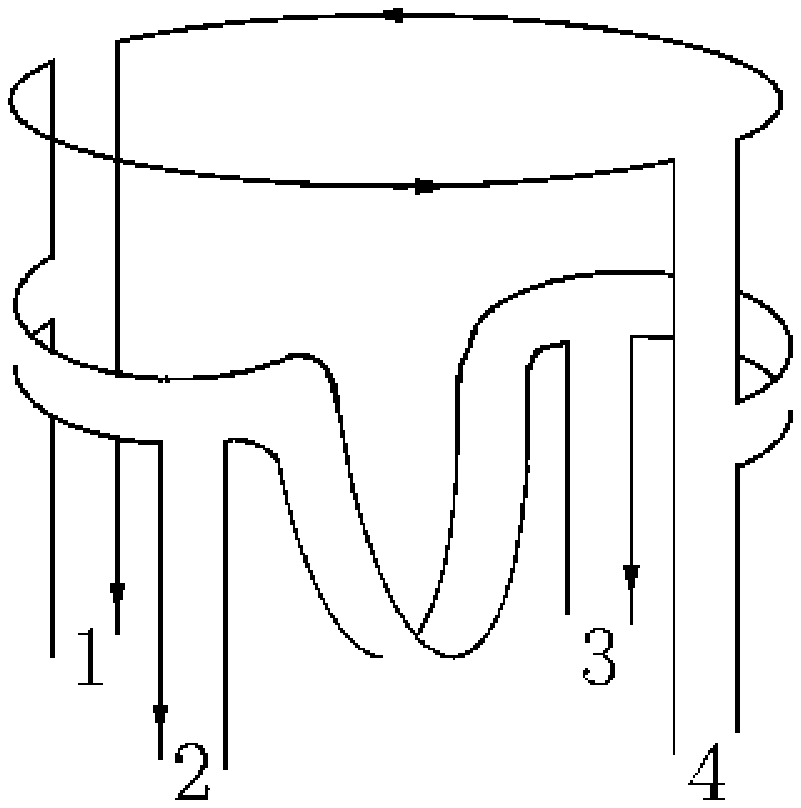}}
  +
  \parbox{2cm}{\includegraphics[width=2cm]{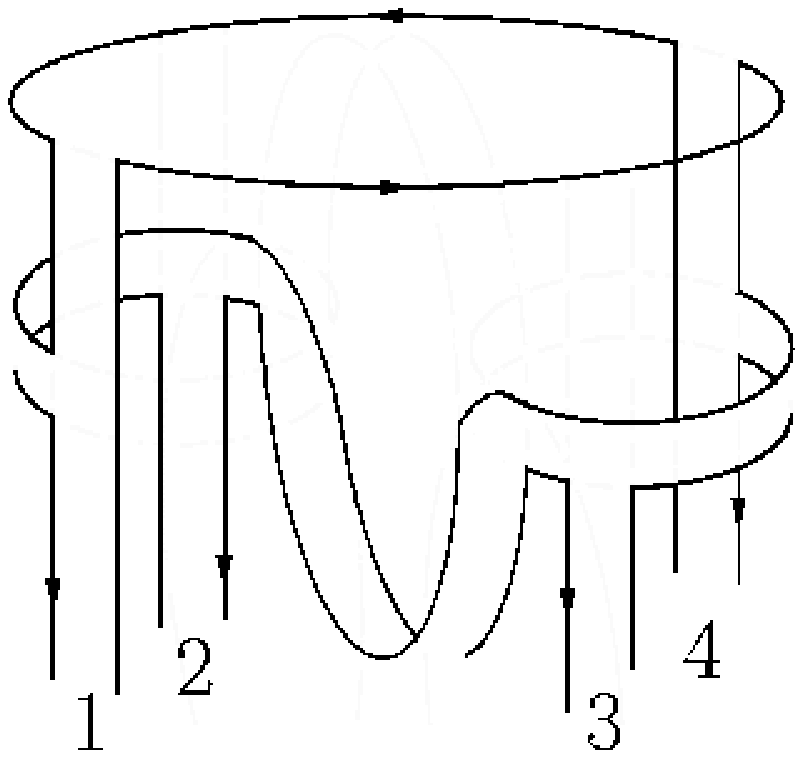}}
  \to
  2{\bar{g}^4}{N_c} K_{2 \to 4} \otimes \mathcal{D}_2(\omega).
\end{align}

Combining these three groups and including the Reggeon-propagators we obtain 
for the amplitude $\mathcal{D}^{\text{NP}}_{4}$:
\begin{align}
  \label{eq:d4pop_inteq}
 (\omega - \sum_i^4 \beta({\bf{k}}_i)  \mathcal{D}^{\text{NP}}_{4} (\omega|  {\bf k}_1,  {\bf k}_2, {\bf k}_3, {\bf k}_4)
&=
        {\bar g}^3 N_c   \sum K_{2 \to 3}  \otimes \mathcal{D}_3^{(123)}(\omega)
        +
        {\bar g}^3 (-N_c)  \sum K_{2 \to 3}  \otimes \mathcal{D}_3^{(321)}(\omega) 
        \notag \\
        +
        2{\bar{g}^4}{N_c}  K_{2 \to 4} \otimes \mathcal{D}_2(\omega)
        +
         \bar{g}^2N_c \bigg[ &
          \sum_{\substack{(12), (13), \\(24), (34) }}  K_{2 \to 2} \otimes \mathcal{D}_4^{(1234)}(\omega) 
           + 
           \sum_{\substack{(12), (13), \\(24), (34) }}  K_{2 \to 2} \otimes \mathcal{D}_4^{(2143)}(\omega)
          \notag \\
           &+
           \sum_{\substack{(12), (23), \\(14), (34) }}  K_{2 \to 2} \otimes \mathcal{D}_4^{(2134)}(\omega)
           +
           \sum_{\substack{(12), (23), \\(14), (34) }}  K_{2 \to 2} \otimes \mathcal{D}_4^{(1243)}(\omega)\bigg].
\end{align}
In the next step we make use of the fact that, according to the results of
Sec.\ref{sec:regg_d3} and Sec.\ref{sec:regg_d4}, all planar three and four
gluon partial amplitudes on the rhs, $\mathcal{D}_3$  and $\mathcal{D}_4$, 
can be written as superpositions of two gluon amplitudes
$\mathcal{D}_2(\omega)$. As a result, the right hand side of
Eq.(\ref{eq:d4pop_inteq}) depends only on
$\mathcal{D}_2(\omega)$, convoluted with the sum of transition kernels:
\begin{align}
  \label{eq:d4pop_inteq_vertex}
(\omega - \sum_{i = 1}^4 \beta({\bf k}_i)) \mathcal{D}_4^{\text{NP}}(\omega|{\bf k}_1,{\bf k}_2,{\bf k}_3,{\bf k}_4)
=&
\frac{\bar{\lambda}^2}{N_c} V_{\text{TPV}} \otimes \mathcal{D}_2(\omega) .
\end{align}
The right hand side of Eq.(\ref{eq:d4pop_inteq_vertex}) defines 
the Triple-Pomeron-Vertex $ V_{\text{TPV}} $ on the pairs-of-pants.
It describes the coupling of the upper BFKL Pomeron, $\mathcal{D}_2(\omega)$, 
to the lower ones, $\mathcal{D}_2(\omega_1)$ and $\mathcal{D}_2(\omega_2)$.

The function $V_{\text{TPV}}$ 
coincides with the $V$ function found in \cite{Bartels:1994jj}.  
In this paper, a finite $N_c$ calculation with $N_c = 3$ has been carried out,
and the following result for the $2 \to 4$ gluon vertex has been obtained:
\begin{align}
  \label{eq:v24_bartelswuesthoff}
V_{2 \to 4}^{a_1a_2a_3a_4}  ({\bf l}_1, {\bf l}_2 | {\bf k}_1,{\bf k}_2,{\bf k}_3,{\bf k}_4) )
 & = 
\delta^{a_1a_2}\delta^{a_3a_4} V ({\bf l}_1, {\bf l}_2 |{\bf k}_1,{\bf k}_2;{\bf k}_3,{\bf k}_4) )  \notag \\
+\, \delta^{a_1a_3}\delta^{a_2a_4}&  V ({\bf l}_1, {\bf l}_2 |{\bf k}_1,{\bf k}_3;{\bf k}_2,{\bf k}_4) ) + \delta^{a_1a_4}\delta^{a_2a_3} V ({\bf l}_1, {\bf l}_2 |{\bf k}_1,{\bf k}_4;{\bf k}_2,{\bf k}_3) .
\end{align}
Here $a_i$, $i = 1, \ldots 4$ denote color indices in the adjoint
representation of the $t$-channel gluons. To compare with our result,
we use the finite $N_c$ version of Eq.(\ref{eq:v24_bartelswuesthoff})
in \cite{Bartels:1999aw} which has been obtained for  arbitrary $N_c$ and we find
\begin{align}
  \label{eq:pp_barwue}
\frac{\bar{\lambda}^2}{N_c} V_{\text{TPV}}({\bf l}_1, {\bf l}_2 |{\bf k}_1,{\bf k}_2;{\bf k}_3,{\bf k}_4) ) = N_c V ({\bf l}_1, {\bf l}_2 |{\bf k}_1,{\bf k}_2;{\bf k}_3,{\bf k}_4) ).
\end{align}
It is easy to verify that this result is in complete agreement with
the large large-$N_c$ limit of \cite{Bartels:1994jj,Bartels:1999aw}.
Namely, if we project, in Eq.(\ref{eq:v24_bartelswuesthoff}), on the
color single state of the gluon-pairs (12)and (34) and then consider
the limit of large $N_c$, only the first term on the right hand side
of Eq.(\ref{eq:v24_bartelswuesthoff}) contributes, and we indeed
obtain agreement of the two results  as required. In \cite{Bartels:1995kf} it has been
shown that the vertex $V$ is invariant under M\"obius transformations.
Furthermore, $V$ can be expressed in terms of another function $G({\bf
  k}_1,{\bf k}_2,{\bf k}_3)$:
\begin{align}
  \label{eq:VbyG}
  V  ( {\bf k}_1,{\bf k}_2,{\bf k}_3,{\bf k}_4) =& \bar{\lambda} \big[
G(1,23,4) + G(2,13,4) + G(1, 24, 3) + G(2,14,3) - G(12,3,4)
\notag \\
&
 - G(12,4,3) - G(1,2,34) - G(2,1,34) + G(12, -,34)
\big],
\end{align}
where we suppressed the dependence on the momenta ${\bf l}_1$ and
${\bf l}_2$. This function $G$ has been first introduced for the
forward case and $N_c=3$ in \cite{Bartels:1994jj}, whereas the above
version was introduced in \cite{Braun:1997nu}. It has the nice property
that it is not only infra-red finite but also by itself invariant
under M\"obius transformations \cite{Braun:1997nu}.

\section{The six-point amplitude on the pair-of-pants surface}
\label{sec:6point_res}
\begin{figure}[htbp]
  \centering
  \parbox{6cm}{\includegraphics[height=4.5cm]{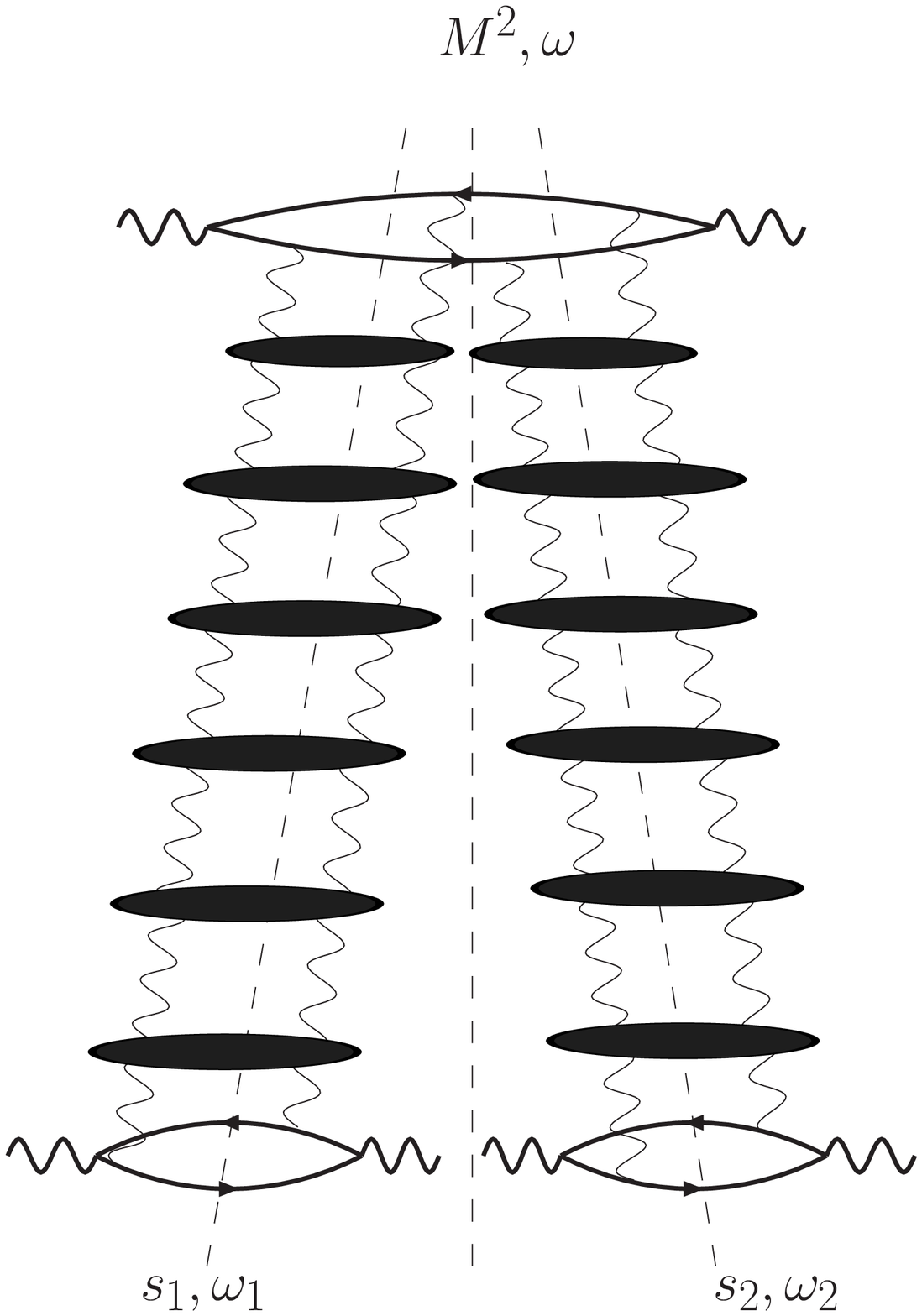}}
 \parbox{4cm}{\includegraphics[height = 3.2cm]{trouser_born.eps}}
  \caption{\small Diagrams where the  two Pomerons couple directly to the  upper quark-loop. 
They contain the lowest order diagram in Fig.\ref{fig:trouser}. The color factors  
can be reduced to the color factor to the right and therefore belong to the planar class.}
  \label{fig:quark_branching}
\end{figure}

With the results of the previous sections, we have almost all
constituents that are needed to build the triple Regge limit of the
six-point amplitude with the pair-of-pants topology.  However before
we add together the amplitudes of the foregoing sections, we need to
take care of a peculiarity inside the planar piece of the upper amplitude. 
To leading order in $\lambda$, the two lower Pomerons couple
directly to the upper quark-loop, as depicted in
Fig.\ref{fig:quark_branching} and the branching vertex is not given by
a real gluon, but by the quark-loop itself. As long as we have, in addition 
to the $q\bar{q}$ pair, one (or more) $s$-channel gluons that contribute to 
the $M^2$ discontinuity, the mass of the $q\bar{q}$-system is integrated over,
and this integration enters into the definition of the impact factors 
$\mathcal{D}_{4;0}^{(\text{ijkl})}$. Without such gluons, the mass 
of the $q\bar{q}$ pair coincides with the diffractive mass $M$ which
is a fixed external parameter. In this case, the coupling of the $t$-channel four gluon system
to the quark-loop is hence given by the triple discontinuity of the
quark loop without the integration over the diffractive mass of the 
$q\bar{q}$-system, and defines an 'unintegrated' four
gluon impact factor, $\bar{\mathcal{D}}_{(4;0)}(M^2)$. Its precise analytic form 
can be obtained from Eq.(2.8) of \cite{Bartels:1994jj}.
Correspondingly, its Mellin transform, which enters the partial wave $F(\omega, \omega_1,
\omega_2)$ will be denoted by $\bar{\mathcal{D}}_{(4;0)}(\omega)$.
When $\bar{\mathcal{D}}_{(4;0)}(M^2)$ is integrated over the squared 
diffractive mass $M^2$, it coincides 
with sum over the impact factors $\mathcal{D}_{4;0}^{(\text{ijkl)}}$. 
 
\begin{figure}[htbp]
  \centering
 \parbox{7cm}{\center \includegraphics[height=4.5cm]{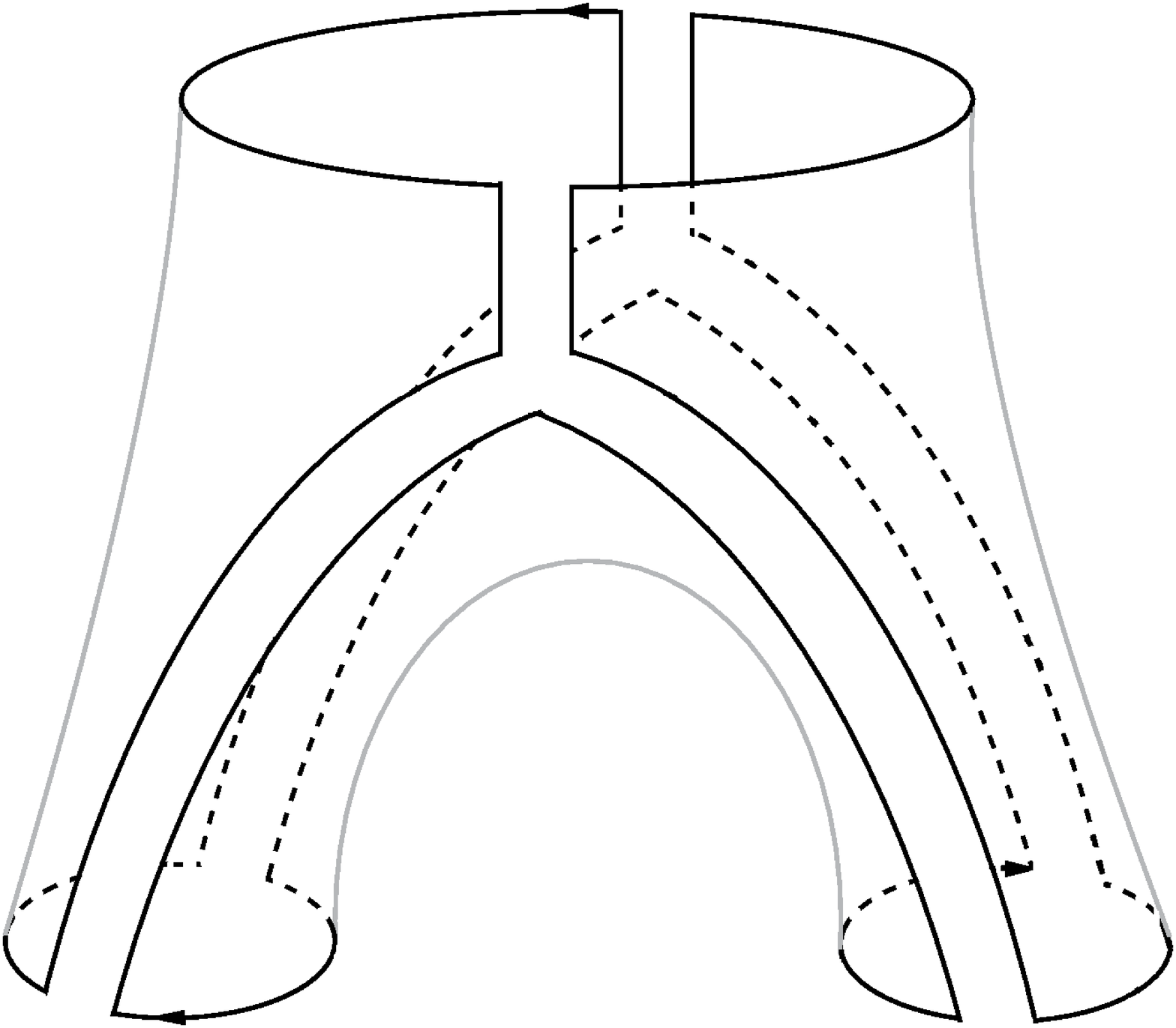}}
 \parbox{7cm}{\center \includegraphics[height=4.5cm]{vpp.eps}}\\
\parbox{7cm}{\center (a)}\parbox{7cm}{\center (b)}
  \caption{\small (a) Planar color graphs  associated with the decay of two reggeized gluons  and  (b) non-planar color graphs which lead to the Triple-Pomeron-Vertex. In both cases a two gluon state decays into a four gluon state.}
  \label{fig:vpp}
\end{figure}

After this remark we write the final result as the sum of two terms:
\begin{align}
  \label{eq:F_sum}
F(\omega, \omega_1, \omega_2)  = F^{(\text{P})}(\omega, \omega_1, \omega_2) 
+  F^{(\text{NP})}(\omega, \omega_1, \omega_2),
\end{align}
where the first term sums the planar diagrams (including  the 
'unintegrated' impact factors), the second term the
non-planar ones. When doing the convolution of the three amplitudes,
special care has to be taken of the counting of interactions inside
the gluon pairs (12) and (34) just below the branching point where the
upper cylinder splits into two cylinders (c.f. the discussion given in
Sec.\ref{sec:below}).  This issue has been addressed, for the
'$N_c$-finite' case, in Sec.4 of \cite{Bartels:1994jj} (see, in
particular Eqs.(4.14) and Eq.(4.15)), and in the following we shall
apply the same line of arguments. We identify, for the $M^2$
discontinuity, the lowest $s$-channel intermediate state which leads
to the 'last' interaction between the two gluon pairs (12) and (34):
below this interaction, the upper cylinder has split into two separate
cylinders.  In particular, this last interaction cannot consist of a
two-to-two kernel acting on the gluon pairs (12) or (34). Since, in
Secs.\ref{sec:inteq_d4} and \ref{sec:trip_pom}, we have defined the
functions $\mathcal{D}_{4}^{(\text{ijkl})}$ and
$\mathcal{D}_{4}^{(\text{NP})}$ in such a way that the last
interaction includes these contributions, we first have to remove
them. Furthermore, as explained above, the coupling of the
$\mathcal{D}^{(12)}_2(\omega_1) $ and $
\mathcal{D}^{(34)}_2(\omega_2)$ directly to the upper quark-loop is
described by $\bar{\mathcal{D}}_{4;0}(\omega)$, and we write this term separately. We therefore arrive at  the full
partial wave in the following form:
\begin{align}
  \label{eq:F_factor}
F(\omega, \omega_1, \omega_2) 
&=
4  \mathcal{D}^{(12)}_2(\omega_1) \otimes_{12} \mathcal{D}^{(34)}_2(\omega_2)\otimes_{34}\notag \\
 &\bigg(  \bar{\mathcal{D}}_{(4;0)}(\omega)
+
\big[\omega - \sum_{i = 1}^4 \beta({\bf k}_i)\big]  \!\!\!\!  
\sum_{(\text{ijkl}) = \substack{(1234), (2143), \\(2134), (1243) }}  \!\!\!\! 
\mathcal{D}_{4}^{(\text{ijkl})}(\omega)
    + \frac{{\bar{\lambda}}^2}{N_c} V_{\text{TPV}} \otimes \mathcal{D}_2(\omega)
\notag \\ 
& - 2{\bar{\lambda}}(  K^{(12)}_{2 \to 2} +  K^{(34)}_{2 \to 2} ) 
\otimes  \!\!\!\!  
\sum_{(\text{ijkl}) = \substack{(1234), (2143), \\(2134), (1243) }}  \!\!\!\! 
\mathcal{D}_{4}^{(\text{ijkl})}(\omega) 
 \bigg) 
\end{align}
In the first line, the indices of the convolution symbols indicate that
$\mathcal{D}^{(12)}_2(\omega_1)$ and $ \mathcal{D}^{(34)}_2(\omega_2)$
are to be contracted with the gluons (12) and (34), resp. 
The last line  subtracts the two-to-two kernels acting on gluon
pairs (12) and (34).
 Using the BFKL-equation 
$$ 2\bar{\lambda}{K}_{2 \to 2}^{(12)}
\mathcal{D}^{(12)}_2(\omega_1) = (\omega_1 - \beta({\bf k}_1) -
\beta({\bf k}_2) ) \mathcal{D}^{(12)}_2(\omega_1) -
\mathcal{D}^{(12)}_{(2;0)} $$ 
(and a similar expression for the gluon pair (34)),
and making further use of Eq.(\ref{eq:d4_planar_vertex}), we
arrive at
\begin{align}
  \label{eq:F_factor_2}
& F(\omega, \omega_1, \omega_2) =
\notag \\
& \quad
 4 \bigg\{ \mathcal{D}^{(12)}_2(\omega_1) \otimes_{12} \!\mathcal{D}^{(34)}_2(\omega_2)\otimes_{34}\!
 \bigg( 
 \bar{\mathcal{D}}_{(4;0)}(\omega)
+
\big[\omega \!-\! \omega_1 \!-\!\omega_2 \big]  V^{\text{R}}\!\otimes\! \mathcal{D}_2(\omega)
   + \frac{\bar{\lambda}^2}{N_c} V_{\text{TPV}} \! \otimes \!\mathcal{D}_2(\omega) \bigg) \notag \\ 
& 
\qquad
 + \mathcal{D}^{(12)}_2(\omega_1) \otimes_{12} \mathcal{D}^{(34)}_{(2;0)}\otimes_{34}
V^{\text{R}}\otimes \mathcal{D}_2(\omega)
+
\mathcal{D}^{(12)}_{(2;0)} \otimes_{12} \mathcal{D}^{(34)}_{2}(\omega_2)\otimes_{34}
V^{\text{R}}\otimes \mathcal{D}_2(\omega)
\bigg\}.
\end{align}
The terms in the last  line are either independent of $\omega_2$ or
$\omega_1$ and the integrals over  $\omega_2$
or $\omega_1$ vanish in Eq.(\ref{tripleregge}).  We therefore drop
these terms and obtain for the two partial waves $F^{(\text{P})}(\omega,
\omega_1, \omega_2)$ and $F^{(\text{NP})}(\omega, \omega_1, \omega_2)$ the
following results:
\begin{align}
  \label{eq:F_factor_3}
F^{(\text{P})}(\omega, \omega_1, \omega_2) 
&= 4 
 \mathcal{D}^{(12)}_2(\omega_1) \otimes_{12} \mathcal{D}^{(34)}_2(\omega_2)\otimes_{34}\bigg( \bar{\mathcal{D}}_{(4;0)}(\omega) +
   \big[\omega - \omega_1 -\omega_2 \big] \frac{\bar{\lambda}}{N_c} V^R \otimes
 \mathcal{D}_{2}(\omega)  \bigg) 
\end{align}
and
\begin{align}
  \label{eq:F_nonplanar}
F^{(\text{NP})}(\omega, \omega_1, \omega_2) 
&=
4  \mathcal{D}^{(12)}_2(\omega_1) \otimes_{12} \mathcal{D}^{(34)}_2(\omega_2)\otimes_{34}
   \frac{\bar{\lambda}^2}{N_c} V_{(\text{TPV})}  \otimes \mathcal{D}_2(\omega).
\end{align}
Both terms are illustrated in Figs.\ref{fig:vpp}a and \ref{fig:vpp}b. 

\section{Conclusion}
\label{sec:concl}
In the present analysis we have considered, in the large-$N_c$ limit,
the triple Regge-limit of the scattering of three virtual photons. Our
emphasis has been on the topology of color factors: we have summed, in
the generalized leading-log approximation, only those diagrams which
fit onto the pair-of-pants surfaces. These diagrams group themselves
naturally into two classes (Fig.\ref{fig:vpp}). The first class
consists of all diagrams which, by contracting closed color loops,
coincide with one of the lowest order graphs (of order $g^8 N_c^3$)
illustrated in Fig.13. Making use of the bootstrap conditions, the sum
of these graphs is shown to reduce to reggeized gluons with a simple
splitting vertex. In addition, starting at the order $g^8 N_c^3 (g^2
N_c)$, a second class of diagrams appears which, by contracting closed
color loops, cannot be drawn as one of the lowest order graphs shown
in Fig.13.  The sum of these graphs can be written as a convolution of
three BFKL amplitudes, connected by the triple Pomeron vertex found in
earlier papers.  This triple Pomeron vertex has been shown to be
invariant under M\"obius transformation \cite{Bartels:1995kf}.

The analytic expression for the six-point amplitude derived in this
paper coincides with the large-$N_c$ limit of the QCD result in
\cite{Bartels:1994jj}. However, the analysis of the present paper
provides another interpretation of gluon reggeization and the
appearance of the M\"obius invariant triple Pomeron vertex: on the
pair-of-pant surface, the reggeizing pieces are completely planar,
whereas the the triple Pomeron vertex belongs to a distinct class of
color diagrams which reflect the non-planar Mandelstam cross.

We believe that the topological approach pursued in this paper is
well-suited for studying the Regge limit of N=4 Super Yang Mills
Theory within the AdS/CFT correspondence.  On the string side,
amplitudes are naturally expanded in terms of topologies. In
particular, the pair-of-pants topology studied in this paper is the
same as that of the vertex which describes the coupling of three
closed strings.

For a systematic study of the AdS/CFT correspondence it is convenient to
make use of $R$-currents: they allow to formulate current correlators
which are well-defined both on the gauge theory side and on the string
side. As a first step of investigating, in the Regge limit, the
duality between N=4 Super Yang Mills Theory and $AdS_5$ string theory,
the 4-point function for such currents has been studied on the gauge
theory side in \cite{Bartels:2008zy}.  This study  confirms that the interaction between two $R$-currents in the
Regge-limit is, indeed, mediated by a BFKL-Pomeron, and it contains the 
two gluon impact factors in N=4 SYM. On the string
side, the scattering of two R-currents has been shown to involve, in
lowest order, the one-graviton exchange \cite{bkms}.  In the present
paper we have addressed, on the gauge theory side, the next term in
the topological expansion. As a start, we have restricted ourselves to
nonsupersymmetric QCD($N_c$). The generalization to the supersymmetric
case where, inside the impact factors, fermions and scalar particles
in the adjoint representation have to be considered, will be presented
elsewhere \cite{behm}.  On the string side, the analogous 6-point
R-current correlator is expected to contain the triple graviton
vertex. This project is currently being studied.

\subsubsection*{Acknowledgements}
We thank L.N.~Lipatov for helpful discussions. M.H. is grateful for financial support from the Graduiertenkolleg 
"Zuk\"unftige Entwicklungen in der Teilchenphysik" and from DESY.

\end{document}